\documentclass[
aps,pra,twocolumn,floatfix,superscriptaddress
]{revtex4-2}

\usepackage[utf8]{inputenc}
\usepackage{multirow}
\usepackage{graphicx}
\usepackage{hyperref}
\usepackage[english]{babel}
\usepackage{braket}
\usepackage{amsmath}
\usepackage{graphicx}
\usepackage{siunitx}
\usepackage{nicefrac}
\usepackage[rgb]{xcolor} 
\usepackage{upgreek}
\usepackage{longtable}
\usepackage{multirow}
\usepackage{cleveref}
\crefname{table}{}{}

\usepackage{siunitx}
\usepackage{xfrac}
\usepackage{xr}

\usepackage{rotating}
\usepackage{dcolumn}
\usepackage{silence}

\makeatletter
\newcommand*{\addFileDependency}[1]{
  \typeout{(#1)}
  \@addtofilelist{#1}
  \IfFileExists{#1}{}{\typeout{No file #1.}}
}
\makeatother


\renewcommand{\thetable}{S\arabic{table}}

\begin{document}

\title{Microwave spectroscopy and multi-channel quantum defect analysis of ytterbium Rydberg states}

\author{Rin Kuroda}
\email{These authors contributed equally to this work.}
\affiliation{Department of Electrical and Computer Engineering, Princeton University, Princeton, NJ 08544, USA}
\affiliation{Department of Physics, Princeton University, Princeton, NJ 08544, USA}
\author{Vernon M. Hughes}
\email{These authors contributed equally to this work.}
\affiliation{Department of Electrical and Computer Engineering, Princeton University, Princeton, NJ 08544, USA}
\affiliation{Department of Physics, Princeton University, Princeton, NJ 08544, USA}
\author{Martin Poitrinal}
\affiliation{Department of Electrical and Computer Engineering, Princeton University, Princeton, NJ 08544, USA}
\author{Michael Peper}
\email{michael.peper@princeton.edu}
\affiliation{Department of Electrical and Computer Engineering, Princeton University, Princeton, NJ 08544, USA}
\author{Jeff D. Thompson}
\email{jdthompson@princeton.edu}
\affiliation{Department of Electrical and Computer Engineering, Princeton University, Princeton, NJ 08544, USA}

\date{\today}

\begin{abstract}
The complex Rydberg structure of ytterbium atoms is shaped by multiple low-lying ion-core-excited states and strong channel interactions, which presents both opportunities and challenges for quantum information processing and precision metrology.
In this work, we extend high-resolution microwave spectroscopy and multichannel quantum defect theory (MQDT) modeling of singly excited $6sn\ell$ Rydberg states in $^{174}$Yb and $^{171}$Yb to include the $\ell = 3$ ($f$) and $\ell = 4$ ($g$) series. Our measurements reveal $p$-$f$ mixing in odd-parity Rydberg states of $^{171}$Yb, which we incorporate by combined MQDT models for $6snp$ and $6snf$ series. Additionally, we observe that for $\ell = 4$ the spin-orbit interaction dominates over the exchange interaction, such that the $6sng$ states are more accurately described in a $jj$-coupled basis. We validate our models by comparing the predicted Land\'e $g$-factors and static dipole polarizabilities with experimental measurements, finding excellent agreement. These results provide important input for designing high-fidelity entangling gates with ytterbium atoms.
\end{abstract}

\maketitle

\section{\label{sec:level1}Introduction}

Ytterbium (Yb) atoms provide a powerful and versatile platform for quantum computing and simulation~\cite{Jenkins2022Ytterbium,Ma2023High,muniz2025high,nakamura2024hybrid,senoo2025high}, and precision metrology~\cite{boulder2021frequency,bothwell2025lattice,niroula2024quantum} due to their unique electronic structure and favorable atomic properties. In particular, the fermionic isotope 
$^{171}$Yb, with nuclear spin $I=1/2$, supports high-fidelity qubits using either the ground-state nuclear spin or the metastable $6s6p\,^3P_0$ state.

Encoding the qubit in the metastable $^3P_0$ state allows for midcircuit operations on atoms in the ground state, enabling efficient erasure error conversion~\cite{Wu2022Erasure,Lis2023Midcircuit,Ma2023High,zhang2025leveraging}. Combined with continuous coherent atom replacement~\cite{li2025fast} and fast, high-fidelity entanglement via single-photon Rydberg excitation~\cite{peper2025spectroscopy,muniz2025high,senoo2025high} ytterbium atoms represent a promising candidate for scalable quantum technologies.

However, the energies and wavefunctions of Rydberg states in divalent atoms are modified by interactions with low-lying doubly excited \textit{perturber} states~\cite{Armstrong1977Bound,Esherick1977Bound,Aymar1978Two}. Furthermore, in isotopes with non-zero nuclear spin, hyperfine coupling in the ionic core leads to additional mixing and level splittings within the Rydberg series~\cite{Rinneberg1982Resonance,Rinneberg1983Hyperfine,Beigang1983Hyperfine}.
These effects can be accurately captured through precision spectroscopy and modeling using multichannel quantum defect theory (MQDT)~\cite{Seaton1966Quantum,Fano1970Quantum,Lee1973Spectroscopy,Cooke1985Multichannel}.

Previous studies of singly-excited $6sn\ell$ Rydberg states in ytterbium have reported low-$\ell$ MQDT models to reproduce the observed energy spectra~\cite{Camus1969spectre,Aymar1980Highly,Aymar1984three,Majewski1985diploma,BiRu1991The,Maeda1992Optical,Ali1999Two,Lehec2017PhD,Lehec2018Laser,Lehec2017PhD,peper2025spectroscopy}. A comprehensive study in Ref.~\cite{peper2025spectroscopy} unified and refined the models for Rydberg series with $\ell\le 2$, enabling the accurate prediction of static dipole polarizablities and interaction potentials of $6sns$ Rydberg states, leading to improved Rydberg gate fidelities. There have been several studies of higher angular momentum Rydberg series ($\ell\geq3$) in $^{174}$Yb~\cite{Ali1999Two,Aymar1984three,Lehec2017PhD,Niyaz2019Microwave}. However, many series with $\ell\geq3$ have not yet been measured, and there are no measurements of states with $\ell\ge3$ in $^{171}$Yb. Completing this picture would enable the prediction of properties of $6snd$ Rydberg states such as static dipole polarizabilities and Rydberg-Rydberg interactions. 

In this article, we present three main results. First, in Sec.~\ref{sec:174Yb_Fstates} and Sec.~\ref{sec:171Fstates} we extend the comprehensive microwave spectroscopy of $^{174}$Yb and $^{171}$Yb to $\ell = 3$ Rydberg series. We validate the accuracy of the presented models by comparison to measurements of $g$-factors. We observe that the $6snp$ and $6snf$ Rydberg series in $^{171}$Yb are mixed for $F = 3/2$ and $F = 5/2$ due to configuration interactions, and demonstrate that combined MQDT models for these states resolve a previously reported deviation between the experimental and MQDT energies in the $6snp\, (F=3/2)$ series~\cite{peper2025spectroscopy}. 

Next, in Sec.~\ref{sec:174_Gstates} and Sec.~\ref{sec:171Gstates} we measure several previously unobserved $\ell = 4$ series in both $^{174}$Yb and $^{171}$Yb and develop MQDT models for these series. We find that for $\ell =4 $ the spin-orbit interaction dominates over the exchange interaction, such that the $6sng$ series are best described in $jj$ instead of $LS$ coupling, unlike the $\ell\le 3$ states.

Finally, in Sec.~\ref{SEC: Dpol}, we validate the MQDT energies and wavefunction predictions for the $6snp$ and $6snf$ series in $^{174}$Yb and $^{171}$Yb by measuring the static dipole polarizabilities of $6snd$ states. We find excellent agreement between the measured polarizabilities and those predicted by the model. This confirms the model accuracy, enabling the predictions of Rydberg–Rydberg interactions involving $6snd$ states, which we discuss in Sec.~\ref{sec:171int}.

\begin{figure}[tb]
    \includegraphics[width=\linewidth]{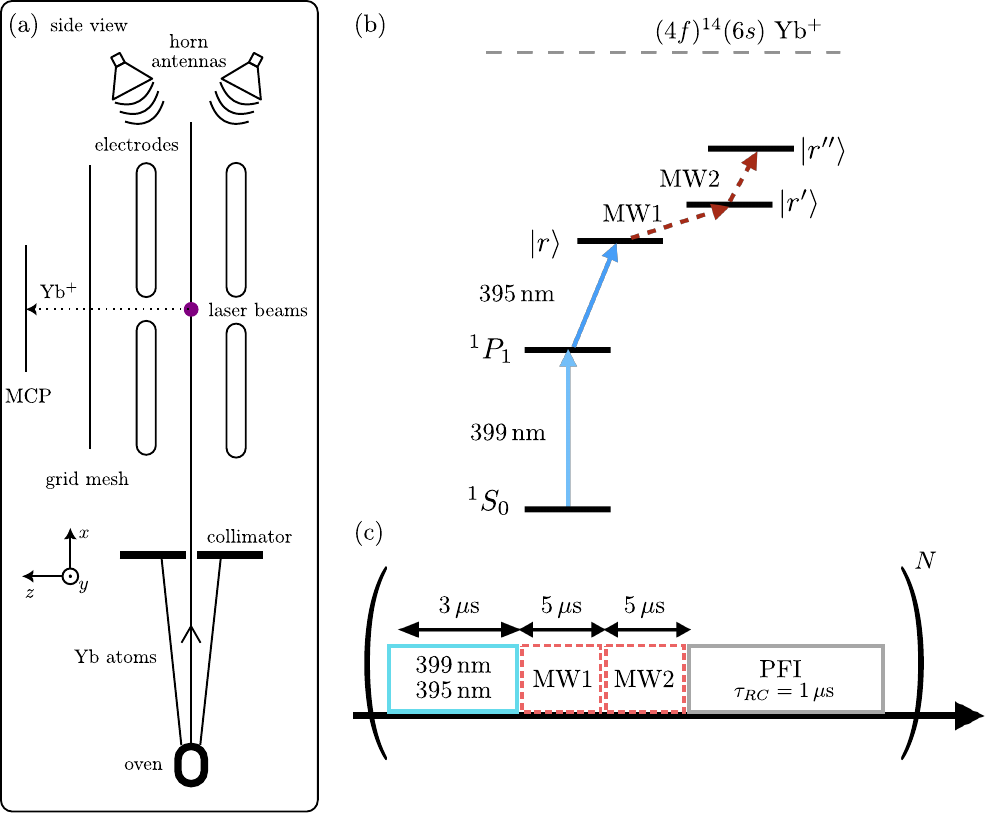}
    \caption{(a) Schematic diagram of the spectroscopy setup adapted from Ref.~\cite{peper2025spectroscopy}. A collimated atomic beam interacts with the laser beams and microwave pulses near the center of the vacuum chamber. The atoms are ionized by PFI and detected by the MCP. (b) Schematic level diagram depicting the typical transition scheme to probe Rydberg states of interest. Details described in the text. (c) Typical pulse sequence. The laser pulses are simultaneously applied for $\SI{3}{\mu s}$, followed by the microwave pulses for $\SI{5}{\mu s}$ each. Subsequently, the Rydberg atoms are ionized by PFI and the time-of-flight of the resulting ions to the MCP is recorded using an oscilloscope. The experimental sequence is repeated at a rate of 1000 cycles per second, with each data point averaged over typically $N = 1500$ repetitions.}
    \label{fig:setup}
\end{figure}

\section{Methods}
\subsection{Experimental setup}
\label{sec:expsetup}

The experimental setup and methods used in this work follow those of Ref.~\cite{peper2025spectroscopy}. A sketch of the experimental setup is presented in Fig.~\ref{fig:setup}(a). In summary, an atomic beam of Yb is generated from an effusive oven and collimated using a pinhole aperture. We excite to $6sns$ and $6snd$ Rydberg states via a two-photon transition through the intermediate $6s6p\,^1P_1$ state using counterpropagating laser beams at \SI{399}{\nm} and $\approx\SI{395}{\nm}$, generated by frequency-doubled Ti:sapphire lasers. The laser beams are applied perpendicular to the atomic beam to minimize Doppler shifts. Using the isotope shift of the $6s6p\,^1P_1$ transition, we selectively probe either $^{171}$Yb $(I=1/2)$ or $^{174}$Yb $(I=0)$ Rydberg states.

Rydberg population is detected via state-selective pulsed-field ionization (PFI)~\cite{Gallagher_1994} followed by time-resolved ion detection on a microchannel plate (MCP). To probe laser-inaccessible $6sn\ell$ Rydberg states, including those with $\ell = \{1,3,4\}$, transitions between Rydberg states are driven using microwaves from 20--\SI{170}{\GHz}, generated via frequency multiplication and applied through horn antennas. Laser and multi-step microwave pulses are applied consecutively in time. The microwave intensity is deliberately kept low to eliminate residual light shifts. A simplified level diagram showing the transition scheme and a typical pulse sequence of the spectroscopy experiment is shown in Fig.~\ref{fig:setup}(b) and (c). 

Static electric fields are controlled by a set of electrodes placed inside the vacuum chamber. Static magnetic fields are controlled by three pairs of Helmholtz coils. Residual electric fields and magnetic fields in the spectroscopy region are compensated as needed via dc Stark and Zeeman shift measurements. 

\subsection{MQDT modeling}
\label{sec:mqdtfitting}

We extend the MQDT framework described in Ref.~\cite{peper2025spectroscopy} to $6snf$ and $6sng$ Rydberg states of ytterbium. The $^{174}$Yb MQDT model parameters are determined as part of a global weighted optimization of all the $\ell\le4$ model parameters using previous and new spectroscopic data of $^{174}$Yb, as compiled in Ref.~\cite{peper2025spectroscopy} and App.~\ref{sec:specdata}, respectively. 

For $^{171}$Yb, we prepare an initial guess based on the $^{174}$Yb MQDT models and include the hyperfine interaction through a frame transformation \cite{Sun1988Hyperfine,Robicheaux2018theory,peper2025spectroscopy}. We then optimize the $^{171}$Yb MQDT parameters in a global weighted fit to all previous and new spectroscopic data of $^{171}$Yb $\ell \le 4$ states, as compiled in Ref.~\cite{peper2025spectroscopy} and App.~\ref{sec:specdata}, respectively.

\begin{figure}[tb] 
    \includegraphics[width=\linewidth]{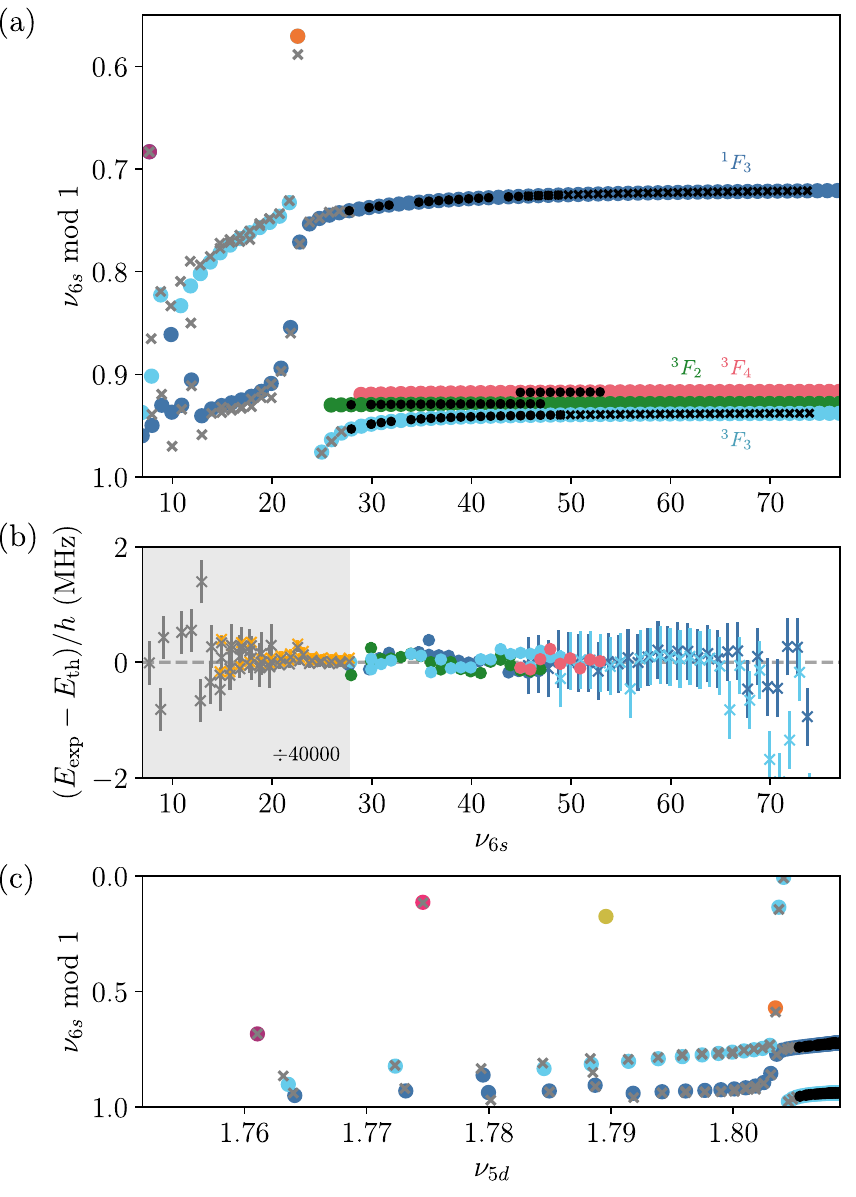}
    \caption{(a) Lu-Fano-type plot of $^{174}$Yb odd-parity $6snf$ Rydberg states.
    The colored circles correspond to MQDT model predictions. Black crosses indicate microwave spectroscopic data obtained in Ref.~\cite{Lehec2017PhD}. The black points are experimentally measured energies by microwave spectroscopy reported in this work. The gray crosses are laser spectroscopy data from Refs.~\cite{Aymar1984three,Ali1999Two}. The effect of a global offset of \SI{-2.4}{\GHz} to the term values reported in Ref.~\cite{Aymar1984three} is discussed in the text. The $6snf\,^{1,3}F_3$ series are mixed due to spin-orbit coupling, but the states are categorized based on their dominant character. (b) Deviation between experimental energies and the MQDT model predictions. Deviations of laser spectroscopy data (gray crosses) are divided by 40000, and the orange crosses are the deviations before the global offset was added. (c) Lu-Fano plot of the $6snf\, ^{1,3}F_3$ Rydberg states as a function of the effective principal quantum number $\nu_{5d}$ with respect to the $ 4f^{13}5d6s$ ionization threshold. The four non-blue-colored points denote the four perturbers discussed in Sec.~\ref{sec:174Yb_13F3}, as reported in Refs.~\cite{Aymar1984three, Ali1999Two}.
    }
    \label{fig: 174}
\end{figure}

\section{$^{174}$Yb $6snf$ States}
\label{sec:174Yb_Fstates}
In $^{174}$Yb, there are four distinct $6snf$ Rydberg series, which in $LS$ coupling are labeled as $^1F_3$, $^3F_3$, $^3F_2$, and $^3F_4$. In this work, we present new microwave spectroscopic data on all four series and analyze them in conjunction with previously reported laser~\cite{Aymar1984three,Ali1999Two} and microwave~\cite{Lehec2017PhD} spectroscopic data using single- and multi-channel quantum defect models. Series-specific details and comparisons are discussed in the following subsections.

An overview of the modeled spectroscopic data of the $6snf$ states of $^{174}$Yb is presented in Fig.~\ref{fig: 174}. 

\begin{figure*}[tb]
    \includegraphics[width=\linewidth]{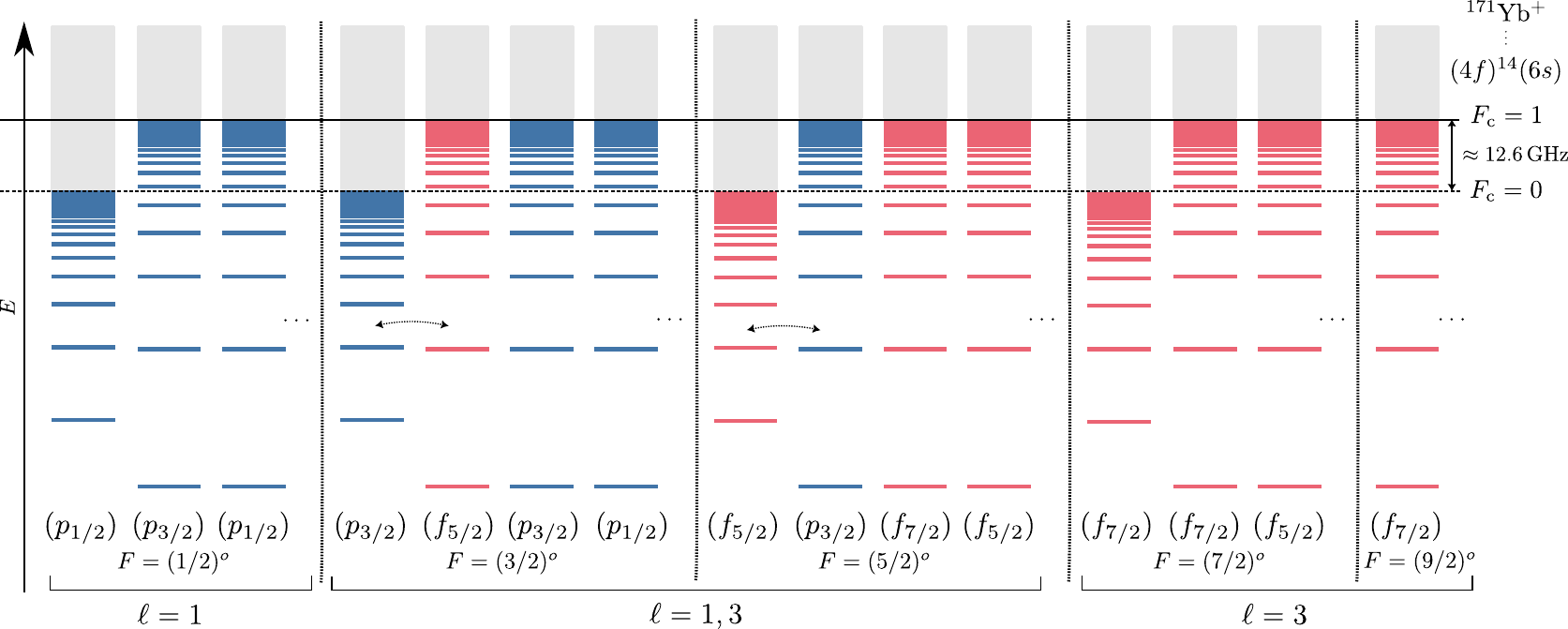}
    \caption{Schematic energy level diagram of the odd-parity $6snp$ (blue) and $6snf$ (red) Rydberg states of $^{171}$Yb. Rydberg series with the same parity and total angular momentum $F$ can mix. Arrows indicate mixing between the $6snp$ and $6snf$ series. Channels converging to ion-core-excited states are omitted, but are included in the MQDT models. The series are labeled by $F$ and the quantum numbers of the Rydberg electron.}
    \label{fig:PFmixingschematic}
\end{figure*}

\subsection{$6snf\,^{1,3}F_3$}
\label{sec:174Yb_13F3}

The $6snf\,^{1,3}F_3$ Rydberg series in $^{174}$Yb exhibit complex perturbations that have been partially characterized in earlier spectroscopic work~\cite{Meggers1970First,Wyart1979Extended,Aymar1984three,Ali1999Two,Lehec2017PhD}. Aymar et al.~\cite{Aymar1984three} reported state energies for $6snf\,^{1,3}F_3$ states in the range of $16 \leq n\leq51$ with \SI{3}{\giga\hertz} accuracy, and identified a single perturbing state at \SI{50228.09}{\per\centi\meter}, which they attributed to a $4f^{13}5d^26s\,(J = 3)$ state. The laser spectroscopic data in Ref.~\cite{Aymar1984three} was modeled by a three channel MQDT model optimized for $6snf$ states in the range of $16 \leq n\leq51$. Ali et al.~\cite{Ali1999Two} reported additional lower lying $6snf\,^{1,3}F_3$ Rydberg states with $9\leq n\leq30$ to \SI{15}{\giga\hertz} accuracy, as well as three additional perturbers at \SI{48584.1}{\per\centi\meter}, \SI{49122.4}{\per\centi\meter}, and \SI{49661.6}{\per\centi\meter}, which were assigned as $4f^{13}5d^26s\,^3D_3$, $4f^{13}5d^26s\,^1F_3$, and $4f^{13}5d^26s\,^3G_3$ states, respectively (see Tab.~\ref{tab:174YbSpec_13F3}). The laser spectroscopic data was modeled by a five-channel MQDT model, which was optimized for states in the range $9\leq n\leq 20$ and does not include the perturbing state observed in by Aymar et al.~\cite{Aymar1984three}.
Lehec~\cite{Lehec2017PhD} performed high-precision microwave spectroscopy of new $6snf\,^{1,3}F_3$ states for $47 \le n \le 75$, with \SI{500}{\kilo\hertz} accuracy.

In this work, we extend the coverage of microwave spectroscopy to lower principal quantum numbers, reporting measurements of $6snf\,^{1,3}F_3$ states in the range $30 \le n \le 50$ with \SI{100}{\kilo\hertz} accuracy. These results bridge the gap between the earlier laser~\cite{Aymar1984three,Ali1999Two} and microwave~\cite{Lehec2017PhD} spectroscopy, and provide more accurate benchmark data for MQDT modeling. The $6snf\, ^{1, 3}F_3$ states reported in this work are probed by a single-photon microwave transition from $6sn'd\, ^{1, 3}D_2$ Rydberg states (see Tab.~\ref{tab:174Spec_Microwave_13F3}) or by a two-photon microwave transition from another $6sn'f\, ^3F_3$ state (see Tab.~\ref{tab:174Spec_Microwave_13F3_twophoton}). 

Based on the previous~\cite{Aymar1984three,Ali1999Two,Lehec2017PhD} and new spectroscopic data, we develop a new seven-channel MQDT model. This model combines the previously developed MQDT models~\cite{Aymar1984three,Ali1999Two} into a single model for all $6snf$ states with $n\ge9$. The MQDT model consists of the two $6snf\,^1F_3$ and $6snf\,^3F_3$ series, as well as five channels converging to the weighted average of the $4f^{13}(^2F_{7/2})5d6s\,^{1,3}D$ ionization thresholds at \SI{83967.70}{\per\centi\meter}. Four of the channels correspond to the perturbers observed in Refs.~\cite{Aymar1984three,Ali1999Two}. To capture the energy dependence of the quantum defect of the $6snf\,^{1,3}F_3$ Rydberg states, we found it necessary to include a fifth perturbing state with an energy above the lowest ionization threshold.

The parameters of the $6snf\,^{1,3}F_3$ MQDT models obtained from the global fit routine (see Sec.~\ref{sec:mqdtfitting}) are tabulated in Tab.~\ref{tab:174MQDT_13F3}. The resulting model is presented as a Lu-Fano-type plot of the quantum defects and fit residuals in Fig.~\ref{fig: 174}.

We observe excellent agreement between the observed and predicted state energies in the range of $30\leq\nu_{6s}\leq65$, with a maximum fit residual of less than \SI{500}{\kHz}. We note that the microwave data with $\nu > 65$ from Ref.~\cite{Lehec2017PhD} exhibits large scattering and is thus excluded from the fit. Additionally, we shift the laser spectroscopy data of Ref.~\cite{Aymar1984three} by \SI{2.4}{\GHz} to align with our microwave data in the range where they overlap. This is within the quoted \SI{3}{\GHz} uncertainty in Ref.~\cite{Aymar1984three}.

We note that despite the inclusion of additional perturbing states, the energy dependence of the $6snf\,^1F_3$ eigenchannel quantum defect $\mu_{^1F_3}$ is unusually large. Together with the significant deviations between the MQDT model predictions and the experimentally determined term values at low effective principal quantum number, this could indicate additional unaccounted perturbing states. 

The singlet-triplet mixing for $6snf\, ^{1, 3}F_3$ states is too small to measure from $g$-factors. We extract a mixing angle of $\theta(\nu) = -0.02084(4) + 0.239(2)/\nu^2$ from the hyperfine structure~\cite{Aymar1984Rydberg,peper2025spectroscopy,Clausen2025Ionization} of $6snf$ states in $^{171}$Yb (see Sec.~\ref{sec:171Fstates}), yielding a $^1F_3$ admixture of $\sim\SI{0.04}{\percent}$ in the $^3F_3$ states. Curiously, this is significantly smaller than the singlet-triplet mixing in high-$n$ $6snp\, P$ ($\sim$\SI{4}{\percent}) and $6snd\, D$ ($\sim$\SI{15}{\percent}) states of ytterbium~\cite{peper2025spectroscopy}. We tentatively attribute this to the large singlet-triplet splitting between the $6snf\, ^{1, 3}F_3$ states.  

\begin{figure*}[tb]
    \centering
    \includegraphics[width=\linewidth]
    {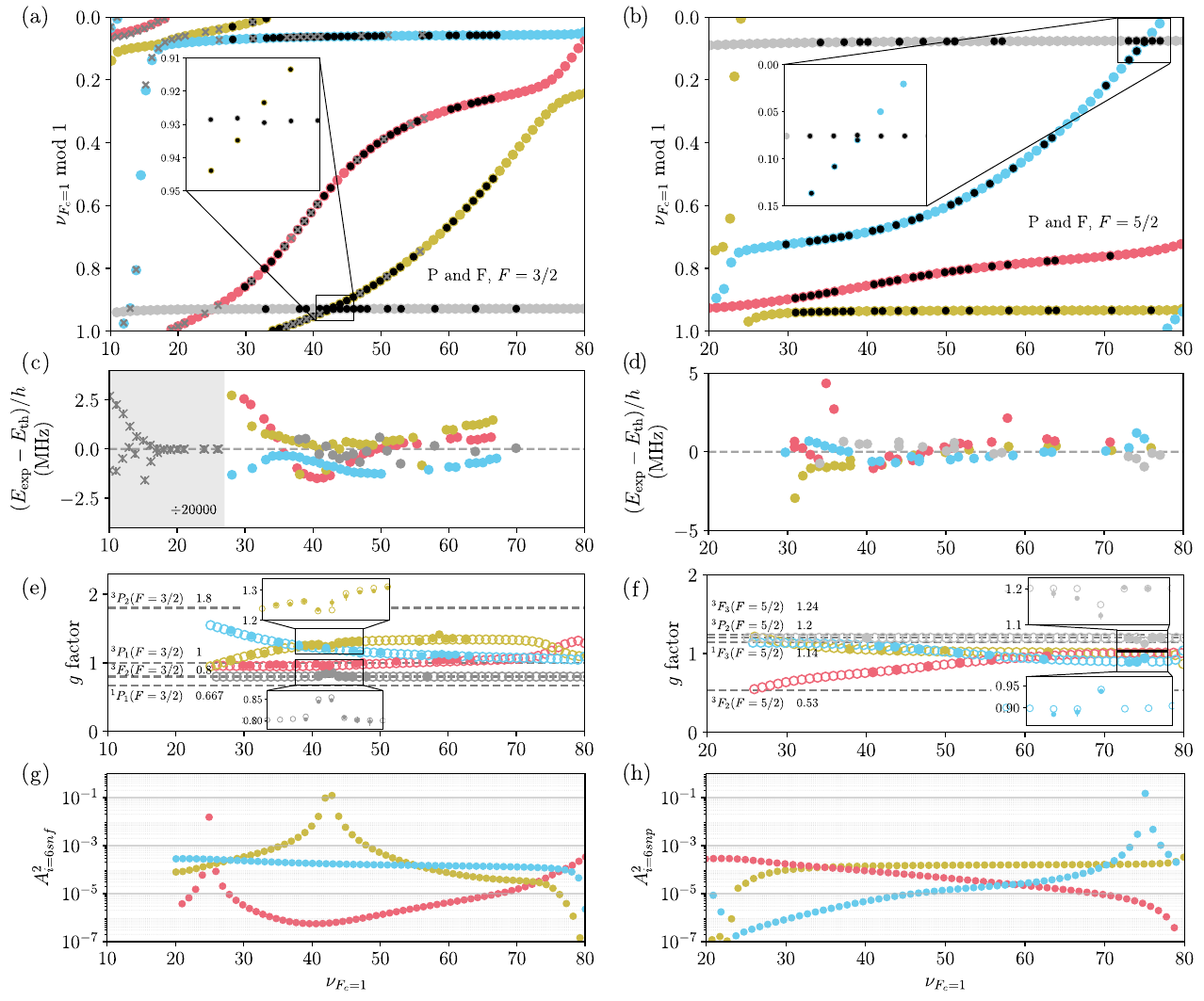}
    \caption{MQDT models for $^{171}$Yb odd-parity Rydberg series. (a) Lu-Fano-type plots of the $F=3/2$ and (b) $F=5/2$ series, respectively. The black points are microwave spectroscopy data measured in this work and in Ref.~\cite{peper2025spectroscopy}. The gray crosses are laser spectroscopic data reported in Ref.~\cite{Majewski1985diploma} and compiled in Ref.~\cite{peper2025spectroscopy}.  In (a), the gray colored series is of dominantly $\ell=3$ character and the rest are dominantly $\ell = 1$. In (b), the gray colored series is of dominantly $\ell=1$ character and the rest are dominantly $\ell = 3$. The insets are zoomed in near the avoided crossing between the $\ell = 1$ and $\ell = 3$ series. (c),(d) Deviation between experimental energies and the MQDT model bound state energies. (e),(f) Measured (filled points) and predicted (open circles) $g$-factors. The insets are zoomed in near the avoided crossing, where we observe admixture in the $g$-factors. The gray dashed lines indicate the analytical $g$-factors in $LS$ coupling. (g) $6snf$ channel fraction of the dominantly $6snp$ series for $F = 3/2$. (h) $6snp$ channel fraction of the dominantly $6snf$ series for $F = 5/2$.}
    \label{fig:171spec3252}
\end{figure*}

\begin{figure*}[tb]
    \centering
    \includegraphics[width=\linewidth]{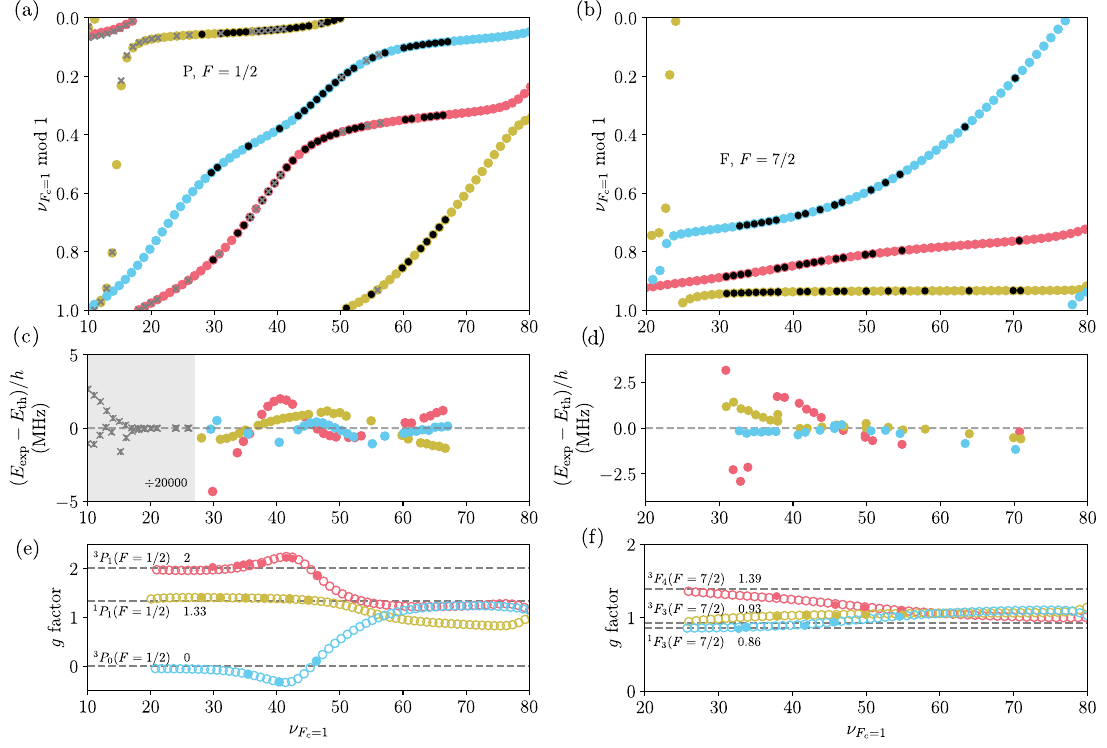}
    \caption{MQDT models for the $^{171}$Yb $6snp\,(F=1/2)$ and $6snf\,(F=7/2)$ Rydberg series. (a),(b) Lu-Fano-type plots. The black points correspond to microwave spectroscopic data, and colored points are MQDT bound state energy predictions. The gray crosses are three-photon laser spectroscopic data reported in Ref.~\cite{Majewski1985diploma}. (c),(d) Deviation between experimental energies and the MQDT model bound state energies. (e),(f) Measured (filled points) and predicted (open circles) $g$-factors. The gray dashed lines indicate the analytical $g$-factors in $LS$ coupling. }
    \label{fig:171spec1272}
\end{figure*}

\subsection{$^{3}F_2$ and $^{3}F_4$}
Low-lying $6snf\, ^3F_2$ and $^3F_4$ states ($5 \leq n \leq 14$) were previously assigned based on discharge-tube spectra recorded on a grating spectrometer~\cite{Wyart1979Extended}. Here, we present the results of microwave spectroscopy of the $6snf\, ^3F_2$ and $6snf\, ^3F_4$ series in the ranges $27<\nu_{6s}<47$ and $44<\nu_{6s}<53$, respectively. The $6snf\, ^3F_2$ states are probed by microwave transitions from $6sn'd\,^{1, 3}D_2$ states (see Tab.~\ref{tab:174Spec_Microwave_3F2}) or two-photon microwave transitions from $6snf\, ^3F_2$ or $^3F_3$ states (see Tab.~\ref{tab:174Spec_Microwave_3F2_twophoton}). For $6snf\, ^3F_4$ states, we record single-photon transitions from $6sn'd\,^3D_3$ states, which we populate by a preceding two-photon microwave transition from a $6sn''d\,^3D_2$ state (see Tab.~\ref{tab:174Spec_Microwave_3F4}).

We restrict the theoretical modeling of the $6snf\, ^3F_2$ and $6snf\, ^3F_4$ series to the microwave spectroscopic data reported in this work. Over the limited energy range of states recorded in this work, the $6snf\, ^3F_2$ and $6snf\, ^3F_4$ series are well described by a single-channel quantum defect model using energy-dependent quantum defects $\mu_\alpha$

\begin{align} \label{eq:qd_energy}
    \mu_\alpha(\nu_{6s}) = \mu_\alpha^{(0)} + \frac{\mu_\alpha^{(2)}}{\nu_{6s}^2} +\frac{\mu_\alpha^{(4)}}{\nu_{6s}^4} \dots\,,
\end{align}  
where $\nu_{6s}$ is the effective principal quantum number with respect to the lowest ionization limit~\cite{Lee1973Spectroscopy,peper2025spectroscopy}. The single-channel quantum defects for $6snf\, ^3F_2$ and $6snf\, ^3F_4$ are presented in Tab.~\cref{tab:174QDTs}. 

A more comprehensive MQDT model of the $6snf\, ^3F_2$ and $6snf\, ^3F_4$ Rydberg series would require additional spectroscopic data, particularly of lower lying states~\cite{Wyart1979Extended}. 

\begin{figure}[!ht]
    \includegraphics[width=\linewidth]{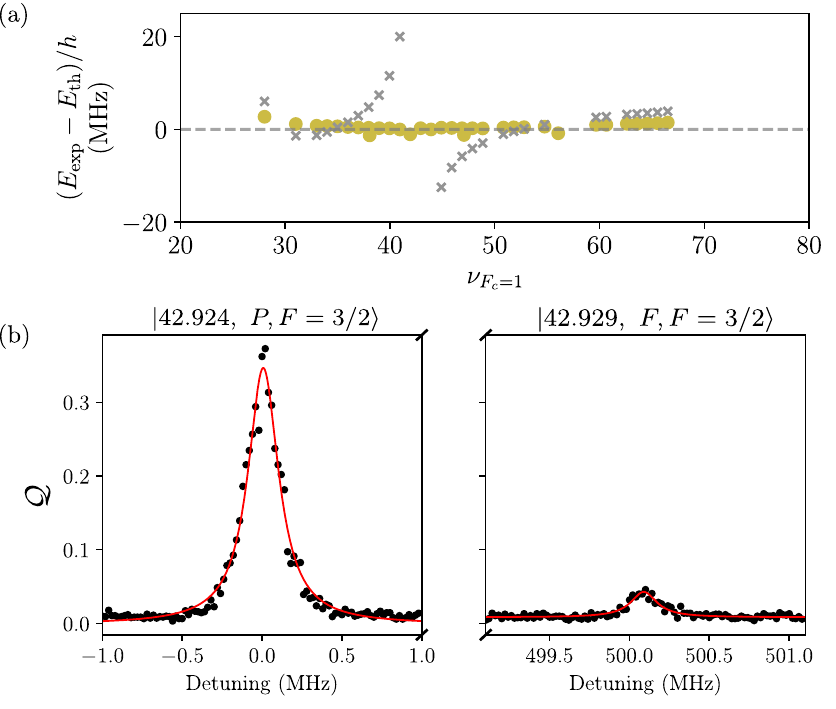}
    \caption{(a) Deviation between experimental energies and MQDT model bound state energies before (gray crosses)~\cite{peper2025spectroscopy} and after (yellow points) the introduction of the $p$-$f$ mixing angle in the MQDT model. (b) Microwave excitation spectrum from the $\ket{\nu_{F_c=1}=41.703, S, F=1/2}$ state to the $\ket{\nu_{F_c=1}=42.924, P, F=3/2}$ and $\ket{\nu_{F_c=1}=42.929, F, F=3/2}$ states, respectively. The experimentally measured ratio between the transition strengths was 11.9(9), which is comparable with the MQDT model prediction of 8.0. We note that this deviation may arise from potentially frequency dependent microwave intensity at the atoms between the two transitions. $\mathcal{Q}$ denotes the fraction of the microwave signal with respect to the original Rydberg laser signal.}
    \label{fig: PFmixing}
\end{figure}

\section{\label{sec:171Fstates}Odd-parity Rydberg states of $^{171}$Yb}

Hyperfine coupling with the $I=1/2$ nucleus doubles the number of Rydberg series in $^{171}$Yb, compared to $^{174}$Yb. There are eight $6snf$ series, which in $LS$ coupling are described as $^3F_2$ ($F=\{3/2, 5/2\}$), $^1F_3$ ($F=\{5/2, 7/2\}$), $^3 F_3$ ($F = \{5/2, 7/2\}$), and $^3F_4$ ($F=\{7/2, 9/2\}$). However, certain $6snf$ series are mixed with $6snp$ series with the same $F$ by a configuration interaction, which is allowed since both series have odd parity (see Fig.~\ref{fig:PFmixingschematic}). Therefore, our model also includes the seven $6snp$ series, which in $LS$ coupling are described as $^3P_0$ ($F=\{1/2\}$), $^1P_1$ ($F=\{1/2, 3/2\}$), $^3 P_1$ ($F = \{1/2, 3/2\}$), and $^3P_2$ ($F=\{3/2, 5/2\}$)~\cite{peper2025spectroscopy}.

Spectroscopy of the $6snp$ Rydberg states of $^{171}$Yb has been reported previously in Refs.~\cite{Majewski1985diploma, peper2025spectroscopy}. In Ref.~\cite{peper2025spectroscopy}, the spectroscopic data were modeled by introducing hyperfine interaction through a frame transformation~\cite{Sun1989Hyperfine,Robicheaux2018theory}, with mostly good agreement particularly for higher lying ($\nu_{F_c=1}>30$) Rydberg states. However, the energy of certain $6snp\,(F=3/2)$ Rydberg states obtained by precision microwave spectroscopy deviated significantly from the theoretical predictions obtained from the presented MQDT model~\cite{peper2025spectroscopy}, with the fit residuals resembling a dispersion-like feature, implying the existence of an unaccounted channel interaction.

The observed discrepancies can be attributed to configuration interaction between Rydberg series with different orbital angular momenta. In multichannel systems like ytterbium, such interactions play a central role in shaping both the energy spectrum and the character of the Rydberg wavefunctions. Although total parity and total angular momentum $F$ remain conserved, coupling between channels with different orbital angular momentum $\ell$---notably with $\Delta \ell = 2$---can occur through effective quadrupole interactions in the ionic core~\cite{Lee1974Spectroscopy}. This mechanism has been observed as $p$–$f$ mixing in molecular hydrogen~\cite{Osterwalder2004high} and  as $s$–$d$ mixing in noble gases~\cite{Worner2003Multichannel,Schafer2006Millimeter}. While such mixing can occur even in the absence of hyperfine structure, the dense level structure of $^{171}$Yb creates near-degeneracies between series of different $\ell$, significantly enhancing sensitivity to these interactions~\cite{Worner2003Multichannel}.

In order to probe the $p$-$f$ mixing in ytterbium, we record microwave spectra of previously unobserved $6snp$ and $6snf$ states. These states are excited via single photons from $6snd$ $F= 3/2$ and $F = 5/2$ states, which are populated by a two-photon laser transition from the ground state via the $6s6p\,^1P_1$ intermediate state. With this scheme, odd-parity Rydberg series with $F=\{1/2,3/2, 5/2,7/2\}$ are accessible. The newly observed experimental transition frequencies are reported in Tabs.~\cref{tab:171Spec_PF32,tab:171Spec_PF52,tab:171Spec_F72} for $F = 3/2, 5/2$, and $7/2$, respectively.

The multi-channel models for odd-parity states with $F = 3/2$ and $F = 5/2$ include contributions from both the $6snp$ and $6snf$ Rydberg series (see Fig.~\ref{fig:PFmixingschematic}), making them susceptible to $p$–$f$ mixing. To account for this, we construct a combined MQDT model for the $6snp$ and $6snf$ states based on the spectroscopic data presented here and in previous studies~\cite{Majewski1985diploma,peper2025spectroscopy}. We find that a single mixing angle between the $6s_{1/2}np_{3/2}$ and $6s_{1/2}nf_{5/2}$ series is sufficient to capture the observed structure. In $LS$ coupling, this  corresponds to a mixing between the $6snp\,^3P_2$ and $6snf\,^3F_2$ channels.

On the other hand, the single- and multi-channel models for odd-parity states with $F=1/2$, $F=7/2$ and $F=9/2$ only contain contributions from either $6snp$ or $6snf$ Rydberg series. Thus, these series do not exhibit $p$-$f$ series mixing. 

Starting from an initial guess based on the $^{174}$Yb MQDT parameters of the $6snp$ and $6snf$ state models, we optimize the $^{171}$Yb MQDT parameters in a global fit of all $\ell \le 4$ spectroscopic data. The resulting MQDT models are presented in App.~\ref{sec:171yb_mqdt}.

The experimental and theoretical energies are presented in Fig.~\ref{fig:171spec3252}(a) and (b) for $F=\{3/2,5/2\}$ and Fig.~\ref{fig:171spec1272}(a) and (b) for $F=\{1/2,7/2\}$. Generally, we observe good agreement between the experimental energies and the MQDT predictions, particularly resolving the dispersion-like feature observed in Ref.~\cite{peper2025spectroscopy} (see Fig.~\ref{fig: PFmixing}(a)), with an RMS deviation of \SI{0.91}{MHz} for all $6snp$ and $6snf$ states measured via microwave transitions.

We additionally probe the accuracy of the MQDT wavefunctions by measuring Land\'e $g$-factors and comparing to the theoretical predictions (Fig.~\ref{fig:171spec3252}(e) and (f) for $F=\{3/2,5/2\}$ and Fig.~\ref{fig:171spec1272}(e) and (f) for $F=\{1/2,7/2\}$). The experimentally observed $g$-factors are reproduced by the MQDT predictions accurately, and both experimental and theory $g$-factors are tabulated in Tabs.~\cref{tab:171Spec_PF32_gfactors,tab:171Spec_PF52_gfactors,tab:171Spec_F72_gfactors}.

As part of our fitting of the MQDT model to spectroscopic data, we extract a $p$-$f$ mixing angle of $\theta_{pf} = 1.842(8)\times10^{-2}$, which governs the strength of coupling between the $6snp$ and $6snf$ channels. The states with predominantly $p$ character therefore have a $\mathcal{O}(\theta_{pf}^2) \approx 10^{-4}$ admixture of $f$ states population, and the predominantly $f$ states have a similar $p$ character (see Fig. \ref{fig:171spec3252}(g) and (h)). Near degeneracies between $6snp$ and $6snf$ series with the same $F$, the states are more strongly mixed, reaching an admixture of $>\SI{10}{\percent}$. An avoided-crossing feature is observed in the energies for $F=3/2$ and $F=5/2$ (see insets in Fig.~\ref{fig:171spec3252}(a) and (b)), and the $g$-factors are also affected (see insets in Fig.~\ref{fig:171spec3252}(e) and (f)). To further demonstrate the mixed wavefunctions, we drive a single-photon microwave transition from $\ket{\nu_{F_c=1}=41.703, S, F=1/2}$ to $\ket{\nu_{F_c=1}=42.929, F, F=3/2}$ (see Fig.~\ref{fig: PFmixing}(b)).

The singlet–triplet mixing angle of the $6snf$ states, fitted to $^{171}$Yb spectroscopic data as $\theta(\nu) = -0.02084(4) + 0.239(2)/\nu^2$, yields a small $^1F_3$ admixture of $\sim\SI{0.04}{\percent}$ in the $^3F_3$ states. We apply this same mixing angle to the $6snf\,^{1,3}F_3$ states of $^{174}$Yb. In addition, $p$–$f$ mixing between the $6snp\,^3P_2$ and $6snf\,^3F_2$ series is also expected in $^{174}$Yb, but the admixture remains inherently small—on the order of $\sin^2\!\left(\theta_{pf}\right) \sim 10^{-4}$—due to the absence of near-degeneracies, despite a similar mixing angle $\theta_{pf}$ as $^{171}$Yb.

\section{$^{174}$Yb $6sng$ states}
\label{sec:174_Gstates}

\begin{figure}[tb]
    \centering
    \includegraphics[width=\linewidth]{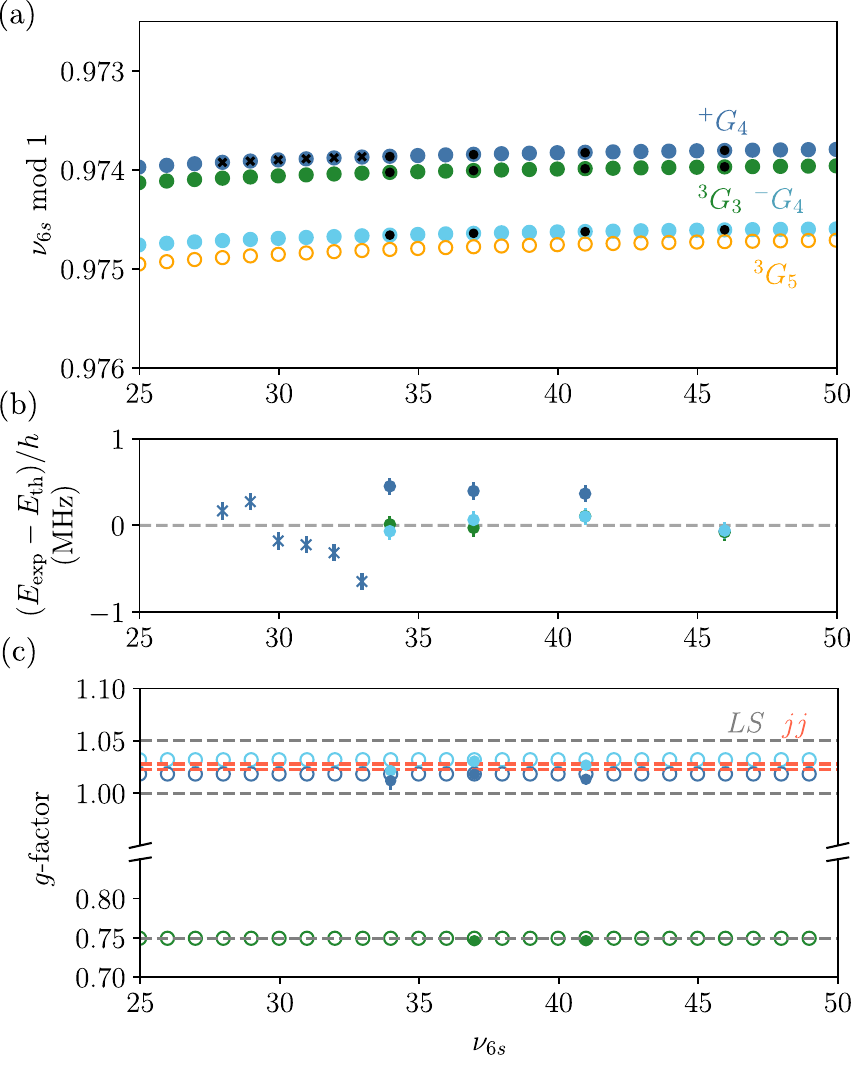}
    \caption{(a) Lu-Fano-type plot of $^{174}$Yb $\ell=4$ Rydberg states: $^+G_4$ (dark blue), $^-G_4$ (light blue), $^3G_3$ (green), and $^3G_5$ (orange). The $J = 4$ series are fit to a single MQDT channel described in $jj$ coupling. The $^3G_5$ MQDT parameters are inferred from the $^{171}$Yb model, as described in the text. The black points and crosses correspond to experimentally observed states by microwave spectroscopy reported in this work and in Ref.~\cite{Niyaz2019Microwave}, respectively.  (b) Deviation between experimental energies and the MQDT bound state energies. Markers as in panel (a). (c) Measured (filled points) and predicted (open circles) $g$-factors. The gray dashed lines correspond to the theoretical $g$-factors for purely $LS$ coupling, while the orange dashed lines correspond to theoretical $g$-factors for $jj$ coupling.}
    \label{fig:G_174}
\end{figure}

\begin{figure}[tb]
    \centering
    \includegraphics[width=\linewidth]{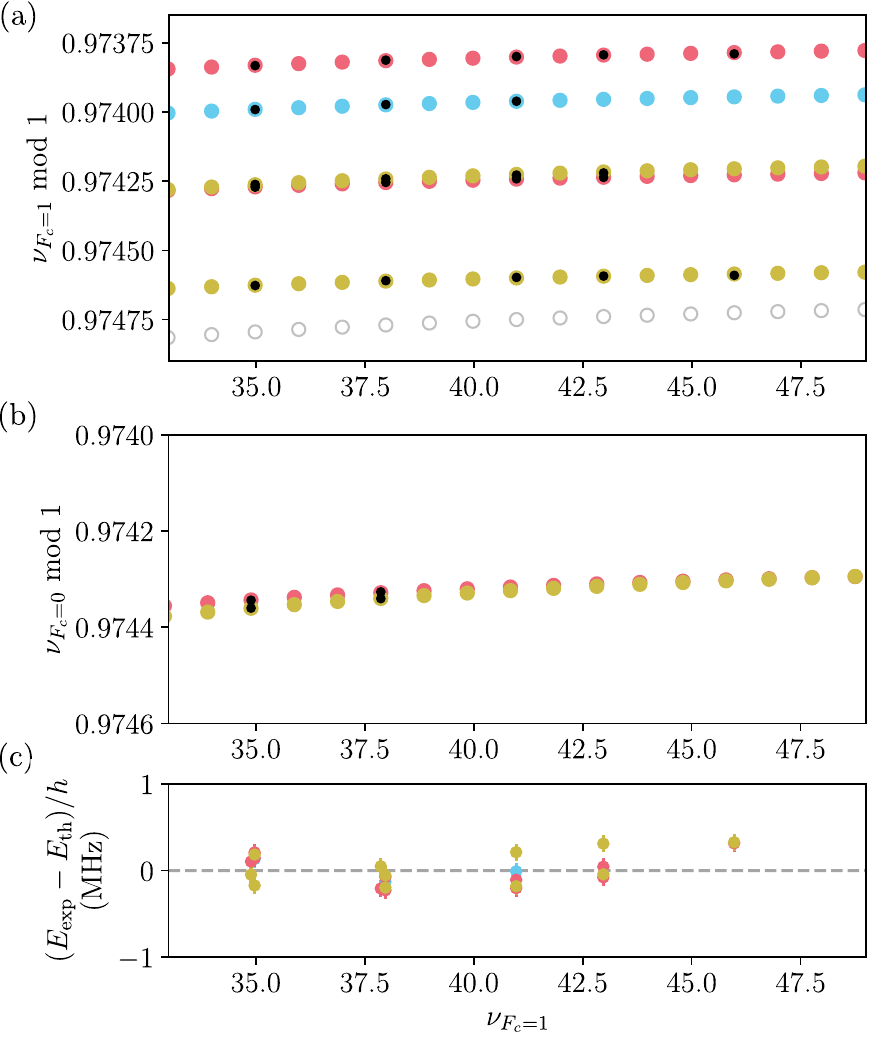}
    \caption{(a),(b) Lu-Fano-type plots of the $G$ $F=5/2$ (yellow), $F=7/2$ (blue), $F=9/2$ (pink), and $F=11/2$ (light gray) series in $^{171}$Yb, converging to the $F_c=1$ and $F_c = 0$ hyperfine thresholds, respectively. The black points correspond to experimentally observed states by microwave spectroscopy reported in this work. (c) Deviation between the experimental energies and predicted MQDT bound state energies, with all energies plotted with respect to the $F_c=1$ hyperfine threshold.}
    \label{fig:G_171}
\end{figure}

In $^{174}$Yb, there are four distinct $6sng$ Rydberg series, which in $LS$ coupling are labeled as $^1G_4$, $^3G_4$, $^3G_3$, and $^3G_5$. Only the $^1G_4$ Rydberg series has been previously observed, using microwave spectroscopy~\cite{Niyaz2019Microwave}.

Here, we present new spectroscopic data and MQDT models of $6sng$ Rydberg states in $^{174}$Yb, which we measure by laser excitation to $6sn''d$ states followed by sequential two-photon excitations via $6sn'f\,^3F_3$ states. The observed transition frequencies are reported in Tabs.~\cref{tab:174Spec_Microwave_pmG4,tab:174Spec_Microwave_3G3}. An overview of the observed $6sng$ states is presented in Fig.~\ref{fig:G_174}.

For the $6sng\,(J=4)$ states, we observe two distinct series with small quantum defects of approximately 0.026, one of which aligns with the previously reported $^1G_4$ states from Ref.~\cite{Niyaz2019Microwave}, while the other represents a newly observed $6sng\,(J=4)$ series. The small quantum defects indicate that these high-$\ell$ states are close to the hydrogenic limit, where the centrifugal barrier strongly suppresses penetration of the Rydberg electron into the Yb$^+$ ion-core. As a result, the spin-orbit interaction (favoring $jj$-coupling) becomes stronger relative to the exchange interaction (favoring $LS$-coupling), and the states are more accurately described in the $jj$-coupled basis. Such behavior has been observed in other alkaline-earth-like atoms including helium, barium, and strontium~\cite{gallagher1982radio,snow2003indirect,fields2021high}.

The precise nature of the $^{174}$Yb $6sng\,(J=4)$ series is probed by measuring the Land\'e $g$-factors (see Fig.~\ref{fig:G_174}(c), Tab.~\ref{tab:171Spec_pmG4_gfactors}). When treating the two series in terms of $LS$-coupled $^{1,3}G_4$, the extracted singlet-triplet mixing angle between the two series is $\theta^{LS}=0.66(4)$. This is close to the full singlet-triplet mixing of $\theta^{LS} = 0.73$, given by the analytic expression $\arctan\left(\sqrt{\ell/(\ell+1)}\right)$, obtained by expressing the $LS$-$jj$ frame transformation~\cite{edmonds1996angular} in terms of a rotation matrix. Instead, when describing the two series in $jj$ coupling, the mixing angle evaluates to $\theta^{jj}=-0.06(3)$. This indicates that the two series are more accurately described in $jj$ coupling. We therefore denote these two series as $6sng\, ^{^\pm}G_4$, corresponding to the state labels $(1/2, \ell \pm 1/2)_4$. The two-channel MQDT model is presented in Tab.~\ref{tab:174MQDT_PMG4}). An overview of the microwave spectroscopic data and MQDT predictions are shown in a Lu-Fano-type plot (Fig.~\ref{fig:G_174}(a)). 
The deviations between the experimental and MQDT bound state energies are shown in Fig.~\ref{fig:G_174}(b).

For the $6sng\,^3G_3$ series, we record energies of four states and model the quantum defects of this series with a single-channel quantum defect model (see Tab.~\ref{tab:174QDTs}). Over the limited energy range of observed states, the quantum defects of $6sng\,^3G_3$ series are well described by the series expansion given in Eq.~\ref{eq:qd_energy}. The deviations between the experimentally observed and predicted energies are shown in Fig.~\ref{fig:G_174}(b). To check the assignment of the observed states, we record $g$-factors, which align with the analytical expression for $6sng\,^3G_3$ states of $g_J=0.75$ (see Fig.~\ref{fig:G_174}(c), Tab.~\ref{tab:171Spec_3G3_gfactors}).

The $6sng\,^3G_5$ states are not accessible via a single photon transition from the $6snf\,^3F_3$ intermediate state. However, we observe $^3G_5$ Rydberg states in $^{171}$Yb, as presented in Sec.~\ref{sec:171Gstates}. Thus, we present a single-channel quantum defect model of $^{174}$Yb $6sng\,^3G_5$ in Tab.~\cref{tab:174QDTs} using the corresponding parameters in $^{171}$Yb.

\section{$^{171}$Yb $6sng$ states}
\label{sec:171Gstates}

There are eight $6sng$ series in $^{171}$Yb, which can be expressed as $^3G_3$ ($F=\{5/2, 7/2\}$), $^+G_4$ ($F=\{7/2, 9/2\}$), $^-G_4$ ($F = \{7/2, 9/2\}$), and $^3G_5$ ($F=\{9/2, 11/2\}$). To our knowledge, no prior spectroscopic data have been reported for the $\ell=4$ states of $^{171}$Yb. 

Here, we report spectroscopic data on the $6sng$ $F=5/2$, $7/2$, and $9/2$ series by two-step microwave transitions starting from laser excited $6sn''d$ states, followed by $6sn'f\,(F=7/2)$ states. Transitions from $6snf$ $F = 3/2$ and $F = 5/2$ states were used to verify the assignment of states based on selection rules. Six of the series (including $6sng\,(F=11/2)$) converge to the $F_c=1$ hyperfine threshold (Fig.~\ref{fig:G_171}(a)), while two converge to the $F_c=0$ hyperfine threshold (Fig.~\ref{fig:G_171}(b)). The measurement range is limited to $\nu = 35$ to $46$ by the available microwave sources and difficulties in state-selective field ionization. The recorded transition frequencies are presented in Tabs.~\cref{tab:171Spec_G52,tab:171Spec_G72,tab:171Spec_G92}.

We observe small quantum defects for each of these series, which are consistent with the $6sng$ quantum defects observed in $^{174}$Yb. Also similarly to $^{174}$Yb, we describe these series in $jj$ coupling, using a frame transformation matrix to achieve this effect in our MQDT model. The MQDT model parameters are fitted in a global optimization as described in Sec.~\ref{sec:mqdtfitting}. The resultant model parameters are tabulated in Tabs.~\cref{tab:171QDT_G52,tab:171MQDT_G72,tab:171MQDT_G92}. In particular, the mixing angle between the $^\pm G_4$ states is determined to be $\theta^{jj} = -0.082(6)$, which is consistent with the inferred mixing angle from the $^{174}$Yb $g$-factors. The sign of the mixing angle cannot be determined from state energies, but is inferred from the $^{174}$Yb $g$-factor measurements. 

We have not directly measured states of the $6sng\,(F=11/2)$ series, but can predict their location using parameters obtained from the $^3G_5\,(F=9/2)$ channel.

\begin{figure}[tb]
    \centering
    \includegraphics[width=\linewidth]{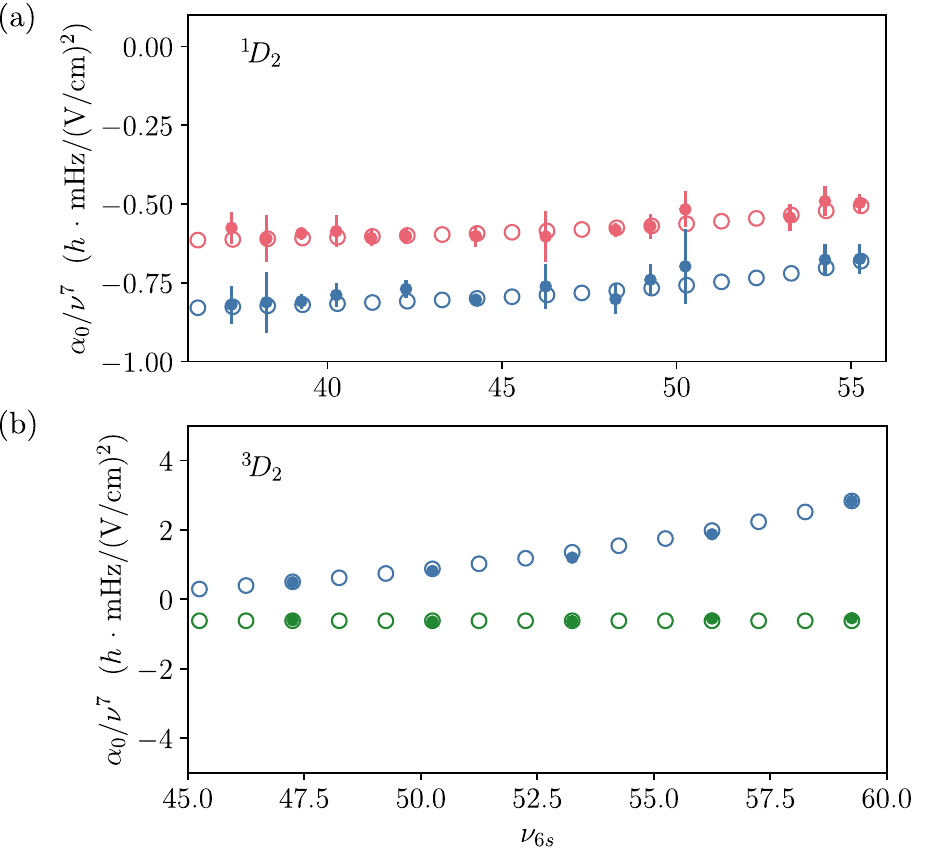}
    \caption{Measured (filled circles) and predicted (open circles) static dipole polarizabilities of (a) $6snd\, ^1D_2$ and (b) $6snd\, ^3D_2$ states in $^{174}$Yb, scaled by $\nu_{6s}^7$. The blue, red, and green points correspond to $m_J=0, 1$, and $2$ sublevels, respectively.}
    \label{fig:Dpol174}
\end{figure}

\begin{figure}[tb]
    \centering
    \includegraphics[width=\linewidth]{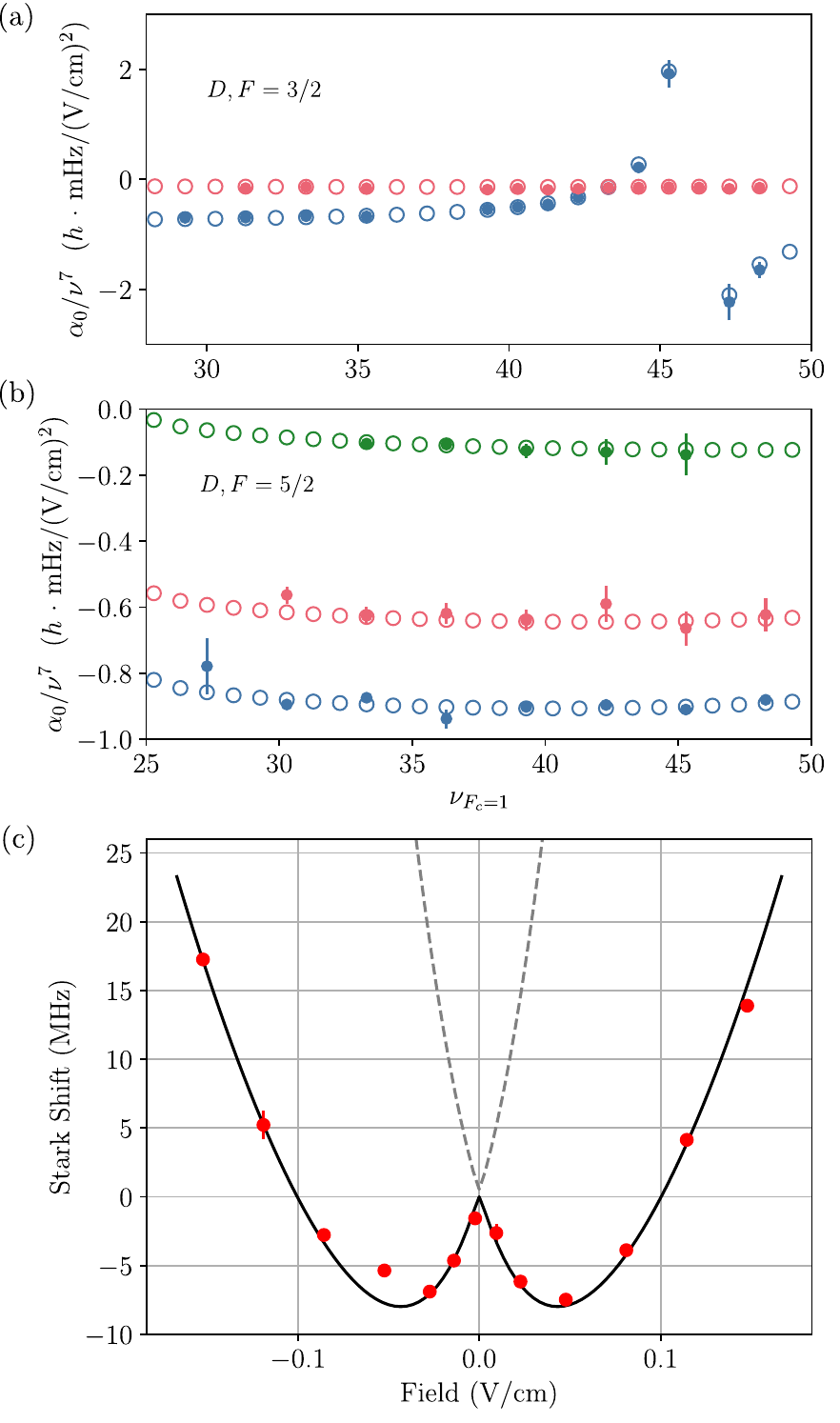}
    \caption{(a),(b) Measured (filled circles) and predicted (open circles) static dipole polarizabilities of (a) $\ket{\nu,D,F=3/2}$ and (b) $\ket{\nu,D,F=5/2}$ states of dominantly $^{1}\!D_2$ character in $^{171}$Yb, scaled by $\nu_{F_c=1}^7$. The blue, red, and green points correspond to $m_F=1/2, 3/2$, and $5/2$, respectively. (c) Experimental (red dots) and theoretical (solid black line) Stark shifts of the $\ket{\nu_{F_c=1}=62.28,D,F=5/2}$ state near a degeneracy with a $\ket{\nu_{F_c=1}=62.28,P,F=3/2}$ state. The Stark shifts of the $\ket{\nu_{F_c=1}=62.28,P,F=3/2}$ state are plotted as the dashed gray line.}
    \label{fig:Dpol171}
\end{figure}

\section{dc Stark shifts of $6snd$ Rydberg states of $^{174}$Yb and $^{171}$Yb} \label{SEC: Dpol}

Knowledge of the $6sns$, $6snp$, $6snd$, $6snf$ and $6sng$ Rydberg state energies and wavefunctions allow us to make accurate predictions on the static dipole polarizabilities of $6snd$ Rydberg states. Here, we test the predictions by measuring energy shifts for a range of $6snd$ states as a function of the static electric field, for both $^{174}$Yb and $^{171}$Yb. 

Experimentally, we measure the Stark shift $\Delta E_\mathrm{Stark}$ over a range of electric fields $F$ chosen so that the maximum Stark shifts are on the order of tens of MHz. We extract the static dipole polarizability $\alpha_0$ by a fit to the quadratic region:

\begin{align}
    \Delta E_\mathrm{Stark} (\mathbf{F}) = -\frac{1}{2}\alpha_0\mathbf{F}^2
\end{align}

\begin{figure*}[htb]
    \centering
    \includegraphics[width=\linewidth]{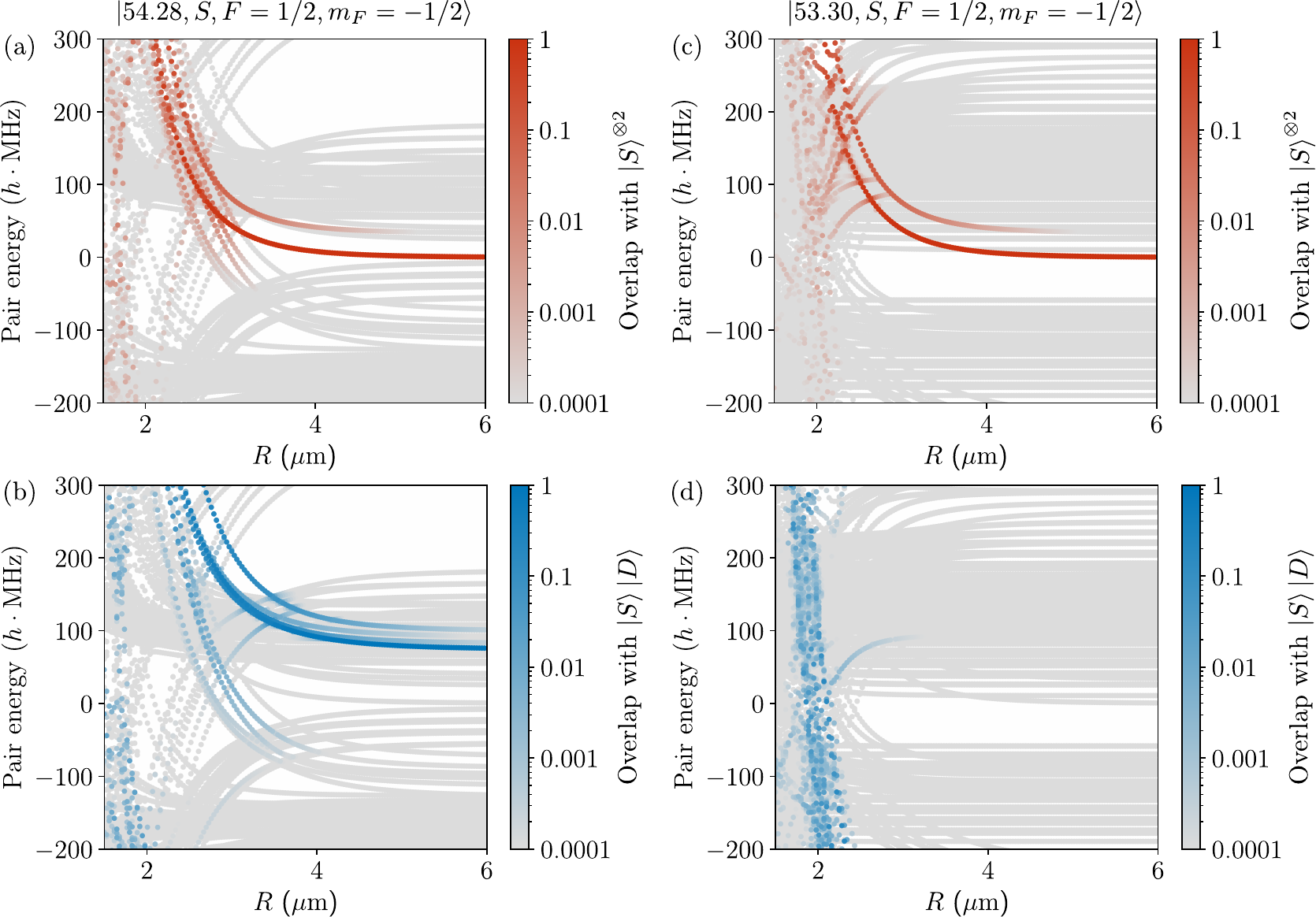}
    \caption{\label{fig:171YbRydints} Predicted $^{171}$Yb pair-interaction potentials for target Rydberg states $\ket{S} = \ket{\nu, L=0, F=1/2, m_F=-1/2}$ at a magnetic field strength of \SI{5}{G} perpendicular to the interatomic axis, shown for (a),(b) $\nu=54.28$ and (c),(d) $\nu=53.30$. Each pair of panels is referenced to the corresponding $\ket{S}^{\otimes 2}$ asymptote. The color of each curve indicates the overlap with product states of $\ket{S}$ and $\ket{D} = \ket{\nu, L=2, F=3/2, m_F=-1/2}$ atoms, with $\nu = 54.29$ in (a),(b) and $\nu = 53.29$ in (c),(d). Panels (a),(c) highlight states with dominant $\ket{S}^{\otimes 2}$ character, while panels (b),(d) show those with dominant $\ket{S}\ket{D}$ character.}

\end{figure*}

For $^{174}$Yb we measure polarizabilities of $^{1}D_2$ and $^{3}D_2$ Rydberg states (see Fig.~\ref{fig:Dpol174}). For $^{171}$Yb, we measure $\ket{\nu, D, F=3/2}$ and $\ket{\nu, D, F=5/2}$ states of dominantly $^1D_2$ character (see Fig.~\ref{fig:Dpol171}). We observe excellent agreement between the experimentally measured polarizabilities and theoretical predictions. The observed and predicted polarizabilities are presented in Tabs.~\cref{tab:174Spec_Pol_1D2,tab:174Spec_Pol_3D2,tab:171Spec_Pol_D32,tab:171Spec_Pol_D52}. Predicted static dipole polarizabilities of additional $6snd$ Rydberg series in both isotopes are presented in App.~\ref{sec:polappendix}. 

When these $6snd$ states are near-degenerate with opposite parity states, their dc Stark shifts can be non-quadratic even for small fields. In particular, the divergent feature of the $\ket{\nu, D, F=3/2, m_F = 1/2}$ polarizability from the expected $\nu_{F_c=1}^7$ scaling near $\nu_{F_c=1} \approx 46$ is caused by a near-degeneracy with $P, F=1/2$ states. More generally, most of the $D, F=\{3/2, 5/2\}$ states in the range $\nu_{F_c=1}>40$ experience a series of near degeneracies with dipole-coupled states, resulting in non-quadratic Stark shifts even for small fields. 

In Fig.~\ref{fig:Dpol171}(c), we present measurements of dc Stark shifts of the $\ket{\nu_{F_c=1}=62.28, D, F=5/2}$ state, which is near-degenerate with the $\ket{\nu_{F_c=1}=62.28, P, F=3/2}$ state. The energy separation between the two states is approximately one MHz, which causes a deviation from a quadratic Stark shift, and we observe a double-well-type Stark shift.

\section{$6snd$ Rydberg-Rydberg interaction potentials}
\label{sec:171int}

Entangling gates in $^{171}$Yb have been implemented via laser excitation to $6sns$ Rydberg states, using single-photon excitation from the metastable qubit state $6s6p\,^3P_0$~\cite{Ma2023High, peper2025spectroscopy, muniz2025high,senoo2025high}. These gate protocols rely on the Rydberg blockade mechanism, in which strong dipole–dipole interactions between nearby Rydberg atoms prevent simultaneous excitation, enabling conditional logic operations between qubits~\cite{Saffman2010Quantum}.

Extensive laser and microwave spectroscopy and modeling of $\ell\le 2$ Rydberg series of $^{171}$Yb Rydberg states in Ref.~\cite{peper2025spectroscopy} enabled the accurate prediction of static dipole polarizablities and interaction potentials of $6sns$ Rydberg states. This revealed the presence of F\"orster resonances in the interaction of $6sns\,(F=3/2)$ Rydberg states, previously used for implementing two-qubit gates~\cite{Ma2023High}. The F\"orster resonances result in a dense manifold of weakly interacting pair states near zero energy at short separations, which can be unintentionally populated within the blockade radius, thereby reducing the fidelity of Rydberg-mediated gates. Recognizing this limitation led to the adoption of a $6sns\,(F=1/2)$ state, whose interaction spectrum is free of F\"orster resonances and thus more favorable for high-fidelity entangling operations.

Another potential source of blockade violations comes from $6snd$ Rydberg states, which are also accessible by a single-photon transition from $6s6p\,^3P_0$. In particular, interactions can shift $SD$ pair states into resonance with the driving laser at particular short atom separations. Accurate predictions of the interactions involving $6snd$ states requires precise models of the $6snf$ state energies and wavefunctions, which are developed in this work.

As an example, in Fig.~\ref{fig:171YbRydints}(a) and (b) we present predicted Rydberg-Rydberg interaction potentials near the $\ket{S}^{\otimes 2} = \ket{54.28, L=0, F=1/2, m_F = -1/2}^{\otimes 2}$ pair state used for entangling gates in our tweezer setup~\cite{peper2025spectroscopy}. The $\ket{S}$ state is detuned from $\ket{D}=\ket{54.29, L=2, F=3/2}$ by only \SI{70}{MHz} at zero magnetic field. In this case, the Rydberg blockade remains strong, because the $\ket{S}\ket{D}$ state lies energetically above the $\ket{S}^{\otimes2}$ state, and both the $\ket{S}^{\otimes 2}$ and $\ket{S}\ket{D}$ pair states exhibit large repulsive interaction shifts. 

In contrast, the Rydberg-Rydberg interaction potentials near the $\ket{S'}^{\otimes 2} = \ket{53.30, L=0, F=1/2, m_F = -1/2}^{\otimes 2}$ asymptote are presented in Fig.~\ref{fig:171YbRydints}(c) and (d). In this case, the corresponding $\ket{D'}=\ket{53.29, L=2, F=3/2}$ state lies \SI{600}{MHz} below the $\ket{S'}$ state. Naively, this large detuning of the $\ket{S'}\ket{D'}$ state should reduce the impact of a Rydberg blockade violation. However, the $\ket{S'}\ket{D'}$ pair state exhibits a strong repulsive interaction, shifting the $\ket{S'}\ket{D'}$ into resonance of the laser at a tweezer separation of $\sim\SI{2}{\mu m}$ (see Fig.~\ref{fig:171YbRydints}(d)).

These two examples illustrate that a detailed analysis is required for specific experimental conditions, including the effects of magnetic field and orientation.

\section{Discussion and Outlook} \label{SEC: discussionoutlook}

The precision microwave spectroscopy presented in this work enabled the development of accurate MQDT models for the $6snp$, $6snf$, and $6sng$ Rydberg series of both $^{174}$Yb and $^{171}$Yb. Particularly, in $^{171}$Yb we identified mixing between $6snp$ and $6snf$ states with total angular momentum $F = 3/2$ and $F = 5/2$, which leads to significant variations in energies and wavefunctions near accidental degeneracies. Incorporating this effect into combined-channel MQDT models resolves previously unexplained deviations in the $6snp\,(F = 3/2)$ series~\cite{peper2025spectroscopy}. In addition, we characterized several previously unobserved $\ell = 4$ Rydberg series and demonstrated that they are better described in a $jj$-coupling framework—unlike the lower-$\ell$ states, which follow $LS$ coupling. Finally, we validated the MQDT models through static polarizability measurements of $6snd$ states, finding excellent agreement with predictions derived from the fitted energies and wavefunctions. These models enable quantitative predictions of atomic properties relevant for quantum computing and precision metrology, including Rydberg–Rydberg interactions, polarizabilities, and $g$-factors, and are made available through an open-source software package \texttt{rydcalc}~\cite{rydcalcgit}. However, several important challenges remain unresolved.

First, while existing MQDT models are constrained primarily by high-$n$ spectroscopic data ($n \gtrsim 20$), the accuracy at low-$n$ remains limited. In some cases, low-$n$ data exist but lack precision; in others, data are missing entirely. Moreover, the large number of accurate high-$n$ measurements skews the fits toward this regime, often at the expense of the model accuracy at low-$n$. Discrepancies at low $n$ may thus arise from either limitations in the available data or missing additional core-excited perturbing channels in the MQDT model. Improved spectroscopy in the low-$n$ regime is therefore essential for refining the models and capturing perturbations from nearby core-excited states, which are critical for calculating matrix elements and state lifetimes.

Second, while energy levels are now predicted to within a few MHz, modeling Rydberg state lifetimes remains challenging due to contributions from core-excited perturber channels~\cite{Vaillant2014Multichannel}. The resulting perturbing states are effectively low-$n$ members of Rydberg series converging to higher ionization thresholds, and their short lifetimes arise from rapid decay within the core-excited configuration. Since the relevant dipole matrix elements of the ionic core are poorly known and cannot be calculated within a single-electron approximation, their impact on Rydberg lifetimes is difficult to predict. Lifetime measurements across a range of $n$ are therefore essential to constrain these contributions and guide improved modeling. Such insights are essential for quantifying leakage pathways from $6sns$ Rydberg states and improving the erasure fraction and overall fidelity of Rydberg-based quantum gates.

Finally, further work is needed in the characterization of doubly excited Rydberg series, many of which remain poorly understood. In current MQDT models, several phenomenological channels are included to reproduce experimental spectra, but their physical identities are unknown. These states belong to Rydberg series converging to core-excited ionization thresholds. Members of these series can couple strongly to the continuum of lower lying ionization limits, resulting in rapid autoionization. Among them, the $6p_{1/2}n\ell$ states are of particular interest, as they can be used for high-fidelity Rydberg-state detection~\cite{Lochead2013Number, Madjarov2020High}, coherent control of population transfer~\cite{Burgers2022Controlling, Pham2022Coherent}, and erasure-based quantum error correction~\cite{Wu2022Erasure}. 

\begin{acknowledgments}
We wish to acknowledge helpful conversations with Yiyi Li, Yicheng Bao, and Chenyuan Li. This work was supported by the Army Research Office (W911NF-24-10358), the Office of Naval Research (N00014-20-1-2426), the National Science Foundation (QLCI grant OMA-2120757, and NSF CAREER PHY-2047620), the Sloan Foundation, and the Gordon and Betty Moore Foundation (Grant DOI 10.37807/gbmf12253). R.K. acknowledges Princeton University’s Office of Undergraduate Research OURSIP Program through the Hewlett Foundation. V.M.H. acknowledges Princeton University’s Office of Undergraduate Research OURSIP Program through the Class of 1952 Fund for Undergraduate Independent Work.
\end{acknowledgments}

\providecommand{\noopsort}[1]{}\providecommand{\singleletter}[1]{#1}%

\clearpage

\appendix

\newpage
\section{Polarizability Trends in $6snd$ Rydberg States}
\label{sec:polappendix}

\begin{figure}[!htb]
    \centering
    \includegraphics[width=\linewidth]{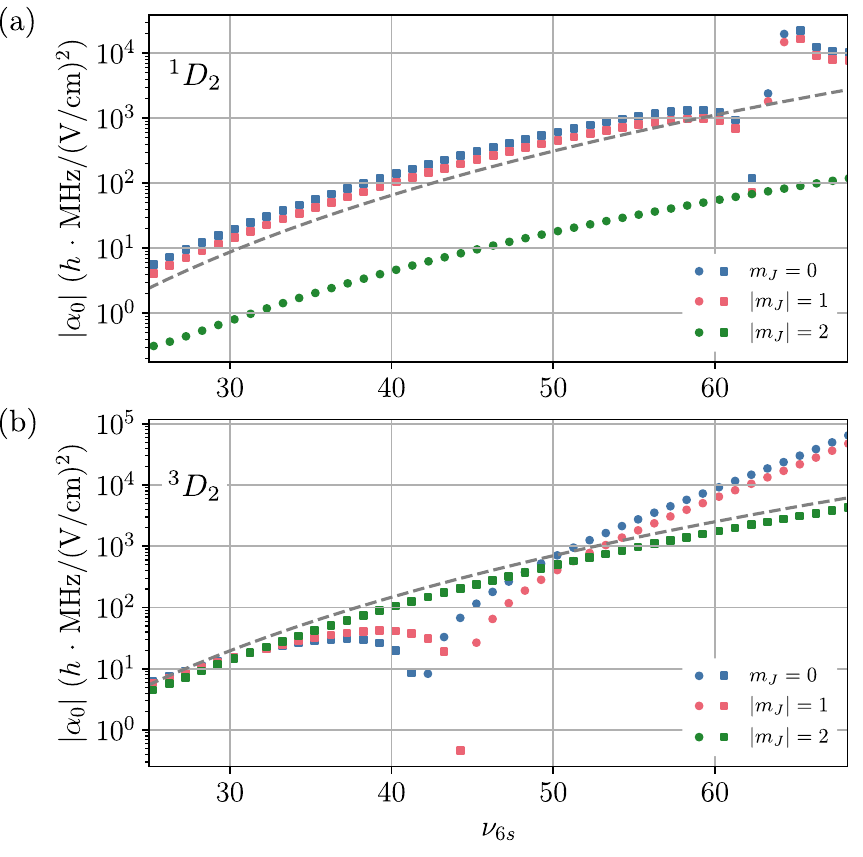}
    \caption{Predicted absolute static dipole polarizabilities of (a) $6snd\, ^1D_2$ and (b) $6snd\, ^3D_2$ states in $^{174}$Yb. The blue, red, and green points correspond to $m_J = 0, 1,$ and 2 sublevels, respectively. Circles indicate positive values of $\alpha_0$, while squares indicate negative values. The gray dashed lines show $\nu_{6s}^7$ scaling as a guide to the eye.}
    \label{fig:Dpol_174_summary}
\end{figure}

\begin{figure*}[tb]
    \centering
    \includegraphics[width=\linewidth]{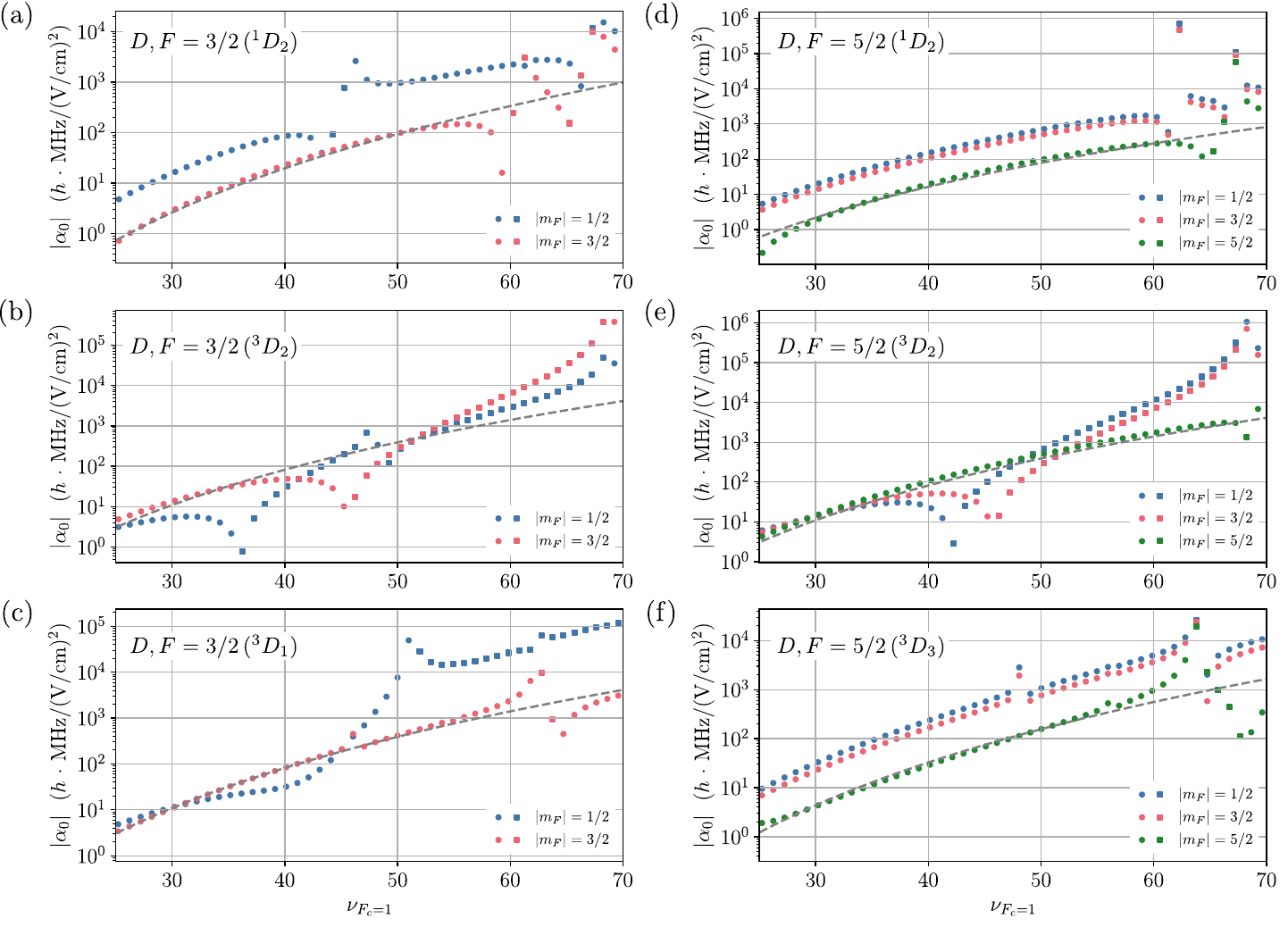}
    \caption{Predicted absolute static dipole polarizabilities of (a),(b),(c) $\ket{D, F=3/2}$ (series of dominantly $^1D_2, \,^3D_2, \, ^3D_1$ character, respectively) and (d),(e),(f) $\ket{D, F=5/2}$ (series of dominantly $^1D_2, \, ^3D_2, \, ^3D_3$ character, respectively) states in $^{171}$Yb. The blue, red, and green points correspond to $m_F = 1/2, 3/2$, and $5/2$ sublevels, respectively. Circles indicate positive values of $\alpha_0$, while squares indicate negative values. The gray dashed lines show $\nu_{F_c = 1}^7$ scaling as a guide to the eye.}
    \label{fig:Dpol_171_summary}
\end{figure*}

In Fig.~\ref{fig:Dpol_174_summary} and Fig.~\ref{fig:Dpol_171_summary}, we present predicted static dipole polarizabilities of the $6snd$ Rydberg series in $^{174}$Yb and $^{171}$Yb, respectively. In $^{174}$Yb, the $6snd\, ^1D_2$ static dipole polarizabilities exhibits a divergent feature near $\nu \approx 64$, due to energy crossings with the $6snp\, ^3P_1$ series. For $6snd\, ^3D_2$, the MQDT model predicts small static dipole polarizabilities near $\nu \approx 43$, giving Rydberg states that are highly insensitive to electric fields. For $^{171}$Yb, we present the predicted polarizabilities for the three $6snd\, (F=3/2)$ and the three $6snd\, (F=5/2)$ series. We note that these polarizabilities exhibit much more complex behavior due to energy crossings with multiple dipole-coupled series. 

\clearpage

\onecolumngrid
\section{MQDT Model Parameters}

\subsection{$^{174}$Yb}

\begin{table*}[ht]
\caption{\label{tab:174QDTs} Single-channel model parameters for the $6snd\,^3D_3$, $6snf\,^3F_2$, $6snf\,^3F_4$, $6sng\,^3G_3$, and $6sng\,^3G_5$ series of $^{174}$Yb described by Eq.~(\ref{eq:qd_energy}), obtained from fit to spectroscopic data presented in Tabs.~\cref{tab:174Spec_Microwave_3D3,tab:174Spec_Microwave_3F2,tab:174Spec_Microwave_3F4,tab:174Spec_Microwave_3G3}. $^\dagger$The $6sng\, ^3G_5$ parameters are obtained from a fit to the $^{171}$Yb $6sng\, (F=9/2)$ spectroscopic data.}
\begin{ruledtabular}
\begin{tabular}{cccccc}
 & $6snd\,^3D_3$ & $6snf\,^3F_2$ & $6snf\,^3F_4$ & $6sng\,^3G_3$ & $6sng\,^3G_5$ \\
\colrule
\rule{0pt}{3.0ex}$\mu_\alpha^{(0)}$ & 0.72902016 & 0.0718252326 & 0.0839027969 & 0.0260964574 & 0.02529201$^\dagger$ \\
$\mu_\alpha^{(2)}$ & $-0.705328923$ & $-1.00091963$ & $-2.91009023$ & $-0.14139526$ & $-0.11588052^\dagger$ \\
$\mu_\alpha^{(4)}$ & 829.238844 & $-106.291066$ & & & \\

\end{tabular}
\end{ruledtabular}
\end{table*}

\begin{table*}[ht]
\caption{\label{tab:174MQDT_1S0} Six-channel MQDT model parameters for the $6sns\,^{1}S_0$ series of $^{174}$Yb, obtained by fitting to spectroscopic data reported in Refs.~\cite{Lehec2018Laser, Aymar1980Highly, Maeda1992Optical}. The electronic configuration of the perturbing Rydberg channels $4f^{13}5d6snp$ is abbreviated as $5d$. The rotations $\mathcal{R}(\theta_{ij})$ are applied in the order $\mathcal{R}(\theta_{12})\mathcal{R}(\theta_{13})\mathcal{R}(\theta_{14})\mathcal{R}(\theta_{34})\mathcal{R}(\theta_{35})\mathcal{R}(\theta_{16})$.
}
\begin{ruledtabular}
\begin{tabular}{ccccccc}
$i$,$\bar\alpha$,$\alpha$ & 1 & 2 & 3& 4& 5 & 6\\
\colrule
\rule{0pt}{3.0ex}$\ket{i}$ & $(6s_{1/2})(ns_{1/2})$& $5d$ \textbf{a} & $(6p_{3/2})(np_{3/2})$ & $5d$ \textbf{b} & $(6p_{1/2})(np_{1/2})$ & $5d$ \textbf{c} \\ 
$I_i$ ($\mathrm{cm}^{-1}$) & $50\,443.070393$ & $83\,967.7$ & $ 80\,835.39$ & $83\,967.7$ & $77\,504.98$& $83\,967.7$\\
$\ket{\bar\alpha}$ & $6sns\,^1S_0$ & $5d$ \textbf{a} & $6pnp\,^1S_0$ & $5d$ \textbf{b} & $6pnp\,^3P_0$ & $5d$ \textbf{c} \\ 
$\mu_\alpha^{(0)}$ & 0.355101645 & 0.204537535 & 0.116393648 & 0.295439966 &0.257664798 & 0.155797119 \\
$\mu_\alpha^{(2)}$ & 0.277673956 & 0.0 & 0.0 & 0.0 & 0.0 & 0.0\\
\colrule
\rule{0pt}{3.0ex}$V_{\bar\alpha\alpha}$  & $\theta_{12}=0.126557575$ & $\theta_{13}=0.300103593$ & $\theta_{14}=0.056987912$ & $\theta_{34}=0.114312578$  & $\theta_{35}=0.0986363362$ & $\theta_{16}=0.142498543$  \\
\colrule
\rule{0pt}{3.0ex}$U_{i\bar\alpha}$   & 1&0&0&0&0&0\\
                    & 0&1&0&0&0&0\\
                    & 0&0&$\sqrt{2/3}$&0&$-\sqrt{1/3}$&0\\
                    & 0&0&0&1&0&0\\
                    & 0&0&$\sqrt{1/3}$&0&$\sqrt{2/3}$&0\\
                    & 0&0&0&0&0&1\\

\end{tabular}
\end{ruledtabular}
\end{table*}

\begin{table*}[ht]
\caption{\label{tab:174MQDT_3P0} Two-channel MQDT model parameters for the $6snp\,^{3}P_0$ series of $^{174}$Yb, obtained by fitting to spectroscopic data reported in Refs.~\cite{peper2025spectroscopy, Aymar1984three}. Following Refs.~\cite{Aymar1984three, peper2025spectroscopy}, we include a perturbing channel of character $4f^{13}5d6snd\,^3P_0$ (labeled $Q$) with an ionization limit of $\SI{83967.7}{\mathrm{cm}^{-1}}$.}
\begin{ruledtabular}
\begin{tabular}{ccc}
$i$,$\bar\alpha$,$\alpha$ & 1 & 2\\
\colrule
\rule{0pt}{3.0ex}$\ket{i}$ & $(6s_{1/2})(np_{1/2})$& $4f^{13}5d6snd$ \\ 
$I_i$ ($\mathrm{cm}^{-1}$) & $50\,443.070393$ & $83\,967.7$\\
$\ket{\bar\alpha}$ & $6snp\,^3P_0$ & $Q$\\
$\mu_\alpha^{(0)}$ & 0.953661478 & 0.198460766 \\
$\mu_\alpha^{(2)}$ & $-0.287531374$ & 0.0 \\
\colrule
\rule{0pt}{3.0ex}$V_{\bar\alpha\alpha}$ & $\theta_{12}=0.163343232$ &  \\
\colrule
\rule{0pt}{3.0ex}$U_{i\bar\alpha}$    & 1 & 0 \\
                     & 0 & 1 \\

\end{tabular}
\end{ruledtabular}
\end{table*}

\begin{table*}[ht]
\caption{\label{tab:174MQDT_13P1} Six-channel MQDT model parameters for the $6snp\,^{1,3}P_1$ series of $^{174}$Yb, obtained by fitting to spectroscopic data reported in Refs.~\cite{peper2025spectroscopy, Aymar1984three, Lehec2017PhD, Camus1969spectre}. The electronic configuration of the perturbing Rydberg channels $4f^{13}5d6snd$ is abbreviated as $5d$. The rotations $\mathcal{R}(\theta_{ij})$ are applied in the order $\mathcal{R}(\theta_{12}(\varepsilon))\mathcal{R}(\theta_{13})\mathcal{R}(\theta_{14})\mathcal{R}(\theta_{15})\mathcal{R}(\theta_{16})\mathcal{R}(\theta_{23})\mathcal{R}(\theta_{24})\mathcal{R}(\theta_{25})\mathcal{R}(\theta_{26})$.}
\begin{ruledtabular}
\begin{tabular}{ccccccc}
$i$,$\bar\alpha$,$\alpha$ & 1 & 2 & 3& 4& 5 & 6\\
\colrule
\rule{0pt}{3.0ex}$\ket{i}$ & $(6s_{1/2})(np_{3/2})$& $(6s_{1/2})(np_{1/2})$ & $5d$ \textbf{a} & $5d$ \textbf{b}& $5d$ \textbf{c}& $5d$ \textbf{d}\\ 
$I_i$ ($\mathrm{cm}^{-1}$) & $50\,443.070393$ & $50\,443.070393$ & $83\,967.7$ & $83\,967.7$ & $83\,967.7$& $83\,967.7$\\
$\ket{\bar\alpha}$ & $6snp\,^1P_1$ & $6snp\,^3P_1$ & $5d$ \textbf{a} & $5d$ \textbf{b}& $5d$ \textbf{c}& $5d$ \textbf{d}\\
$\mu_\alpha^{(0)}$ & 0.922709076 & 0.982084772 & 0.228518316& 0.206081775 & 0.193527605 & 0.181533031 \\
$\mu_\alpha^{(2)}$ & 2.60055203 & $-5.45063476$ & 0.0 & 0.0 & 0.0 & 0.0\\
\colrule
\rule{0pt}{3.0ex}$V_{\bar\alpha\alpha}$  & $\theta_{12}^{(0)}=-0.08410871$ & $\theta_{12}^{(2)}=120.37555$  & $\theta_{12}^{(4)}=-9\,314.23$  &$\theta_{13}=-0.07317986$ & $\theta_{14}=-0.06651879$ & $\theta_{15}=-0.02212194$    \\
    & $\theta_{16}=-0.10452109$ & $\theta_{23}=0.02477464$ &  $\theta_{24}=0.05763934$ & $\theta_{25}=0.0860644$ & $\theta_{26}=0.04993818$  \\  
\colrule
\rule{0pt}{3.0ex}$U_{i\bar\alpha}$   & $\sqrt{2/3}$&$\sqrt{1/3}$&0&0&0&0\\
                    & $-\sqrt{1/3}$&$\sqrt{2/3}$&0&0&0&0\\
                    & 0&0&1&0&0&0\\
                    & 0&0&0&1&0&0\\
                    & 0&0&0&0&1&0\\
                    & 0&0&0&0&0&1\\

\end{tabular}
\end{ruledtabular}
\end{table*}

\begin{table*}[ht]
\caption{\label{tab:174MQDT_3P2} Four-channel MQDT model parameters for the $6snp\,^{3}P_2$ series of $^{174}$Yb, obtained by fitting to spectroscopic data reported in Refs.~\cite{Aymar1984three, Lehec2017PhD, Martin1978Atomic, BiRu1991The, Ali1999Two, Wyart1979Extended}. The electronic configuration of the perturbing Rydberg channels $4f^{13}5d6snd$ is abbreviated as $5d$. The rotations $\mathcal{R}(\theta_{ij})$ are applied in the order $\mathcal{R}(\theta_{12})\mathcal{R}(\theta_{13})\mathcal{R}(\theta_{14})$}
\begin{ruledtabular}
\begin{tabular}{ccccc}
$i$,$\bar\alpha$,$\alpha$ & 1 & 2 & 3 & 4\\
\colrule
\rule{0pt}{3.0ex}$\ket{i}$ & $(6s_{1/2})(np_{3/2})$& $5d$ \textbf{a} & $5d$ \textbf{b} & $5d$ \textbf{c} \\ 
$I_i$ ($\mathrm{cm}^{-1}$) & $50\,443.070393$ & $83\,967.7$ & $83\,967.7$ & $83\,967.7$\\
$\ket{\bar\alpha}$ & $6snp\,^3P_2$ & $5d$ \textbf{a} & $5d$ \textbf{b} & $5d$ \textbf{c}\\
$\mu_\alpha^{(0)}$ & 0.925150932 & 0.230028034 & 0.209224174 & 0.186236574   \\
$\mu_\alpha^{(2)}$ & $-2.69197178$ & 0.0 & 0.0 & 0.0  \\
$\mu_\alpha^{(4)}$ & 66.7159709 & 0.0 & 0.0 & 0.0  \\
\colrule
\rule{0pt}{3.0ex}$V_{\bar\alpha\alpha}$ & $\theta_{12}=0.0706189664$ & $\theta_{13}=0.0231221428$ & $\theta_{14}=-0.0291730345$ &  \\
\colrule
\rule{0pt}{3.0ex}$U_{i\bar\alpha}$    & 1 & 0 & 0 & 0 \\
                     & 0 & 1 & 0 & 0 \\
                     & 0 & 0 & 1 & 0 \\
                     & 0 & 0 & 0 & 1 \\
\end{tabular}
\end{ruledtabular}
\end{table*}

\begin{table}[ht]
\caption{\label{tab:174MQDT_13D2} Five-channel MQDT model parameters for the $6snd\,^{1,3}D_2$ series of $^{174}$Yb, obtained by fitting to spectroscopic data reported in Refs.~\cite{Aymar1984three, Lehec2018Laser, Maeda1992Optical}. The electronic configuration of the perturbing Rydberg channels $4f^{13}5d6snp$ is abbreviated as $5d$. The rotations $\mathcal{R}(\theta_{ij})$ are applied in the order $\mathcal{R}(\theta_{12}(\varepsilon))\mathcal{R}(\theta_{13})\mathcal{R}(\theta_{14})\mathcal{R}(\theta_{24})\mathcal{R}(\theta_{15})\mathcal{R}(\theta_{25})$. The channel-$g$-factors $g^*$ are as given in Ref.~\cite{peper2025spectroscopy}.}
\begin{ruledtabular}
\begin{tabular}{cccccc}
$i$,$\bar\alpha$,$\alpha$ & 1 & 2 & 3& 4& 5 \\
\colrule
\rule{0pt}{3.0ex}$\ket{i}$ & $(6s_{1/2})(nd_{5/2})$&  $(6s_{1/2})(nd_{3/2})$ & $5d$ \textbf{a} & $5d$ \textbf{b} & $(6p)(np)$ \\ 
$I_i$ ($\mathrm{cm}^{-1}$) & $50\,443.070393$ & $50\,443.070393$ & $83\,967.7$ &  $83\,967.7$ & $79\,725.35$\\
$\ket{\bar\alpha}$ & $6snd\,^1D_2$ &  $6snd\,^3D_2$ & $5d$ \textbf{a} &  $5d$ \textbf{b} & $6pnp\,^1D_2$  \\ 
$\mu_\alpha^{(0)}$ & 0.729513646 & 0.752292223 & 0.19612036& 0.233752026 &0.152911249  \\
$\mu_\alpha^{(2)}$ & $-0.0377841183$ & 0.104072325 & 0.0 & 0.0 & 0.0\\
\colrule
\rule{0pt}{3.0ex}$V_{\bar\alpha\alpha}$  & $\theta_{12}^{(0)}=0.21157531$ & $\theta_{13}=0.00521559431$ & $\theta_{14}=0.0398131577$ & $\theta_{24}=-0.0071658109$  & $\theta_{15}=0.10481227$   \\
& $\theta_{12}^{(2)}=-15.3844$&  $\theta_{25}=0.0721660042$ & & & \\
\colrule
\rule{0pt}{3.0ex}$g^*$ & 1  & 1.1670 &1.1846 & 0.9390& 1.2376 \\
\colrule
\rule{0pt}{3.0ex}$U_{i\bar\alpha}$   & $\sqrt{3/5}$&$\sqrt{2/5}$&0&0&0\\
                    & $-\sqrt{2/5}$&$\sqrt{3/5}$&0&0&0\\
                    & 0&0&1&0&0\\
                    & 0&0&0&1&0\\
                    & 0&0&0&0&1\\

\end{tabular}
\end{ruledtabular}
\end{table}

\begin{sidewaystable}[ht]
\caption{\label{tab:174MQDT_13F3} Seven-channel MQDT model parameters for the $6snf\,^{1,3}F_3$ series of $^{174}$Yb, obtained by fitting to spectroscopic data presented in Tabs.~\cref{tab:174YbSpec_13F3,tab:174Spec_Microwave_13F3,tab:174Spec_Microwave_13F3_twophoton}. The electronic configuration of the perturbing Rydberg channels $4f^{13}5d6snd$ is abbreviated as $5d$. The rotations $\mathcal{R}(\theta_{ij})$ are applied in the order $\mathcal{R}(\theta_{12}(\varepsilon))\mathcal{R}(\theta_{13})\mathcal{R}(\theta_{14})\mathcal{R}(\theta_{15})\mathcal{R}(\theta_{16})\mathcal{R}(\theta_{17})\mathcal{R}(\theta_{23})\mathcal{R}(\theta_{24})\mathcal{R}(\theta_{25})\mathcal{R}(\theta_{26})\mathcal{R}(\theta_{27})$. }
\begin{ruledtabular}
\begin{tabular}{cccccccc}
$i$,$\bar\alpha$,$\alpha$ & 1 & 2 & 3& 4& 5 & 6 & 7 \\
\colrule
\rule{0pt}{3.0ex}$\ket{i}$ & $(6s_{1/2})(nf_{7/2})$& $(6s_{1/2})(nf_{5/2})$ & $5d$ \textbf{a} & $5d$ \textbf{b}& $5d$ \textbf{c}& $5d$ \textbf{d} & $5d$ \textbf{e}\\ 
$I_i$ ($\mathrm{cm}^{-1}$) & $50\,443.070393$ & $50\,443.070393$ & $83\,967.7$ & $83\,967.7$ & $83\,967.7$& $83\,967.7$ & $83\,967.7$ \\
$\ket{\bar\alpha}$ & $6snf\,^1F_3$ & $6snf\,^3F_3$ & $5d$ \textbf{a} & $5d$ \textbf{b}& $5d$ \textbf{c}& $5d$ \textbf{d} & $5d$ \textbf{e}\\
$\mu_\alpha^{(0)}$ & 0.276158949 & 0.0715123712 & 0.239015576 & 0.226770354 & 0.175354845 & 0.196660618 & 0.21069642 \\
$\mu_\alpha^{(2)}$ & $-12.7258012$ & $-0.768462937$ & 0.0 & 0.0 & 0.0 & 0.0  & 0.0 \\
\colrule
\rule{0pt}{3.0ex}$V_{\bar\alpha\alpha}$  & $\theta_{12}^{(0)}$ & $\theta_{12}^{(2)}$  &$\theta_{13}$ & $\theta_{14}$ & $\theta_{15}$  & $\theta_{16}$  \\
& $-0.0208481417$ & $0.239045493$ & $-0.00411835457$ & $-0.0962784945$ & 0.132826901 & $-0.0439244317$ \\ 
\colrule
\rule{0pt}{3.0ex}& $\theta_{17}$ & $\theta_{23}$ &  $\theta_{24}$ & $\theta_{25}$ & $\theta_{26}$ & $\theta_{27}$ \\ 
& 0.0508460294 & $-0.0376574252$ & 0.026944623 & $-0.0148474857$ & $-0.0521244126$ & 0.0349516329 \\ 
\colrule
\rule{0pt}{3.0ex}$U_{i\bar\alpha}$ & $\sqrt{4/7}$ & $\sqrt{3/7}$ & 0 & 0 & 0 & 0 & 0 \\ 
& $-\sqrt{3/7}$ & $\sqrt{4/7}$ & 0 & 0 & 0 & 0 & 0 \\ 
&0 & 0 & 1 & 0 & 0 & 0 & 0 \\ 
&0 & 0 & 0 & 1 & 0 & 0 & 0 \\ 
&0 & 0 & 0 & 0 & 1 & 0 & 0 \\
&0 & 0 & 0 & 0 & 0 & 1 & 0 \\ 
&0 & 0 & 0 & 0 & 0 & 0 & 1 \\ 

\end{tabular}
\end{ruledtabular}
\end{sidewaystable}

\begin{table*}[ht]
\caption{\label{tab:174MQDT_PMG4} Two-channel MQDT model parameters for the $6sng\,^\pm G_4$ series of $^{174}$Yb, obtained by fitting to spectroscopic data presented in Tab.~\ref{tab:174Spec_Microwave_pmG4}.}
\begin{ruledtabular}
\begin{tabular}{ccc}
$i$,$\bar\alpha$,$\alpha$ & 1 & 2 \\
\colrule
\rule{0pt}{3.0ex}$\ket{i}$ & $(6s_{1/2})(ng_{9/2})$& $(6s_{1/2})(ng_{7/2})$ \\ 
$I_i$ ($\mathrm{cm}^{-1}$) & $50\,443.070393$ & $50\,443.070393$ \\
$\ket{\bar\alpha}$ & $6sng\,^+G_4$ & $6sng\,^-G_4$ \\
$\mu_\alpha^{(0)}$ & 0.0262659964 & $-0.148808463$ \\
$\mu_\alpha^{(2)}$ & 0.0254568575 & $-0.134219071$ \\
\colrule
\rule{0pt}{3.0ex}$V_{\bar\alpha\alpha}$  & $\theta_{12}^{(0)}=-0.08222676$ & \\  
\colrule
\rule{0pt}{3.0ex}$U_{i\bar\alpha}$  & 1 & 0 \\ 
& 0 & 1 \\
\end{tabular}
\end{ruledtabular}
\end{table*}

\clearpage
\subsection{$^{171}$Yb}
\label{sec:171yb_mqdt}

\begin{sidewaystable}[ht]
\caption{\label{tab:171MQDT_S_F12} Seven-channel MQDT model parameters for the $\ket{\nu,L=0,F=1/2}$ series of $^{171}$Yb, obtained by fitting to spectroscopic data reported in Ref.~\cite{peper2025spectroscopy}. The electronic configuration of the perturbing Rydberg channels $4f^{13}5d6snp$ is abbreviated as $5d$. The rotations $\mathcal{R}(\theta_{ij})$ are applied in the order $\mathcal{R}(\theta_{12})\mathcal{R}(\theta_{13})\mathcal{R}(\theta_{14})\mathcal{R}(\theta_{34})\mathcal{R}(\theta_{35})\mathcal{R}(\theta_{16})$.}
\begin{ruledtabular}
\begin{tabular}{cccccccc}
$i$,$\bar\alpha$,$\alpha$ & 1 $(F_c=0)$& 2 & 3& 4& 5 & 6 & 7 $(F_c=1)$ \\
\colrule
\rule{0pt}{3.0ex}$\ket{i_F}$ & $(6s_{1/2})(ns_{1/2})$ & $5d$ \textbf{a} & $(6p_{3/2})(np_{3/2})$ & $5d$ \textbf{b} & $(6p_{1/2})(np_{1/2})$ & $5d$ \textbf{c} & $(6s_{1/2})(ns_{1/2})$ \\ 
$I_i$ ($\mathrm{cm}^{-1}$) & $50\,442.795744$ & $83\,967.7$ & $ 80\,835.39$ & $83\,967.7$ & $77\,504.98$& $83\,967.7$ & $50\,443.217463$\\
$\ket{\bar\alpha}$ & $6sns\,^1S_0$ & $5d$ \textbf{a} & $6pnp\,^1S_0$ & $5d$ \textbf{b} & $6pnp\,^3P_0$ & $5d$ \textbf{c} & $6sns\,^3S_1$ \\ 
$\mu_\alpha^{(0)}$ & 0.357488757 & 0.203917828 & 0.116813499& 0.287210377&0.247550262 & 0.148686263 &  $\mu_{^3S_1}^{rr}$\\
$\mu_\alpha^{(2)}$ & 0.163255076 & 0.0 & 0.0 & 0.0 & 0.0 & 0.0 &\\
\colrule
\rule{0pt}{3.0ex}$\mu_{^3S_1}^{rr}$ & $\mu_{^3S_1}^{(0)}=0.438803844$ & $\mu_{^3S_1}^{(2)}=3.54561559$ &  $\mu_{^3S_1}^{(4)}=-10673.2496$& $\mu_{^3S_1}^{(6)}=7702824.55$& $\mu_{^3S_1}^{(8)}=-2255973430$& &\\[1.0ex]
\colrule
\rule{0pt}{3.0ex}$V_{\bar\alpha\alpha}$  & $\theta_{12}=0.13179534$ & $\theta_{13}=0.29748039$ & $\theta_{14}=0.0553920359$ & $\theta_{34}=0.100843905$  & $\theta_{35}=0.10317753$ & $\theta_{16}=0.137709223$ &  \\
\colrule
\rule{0pt}{3.0ex}$U_{i\bar\alpha}$   & $1/2$&0&0&0&0&0&$\sqrt{3}/2$\\
                    & 0&1&0&0&0&0&0\\
                    & 0&0&$\sqrt{2/3}$&0&$-\sqrt{1/3}$&0&0\\
                    & 0&0&0&1&0&0&0\\
                    & 0&0&$\sqrt{1/3}$&0&$\sqrt{2/3}$&0&0\\
                    & 0&0&0&0&0&1&0\\
                    & $\sqrt{3}/2$&0&0&0&0&0&$-1/2$\\

\end{tabular}
\end{ruledtabular}
\end{sidewaystable}

\begin{sidewaystable}[ht]
\caption{\label{tab:171MQDT_P12} Eight-channel MQDT model parameters for the $\ket{\nu,\ell=1,F=1/2}$ series of $^{171}$Yb, obtained by fitting to spectroscopic data reported in Refs.~\cite{Majewski1985diploma, peper2025spectroscopy}. The electronic configuration of the perturbing Rydberg channels $4f^{13}5d6snd$ is abbreviated as $5d$. The rotations $\mathcal{R}(\theta_{ij})$ are applied in the order $\mathcal{R}(\theta_{12}(\varepsilon))\mathcal{R}(\theta_{27})\mathcal{R}(\theta_{13})\mathcal{R}(\theta_{14})\mathcal{R}(\theta_{15})\mathcal{R}(\theta_{16})\mathcal{R}(\theta_{23})\mathcal{R}(\theta_{24})\mathcal{R}(\theta_{25})\mathcal{R}(\theta_{26})\mathcal{R}(\theta_{78})$.}
\begin{ruledtabular}
\begin{tabular}{ccccccccc}
$i$,$\bar\alpha$,$\alpha$ & 1 $(F_c=1)$ & 2 $(F_c=1)$& 3& 4& 5 & 6 & 7 $(F_c=0)$& 8\\
\colrule
\rule{0pt}{3.0ex}$\ket{i}$ & $(6s_{1/2})(np_{3/2})$& $(6s_{1/2})(np_{1/2})$& $5d$ \textbf{a} & $5d$ \textbf{b}& $5d$ \textbf{c}& $5d$ \textbf{d}& $(6s_{1/2})(np_{1/2})$&$5d$ \textbf{e} \\ 
$I_i$ ($\mathrm{cm}^{-1}$) & $50\,443.217463$ & $50\,443.217463$ & $83\,967.7$ & $83\,967.7$ & $83\,967.7$& $83\,967.7$ &$50\,442.795744$ &$83\,967.7$\\
$\ket{\bar\alpha}$ & $6snp\,^1P_1$ & $6snp\,^3P_1$ & $5d$ \textbf{a} & $5d$ \textbf{b}& $5d$ \textbf{c}& $5d$ \textbf{d} & $6snp\,^3P_0$ & $5d$ \textbf{e}\\
$\mu_\alpha^{(0)}$ & 0.922046003 & 0.981076381 & 0.229019212& 0.205999445&0.19352756 & 0.181271472 & 0.953288122 & 0.198451139 \\
$\mu_\alpha^{(2)}$ & 2.17932092 & $ -4.48995034$ & 0.0 & 0.0 & 0.0 & 0.0  & $-0.0298396847$& 0.0\\
\colrule
\rule{0pt}{3.0ex}$V_{\bar\alpha\alpha}$  & $\theta_{12}^{(0)}$& $\theta_{12}^{(2)}$ & $\theta_{12}^{(4)}$ & $\theta_{27}$ &$\theta_{13}$ &$\theta_{14}$ & $\theta_{15}$ & $\theta_{16}$  \\ 
 & $-0.100292816$ & $149.140925$& $-13487.7692$& $-0.0016592076$ &$-0.0727917308$ & $-0.0669120237$ & $-0.0221321759$  & $-0.107302569$    \\
\colrule

\rule{0pt}{3.0ex} &$\theta_{23}$&$\theta_{24}$ &$\theta_{25}$ &$\theta_{26}$ & $\theta_{78}$& & &   \\

   & $0.0396527798$  & $0.0596597186$ & $0.0861416497$ & $0.0565415641$ & $0.163175562$ & & \\ 
\colrule
\rule{0pt}{3.0ex}$U_{i\bar\alpha}$   & $-\sqrt{2/3}$&$-\sqrt{1/3}$&0&0&0&0& 0&0\\
                    &$1/\left(2\sqrt{3}\right)$ &$-\sqrt{1/6}$&0&0&0&0& $\sqrt{3}/2$&0\\
                    & 0&0&1&0&0&0&0&0\\
                    & 0&0&0&1&0&0&0&0\\
                    & 0&0&0&0&1&0&0&0\\
                    & 0&0&0&0&0&1&0&0\\
                    & $-1/2$&$1/\sqrt{2}$&0&0&0&0& $1/2$&0\\
                    & 0&0&0&0&0&0& 0&1\\
\end{tabular}
\end{ruledtabular}
\end{sidewaystable}

\begin{sidewaystable}[ht]
\caption{\label{tab:171MQDT_F72} Eight-channel MQDT model parameters for the $\ket{\nu,\ell=3,F=7/2}$ series of $^{171}$Yb, obtained by fitting to spectroscopic data presented in Tab.~\ref{tab:171Spec_F72}. The electronic configuration of the perturbing Rydberg channels $4f^{13}5d6snd$ is abbreviated as $5d$. The rotations $\mathcal{R}(\theta_{ij})$ are applied in the order $\mathcal{R}(\theta_{12}(\varepsilon))\mathcal{R}(\theta_{13})\mathcal{R}(\theta_{14})\mathcal{R}(\theta_{15})\mathcal{R}(\theta_{16})\mathcal{R}(\theta_{23})\mathcal{R}(\theta_{24})\mathcal{R}(\theta_{25})\mathcal{R}(\theta_{26})\mathcal{R}(\theta_{27})$.
}
\begin{ruledtabular}
\begin{tabular}{ccccccccc}
$i$,$\bar\alpha$,$\alpha$ & 1 $(F_c=1)$ & 2 $(F_c=1)$& 3& 4& 5 & 6 & 7 & 8 $(F_c=0)$\\
\colrule
\rule{0pt}{3.0ex}$\ket{i}$ & $(6s_{1/2})(nf_{7/2})$& $(6s_{1/2})(nf_{5/2})$& $5d$ \textbf{a} & $5d$ \textbf{b}& $5d$ \textbf{c}& $5d$ \textbf{d}& $5d$ \textbf{e} & $(6s_{1/2})(nf_{7/2})$ \\ 
$I_i$ ($\mathrm{cm}^{-1}$) & $50\,443.217463$ & $50\,443.217463$ & $83\,967.7$ & $83\,967.7$ & $83\,967.7$& $83\,967.7$ &$83\,967.7$ &$50\,442.795744$ \\
$\ket{\bar\alpha}$ & $6snf\,^1F_3$ & $6snf\,^3F_3$ & $5d$ \textbf{a} & $5d$ \textbf{b}& $5d$ \textbf{c}& $5d$ \textbf{d} & $5d$ \textbf{e} & $6snf\,^3F_4$ \\
$\mu_\alpha^{(0)}$ & $0.27707535$ & $0.0719827367$ & $0.250829254$ & $0.227042016$ & $0.17576256$ & $0.19654694$  & $0.214351625$  &  $\mu_{^3F_4}^{rr}$\\
$\mu_\alpha^{(2)}$ & $-13.2829133$ & $-0.741492064$  &  &  & &  & & \\
\colrule
\rule{0pt}{3.0ex}$\mu_{^3F_4}^{rr}$ & $\mu_{^3F_4}^{(0)} = 0.0834066138$ & $\mu_{^3F_4}^{(2)} = -1.11349283$ & $\mu_{^3F_4}^{(4)} = -1539.63739$ \\[1.0ex]
\colrule
\rule{0pt}{3.0ex}$V_{\bar\alpha\alpha}$  & $\theta_{12}^{(0)}$& $\theta_{12}^{(2)}$ &$\theta_{13}$ &$\theta_{14}$ & $\theta_{15}$ & $\theta_{16}$ & $\theta_{17}$ \\ 
 & $-0.0208481417$ & $0.239045493$ & $-0.0593782568$ & $-0.0755947274$ & $0.122678758$ & $-0.0400418576$ & $0.0646373252$ \\
\colrule
\rule{0pt}{3.0ex}&$\theta_{23}$&$\theta_{24}$ &$\theta_{25}$ &$\theta_{26}$ & $\theta_{27}$& & &   \\
   & $-0.0684538786$ & $0.0352130279$ & $-0.0326572035$ & $-0.050215299$ & $0.0453892695$ & & \\ 
\colrule
\rule{0pt}{3.0ex}$U_{i\bar\alpha}$  
& $3/({2\sqrt7})$ & $3\sqrt{3}/(4\sqrt{7})$ & 0 & 0 & 0 & 0 & 0 & $-\sqrt{7}/4$ \\
& $-\sqrt{3/7}$ & $\sqrt{2/7}$ & 0 & 0 & 0 & 0 & 0 & 0 \\
& 0 & 0 & 1 & 0 & 0 & 0 & 0 & 0 \\
& 0 & 0 & 0 & 1 & 0 & 0 & 0 & 0 \\
& 0 & 0 & 0 & 0 & 1 & 0 & 0 & 0 \\
& 0 & 0 & 0 & 0 & 0 & 1 & 0 & 0 \\
& 0 & 0 & 0 & 0 & 0 & 0 & 1 & 0 \\
& $1/2$ & $\sqrt{3}/4$ & 0 & 0 & 0 & 0 & 0 & $3/4$ \\

\end{tabular}
\end{ruledtabular}
\end{sidewaystable}

\begin{sidewaystable}[ht]
\caption{\label{tab:171MQDT_D32} Six-channel MQDT model parameters for the $\ket{\nu,L=2,F=3/2}$ series of $^{171}$Yb, obtained by fitting to spectroscopic data presented in Ref.~\cite{peper2025spectroscopy}. The electronic configuration of the perturbing Rydberg channels $4f^{13}5d6snp$ is abbreviated as $5d$. The rotations $\mathcal{R}(\theta_{ij})$ are applied in the order $\mathcal{R}(\theta_{12})\mathcal{R}(\theta_{13})\mathcal{R}(\theta_{14})\mathcal{R}(\theta_{24})\mathcal{R}(\theta_{15})\mathcal{R}(\theta_{25})$.
}
\begin{ruledtabular}
\begin{tabular}{ccccccc}
$i$,$\bar\alpha$,$\alpha$ & 1 $(F_c=1)$ & 2 $(F_c=1)$& 3& 4& 5 & 6 $(F_c=0)$\\
\colrule
\rule{0pt}{3.0ex}$\ket{i}$ & $(6s_{1/2})(nd_{5/2})$&  $(6s_{1/2})(nd_{3/2})$ & $5d$ \textbf{a} & $5d$ \textbf{b} & $(6p)(np)$ &$(6s_{1/2})(nd_{3/2})$\\ 
$I_i$ ($\mathrm{cm}^{-1}$) & $50\,443.217463$ & $50\,443.217463$ & $83\,967.7$ &  $83\,967.7$ & $79\,725.35$ & $50\,442.795744$\\
$\ket{\bar\alpha}$ & $6snd\,^1D_2$ &  $6snd\,^3D_2$ & $5d$ \textbf{a} &  $5d$ \textbf{b} & $6pnp\,^1D_2$ & $6snd\,^3D_1$  \\ 
$\mu_\alpha^{(0)}$ & 0.730541589 & 0.751542685 & 0.195864083& 0.235944408&0.147483609 & $\mu_{^3D_1}^{rr}$  \\
$\mu_\alpha^{(2)}$ & $-0.0967938662$ & $0.00038836127$ & 0.0 & 0.0 & 0.0 &\\
\colrule
\rule{0pt}{3.0ex}$\mu_{^3D_1}^{rr}$ &$\mu_{^3D_1}^{(0)}=0.75334159$  &$\mu_{^3D_1}^{(2)}=-1.80187083$  &$\mu_{^3D_1}^{(4)}=986.918851$  &  & &  \\[1.0ex]
\colrule
\rule{0pt}{3.0ex}$V_{\bar\alpha\alpha}$  & $\theta_{12}^{(0)}=0.220048245$ & $\theta_{13}=0.00427599$ & $\theta_{14}=0.0381563093$ & $\theta_{24}=-0.00700797918$  & $\theta_{15}=0.109380331$ & $\theta_{25}=0.0635544456$  \\
& $\theta_{12}^{(2)} = -14.9486$\\
\colrule
\rule{0pt}{3.0ex}$U_{i\bar\alpha}$   & $-\sqrt{3/5}$&$-\sqrt{2/5}$&0&0&0&0\\
                    & $\sqrt{3/5}/2$&$-3/\left(2\sqrt{10}\right)$&0&0&0&$\sqrt{5/2}/2$\\
                    & 0&0&1&0&0&0 \\
                    & 0&0&0&1&0&0\\
                    & 0&0&0&0&1&0\\
                    & $-1/2$&$\sqrt{3/2}/2$&0&0&0&$\sqrt{3/2}/2$\\

\end{tabular}
\end{ruledtabular}
\end{sidewaystable}

\begin{sidewaystable}[ht]
\caption{\label{tab:171MQDT_D52} Six-channel MQDT model parameters for the $\ket{\nu,L=2,F=5/2}$ series of $^{171}$Yb, obtained by fitting to spectroscopic data reported in Ref.~\cite{peper2025spectroscopy}. The electronic configuration of the perturbing Rydberg channels $4f^{13}5d6snp$ is abbreviated as $5d$. The rotations $\mathcal{R}(\theta_{ij})$ are applied in the order $\mathcal{R}(\theta_{12})\mathcal{R}(\theta_{13})\mathcal{R}(\theta_{14})\mathcal{R}(\theta_{24})\mathcal{R}(\theta_{15})\mathcal{R}(\theta_{25})$.
}
\begin{ruledtabular}
\begin{tabular}{ccccccc}
$i$,$\bar\alpha$,$\alpha$ & 1 $(F_c=1)$ & 2 $(F_c=1)$& 3& 4& 5 & 6 $(F_c=0)$\\
\colrule
\rule{0pt}{3.0ex}$\ket{i}$ & $(6s_{1/2})(nd_{5/2})$&  $(6s_{1/2})(nd_{3/2})$ & $5d$ \textbf{a} & $5d$ \textbf{b} & $(6p)(np)$ &$(6s_{1/2})(nd_{5/2})$\\ 
$I_i$ ($\mathrm{cm}^{-1}$) & $50\,443.217463$ & $50\,443.217463$ & $83\,967.7$ &  $83\,967.7$ & $79\,725.35$ & $50\,442.795744$\\
$\ket{\bar\alpha}$ & $6snd\,^1D_2$ &  $6snd\,^3D_2$ & $5d$ \textbf{a} &  $5d$ \textbf{b} & $6pnp\,^1D_2$ & $6snd\,^3D_3$  \\ 
$\mu_\alpha^{(0)}$ & 0.730541589 & 0.751542685 & 0.195864083& 0.235944408&0.147483609 & $\mu_{^3D_3}^{rr}$  \\
$\mu_\alpha^{(2)}$ & $-0.0967938662$ & $0.00038836127$ & 0.0 & 0.0 & 0.0 &\\
\colrule
\rule{0pt}{3.0ex}$\mu_{^3D_3}^{rr}$ &$\mu_{^3D_3}^{(0)}=0.72865616$  &$\mu_{^3D_3}^{(2)}=0.793185994$  &$\mu_{^3D_3}^{(4)}=-523.858959$  &  & &  \\[1.0ex]
\colrule
\rule{0pt}{3.0ex}$V_{\bar\alpha\alpha}$  & $\theta_{12}^{(0)}=0.220048245$ & $\theta_{13}=0.00427599$ & $\theta_{14}=0.0381563093$ & $\theta_{24}=-0.00700797918$  & $\theta_{15}=0.109380331$ & $\theta_{25}=0.0635544456$  \\
& $\theta_{12}^{(2)} = -14.9486$\\[1.0ex]
\colrule
\rule{0pt}{3.0ex}$U_{i\bar\alpha}$   & $\sqrt{7/5}/2$&$\sqrt{7/30}$&0&0&0&$-\sqrt{5/3}/2$\\
                    & $-\sqrt{2/5}$&$\sqrt{3/5}$&0&0&0&0\\
                    & 0&0&1&0&0&0 \\
                    & 0&0&0&1&0&0\\
                    & 0&0&0&0&1&0\\
                    & $1/2$&$\sqrt{1/6}$&0&0&0&$\sqrt{7/3}/2$\\

\end{tabular}
\end{ruledtabular}
\end{sidewaystable}

\begin{table*}[ht]
\caption{\label{tab:171MQDT_PF32} 11-channel MQDT model parameters for the odd-parity $F = 3/2$ series of $^{171}$Yb, obtained by spectroscopic data presented in Tab.~\ref{tab:171Spec_PF32}. The electronic configuration of the perturbing Rydberg channels $4f^{13}5d6snp$ is abbreviated as $5d$. The rotations $\mathcal{R}(\theta_{ij})$ are applied in the order $\mathcal{R}(\theta_{12}(\varepsilon))\mathcal{R}(\theta_{13})\mathcal{R}(\theta_{14})\mathcal{R}(\theta_{15})\mathcal{R}(\theta_{16})\mathcal{R}(\theta_{23})\mathcal{R}(\theta_{24})\mathcal{R}(\theta_{25})\mathcal{R}(\theta_{26})\mathcal{R}(\theta_{78})\mathcal{R}(\theta_{79})\mathcal{R}(\theta_{7,10})\mathcal{R}(\theta_{pf})$.
}
\begin{ruledtabular}
\begin{tabular}{cccccccc}
$i$,$\bar\alpha$,$\alpha$ & 1 $(F_c=1)$ & 2 $(F_c=1)$& 3& 4& 5 & 6 & \\
\cline{1-7}
\rule{0pt}{3.0ex}$\ket{i}$ & $(6s_{1/2})(np_{3/2})$&  $(6s_{1/2})(np_{1/2})$ & $5d$ \textbf{a} & $5d$ \textbf{b} & $5d$ \textbf{c} & $5d$ \textbf{d} &\\ 
$I_i$ ($\mathrm{cm}^{-1}$) & $50\,443.217463$ & $50\,443.217463$ & $83\,967.7$ &  $83\,967.7$ & $83\,967.7$ & $83\,967.7$ &\\
$\ket{\bar\alpha}$ & $6snp\,^1P_1$ &  $6snp\,^3P_1$ & $5d$ \textbf{a} &  $5d$ \textbf{b} & $5d$ \textbf{c} & $5d$ \textbf{d} & \\ 
$\mu_\alpha^{(0)}$ & 0.922046003 & 0.981076381 & 0.229019212 & 0.205999445 & 0.19352756 & 0.181271472 & \\
$\mu_\alpha^{(2)}$ & 2.17932092 & $-4.48995034$ & 0.0 & 0.0 & 0.0 & 0.0 & \\
\cline{1-7}
\rule{0pt}{3.0ex}\multirow{4}{*}{$V_{\bar\alpha\alpha}$} & $\theta_{12}^{(0)}$ & $\theta_{12}^{(2)}$ & $\theta_{12}^{(4)}$ & $\theta_{13}$ & $\theta_{14}$ & $\theta_{15}$ & \\
& $-0.100292816$ & 149.140925 & $-13487.7692$ & $-0.0727917308$ & $-0.0669120237$ & $-0.0221321759$ &\\ 
\cline{2-7}
\rule{0pt}{3.0ex}& $\theta_{16}$ & $\theta_{23}$ & $\theta_{24}$ & $\theta_{25}$ & $\theta_{26}$ &  &\\ 
& $-0.107302569$ & 0.0396527798 & 0.0596597186 & 0.0861416497 & 0.0565415641 &  & $\dots$\\
\cline{1-7}
\rule{0pt}{3.0ex}& $\sqrt{5/3}/2$ & $\sqrt{5/6}/2$ & 0 & 0 & 0 & 0&\\
& $-1/\sqrt{3}$ & $\sqrt{2/3}$ & 0 & 0 & 0 & 0 &\\ 
& 0 & 0 & 1 & 0 & 0 & 0 &\\ 
& 0 & 0 & 0 & 1 & 0 & 0 &\\ 
& 0 & 0 & 0 & 0 & 1 & 0 &\\ 
& 0 & 0 & 0 & 0 & 0 & 1 &\\ 
$U_{i\bar\alpha}$& $1/2$ & $1/(2\sqrt{2})$ & 0 & 0 & 0 & 0 &\\
& 0 & 0 & 0 & 0 & 0 & 0 &\\ 
& 0 & 0 & 0 & 0 & 0 & 0 &\\ 
& 0 & 0 & 0 & 0 & 0 & 0 &\\ 
& 0 & 0 & 0 & 0 & 0 & 0 &\\ 
\end{tabular}
\end{ruledtabular}

\vspace{1em}
\begin{flushleft}
\small{Table~\ref{tab:171MQDT_PF32} (\textit{continued})}
\end{flushleft}
\vspace{-1em}

\begin{ruledtabular}
\begin{tabular}{ccccccc}
&$i$,$\bar\alpha$,$\alpha$ & 7 $(F_c=0)$ & 8 & 9 & 10 & 11 $(F_c=0)$\\
\cline{2-7}
\rule{0pt}{3.0ex}&$\ket{i}$ & $(6s_{1/2})(np_{3/2})$&  $5d$ \textbf{e} & $5d$ \textbf{f} & $5d$ \textbf{g} & $(6s_{1/2})(nf_{5/2})$ \\ 
&$I_i$ ($\mathrm{cm}^{-1}$) & $50\,442.795744$ & $83\,967.7$ & $83\,967.7$ &  $83\,967.7$ & $50\,443.217463$ \\
&$\ket{\bar\alpha}$ & $6snp\,^3P_2$ &  $5d$ \textbf{e} & $5d$ \textbf{f} & $5d$ \textbf{g} & $6snf\,^3F_2$ \\ 
&$\mu_\alpha^{(0)}$ & $\mu_{^3P_2}$ & 0.223859222 & 0.240057861 & 0.184824687 & $\mu_{^3F_2}^{rr}$\\
&$\mu_\alpha^{(2)}$ & & 0.0 & 0.0 & 0.0 \\
\cline{2-7}
\rule{0pt}{3.0ex}&$\mu_{^3P_2}$ & $\mu_{^3P_2}^{(0)} = 0.926908753$ & $\mu_{^3P_2}^{(2)} = -4.6228148$ & $\mu_{^3P_2}^{(4)} = 453.565911$ \\[1ex]
\cline{2-7}
\rule{0pt}{3.0ex}&$\mu_{^3F_2}^{rr}$ & $\mu_{^3F_2}^{(0)} =0.0718810048$ & $\mu_{^3F_2}^{(2)} = -1.08216233$ & $\mu_{^3F_2}^{(4)} = -38.2507093$ \\[1ex]
\cline{2-7}
\rule{0pt}{3.0ex}&\multirow{2}{*}{$V_{\bar\alpha\alpha}$} &  & $\theta_{78}$ & $\theta_{79}$ & $\theta_{7,10}$ & $\theta_{pf}$ \\
$\dots$&&  & 0.0657620037 & 0.0215995148 & $-0.028252844$ & 0.0184267375 \\ 
\cline{2-7}
\rule{0pt}{1.0ex}&& $-\sqrt{3/2}/2$ & 0 & 0 & 0 & 0 \\ 
&& 0 & 0 & 0 & 0 & 0 \\ 
&& 0 & 0 & 0 & 0 & 0 \\ 
&& 0 & 0 & 0 & 0 & 0 \\ 
&& 0 & 0 & 0 & 0 & 0 \\ 
&$U_{i\bar\alpha}$& 0 & 0 & 0 & 0 & 0 \\ 
&& $\sqrt{5/2}/2$ & 0 & 0 & 0 & 0 \\ 
&& 0 & 1 & 0 & 0 & 0 \\ 
&& 0 & 0 & 1 & 0 & 0 \\ 
&& 0 & 0 & 0 & 1 & 0 \\ 
&& 0 & 0 & 0 & 0 & $-1$ \\ 
\end{tabular}
\end{ruledtabular}
\end{table*}

\begin{table*}[ht]
\caption{\label{tab:171MQDT_PF52} 12-channel MQDT model parameters for the odd-parity $F = 5/2$ series of $^{171}$Yb, obtained by fitting to spectroscopic data presented in Tab.~\ref{tab:171Spec_PF52}. The electronic configuration of the perturbing Rydberg channels $4f^{13}5d6snp$ is abbreviated as $5d$. The rotations $\mathcal{R}(\theta_{ij})$ are applied in the order $\mathcal{R}(\theta_{12})\mathcal{R}(\theta_{13})\mathcal{R}(\theta_{14})\mathcal{R}(\theta_{15})\mathcal{R}(\theta_{16})\mathcal{R}(\theta_{17})\mathcal{R}(\theta_{23})\mathcal{R}(\theta_{24})\mathcal{R}(\theta_{25})\mathcal{R}(\theta_{26})\mathcal{R}(\theta_{27})\mathcal{R}(\theta_{89})\mathcal{R}(\theta_{8,10})\mathcal{R}(\theta_{8,11})\mathcal{R}(\theta_{pf})$.
}
\begin{ruledtabular}
\begin{tabular}{cccccccc}
$i$,$\bar\alpha$,$\alpha$ & 1 $(F_c=1)$ & 2 $(F_c=1)$& 3& 4& 5 & 6 \\
\cline{1-7}
\rule{0pt}{3.0ex}$\ket{i}$ & $(6s_{1/2})(nf_{7/2})$&  $(6s_{1/2})(nf_{5/2})$ & $5d$ \textbf{a} & $5d$ \textbf{b} & $5d$ \textbf{c} & $5d$ \textbf{d}\\ 
$I_i$ ($\mathrm{cm}^{-1}$) & $50\,443.217463$ & $50\,443.217463$ & $83\,967.7$ &  $83\,967.7$ & $83\,967.7$ & $83\,967.7$\\
$\ket{\bar\alpha}$ & $6snf\,^1F_3$ &  $6snf\,^3F_3$ & $5d$ \textbf{a} &  $5d$ \textbf{b} & $5d$ \textbf{c} & $5d$ \textbf{d} \\ 
$\mu_\alpha^{(0)}$ & 0.27707535 & 0.0719827367 & 0.250829254 & 0.227042016 & 0.17576256 & 0.19654694  \\
$\mu_\alpha^{(2)}$ & $-13.2829133$ & $-0.741492064$ & 0.0 & 0.0 & 0.0 & 0.0 \\
\cline{1-7}
\rule{0pt}{3.0ex}\multirow{4}{*}{$V_{\bar\alpha\alpha}$}  & $\theta_{12}^{(0)}$ & $\theta_{12}^{(2)}$ & $\theta_{13}$ & $\theta_{14}$ & $\theta_{15}$ & $\theta_{16}$ \\
& $-0.0208481417$ & 0.239045493 & $-0.0593782568$ & $-0.0755947274$ & 0.122678758 & $-0.0400418576$ \\ 
\cline{2-7}
\rule{0pt}{3.0ex}& $\theta_{17}$ & $\theta_{23}$ & $\theta_{24}$ & $\theta_{25}$ & $\theta_{26}$ & $\theta_{27}$ \\ 
& 0.0646373252 & $-0.0684538786$ & 0.0352130279 & $-0.0326572035$ & $-0.050215299$ & 0.0453892695 & \dots \\ 
\cline{1-7}
\rule{0pt}{3.0ex}\multirow{12}{*}{$U_{i\bar\alpha}$} 
& $-2/\sqrt{7}$ & $-\sqrt{3/7}$ & 0 & 0 & 0 & 0 \\ 
& $\sqrt{5/7}/2$ & $-\sqrt{5/21}$ & 0 & 0 & 0 & 0 \\ 
& 0 & 0 & 1 & 0 & 0 & 0 \\ 
& 0 & 0 & 0 & 1 & 0 & 0 \\ 
& 0 & 0 & 0 & 0 & 1 & 0 \\ 
& 0 & 0 & 0 & 0 & 0 & 1 \\ 
& 0 & 0 & 0 & 0 & 0 & 0 \\ 
& 0 & 0 & 0 & 0 & 0 & 0 \\ 
& 0 & 0 & 0 & 0 & 0 & 0 \\ 
& 0 & 0 & 0 & 0 & 0 & 0 \\ 
& 0 & 0 & 0 & 0 & 0 & 0 \\ 
& $-1/2$ & $1/\sqrt{3}$ & 0 & 0 & 0 & 0 \\ 
\end{tabular}
\end{ruledtabular}

\vspace{1em}
\begin{flushleft}
\small{Table~\ref{tab:171MQDT_PF52} (\textit{continued})}
\end{flushleft}
\vspace{-1em}

\begin{ruledtabular}
\begin{tabular}{cccccccc}
&$i$,$\bar\alpha$,$\alpha$ & 7 & 8 $(F_c = 1)$ & 9 & 10 & 11 & 12 $(F_c=0)$\\
\cline{2-8}
\rule{0pt}{3.0ex}&$\ket{i}$ & $5d$ \textbf{e} & $(6s_{1/2})(np_{3/2})$ & $5d$ \textbf{f} & $5d$ \textbf{g} & $5d$ \textbf{h} & $(6s_{1/2})(nf_{5/2})$ \\ 
&$I_i$ ($\mathrm{cm}^{-1}$) & $83\,967.7$ & $50\,443.217463$ & $83\,967.7$ &  $83\,967.7$ &  $83\,967.7$ &  $50\,442.795744$ \\
&$\ket{\bar\alpha}$ & $5d$ \textbf{e} & $6snp\,^3P_2$ &   $5d$ \textbf{f} & $5d$ \textbf{g} & $5d$ \textbf{h} &  $6snf\,^3F_2$ \\ 
&$\mu_\alpha^{(0)}$ & 0.214351625 & $\mu_{^3P_2}$ & 0.223859222 & 0.240057861 & 0.184824687 & $\mu_{^3F_2}^{rr}$ \\
&$\mu_\alpha^{(2)}$ & 0.0 & & 0.0 & 0.0 & 0.0 \\
\cline{2-8}
\rule{0pt}{3.0ex}&$\mu_{^3P_2}$ & $\mu_{^3P_2}^{(0)} = 0.926908753$ & $\mu_{^3P_2}^{(2)} = -4.6228148$ & $\mu_{^3P_2}^{(4)} = 453.565911$ \\[1.0ex]
\cline{2-8}
\rule{0pt}{3.0ex}&$\mu_{^3F_2}^{rr}$ & $\mu_{^3F_2}^{(0)} = 0.0718810048$ & $\mu_{^3F_2}^{(2)} = -1.08216233$ & $\mu_{^3F_2}^{(4)} = -38.2507093$ \\[1.0ex]
\cline{2-8}
&\rule{0pt}{3.0ex}\multirow{2}{*}{$V_{\bar\alpha\alpha}$} & & $\theta_{89}$ & $\theta_{8,10}$ & $\theta_{8,11}$ & & $\theta_{pf}$ \\
\dots & & & 0.0657620037 & 0.0215995148 & $-0.028252844$ & & 0.0184267375 \\ 
\cline{2-8}
\rule{0pt}{3.0ex}&\multirow{12}{*}{$U_{i\bar\alpha}$}   
& 0 & 0 & 0 & 0 & 0 & 0 \\ 
&& 0 & 0 & 0 & 0 & 0 & $\sqrt{7/3}/2$ \\
&& 0 & 0 & 0 & 0 & 0 & 0 \\ 
&& 0 & 0 & 0 & 0 & 0 & 0 \\ 
&& 0 & 0 & 0 & 0 & 0 & 0 \\ 
&& 0 & 0 & 0 & 0 & 0 & 0 \\ 
&& 1 & 0 & 0 & 0 & 0 & 0 \\ 
&& 0 & 1 & 0 & 0 & 0 & 0 \\ 
&& 0 & 0 & 1 & 0 & 0 & 0 \\ 
&& 0 & 0 & 0 & 1 & 0 & 0 \\ 
&& 0 & 0 & 0 & 0 & 1 & 0 \\ 
&& 0 & 0 & 0 & 0 & 0 & $\sqrt{5/12}$ \\ 
\end{tabular}
\end{ruledtabular}
\end{table*}

\begin{table*}[ht]
\centering
\begin{minipage}{0.5\textwidth}
\caption{\label{tab:171QDT_G52} Single-channel model parameters for the $6sng\,^3G_3(F=5/2)$ series of $^{171}$Yb described by Eq.~(\ref{eq:qd_energy}), obtained by fitting to spectroscopic data presented in Tab.~\ref{tab:171Spec_3G3_gfactors}.}
\begin{ruledtabular}
\begin{tabular}{cc}
$\mu_\alpha^{(0)}$ & 0.02611627 \\
$\mu_\alpha^{(2)}$ & $-0.13079392$ \\
\end{tabular}
\end{ruledtabular}
\end{minipage}
\end{table*}

\begin{table*}[ht]
\caption{\label{tab:171MQDT_G72} Three-channel MQDT model parameters for the $\ket{\nu,\ell=4,F=7/2}$ series of $^{171}$Yb, obtained by fitting to spectrscopic data presented in Tab.~\ref{tab:171Spec_G72}.}
\begin{ruledtabular}
\begin{tabular}{cccc}
$i$,$\bar\alpha$,$\alpha$ & 1 $(F_c = 1)$ & 2 $(F_c = 1)$ & 3 $(F_c = 0)$\\
\colrule
\rule{0pt}{3.0ex}$\ket{i}$ & $(6s_{1/2})(ng_{9/2})$& $(6s_{1/2})(ng_{7/2})$ & $(6s_{1/2})(ng_{7/2})$ \\ 
$I_i$ ($\mathrm{cm}^{-1}$) & $50\,443.217463$ & $50\,443.217463$ & $50\,442.795744$ \\
$\ket{\bar\alpha}$ & $6sng\,^+G_4$ & $6sng\,^-G_4$ & $6sng\, ^3G_3$\\
$\mu_\alpha^{(0)}$ & 0.02627822 & 0.02536987 & 0.02611627 \\
$\mu_\alpha^{(2)}$ & $-0.13288481$ & $-0.11668088$ & $-0.13079392$ \\
\colrule
\rule{0pt}{3.0ex}$V_{\bar\alpha\alpha}$ & $\theta_{12} = -0.08222676$ & & \\  
\colrule
\rule{0pt}{3.0ex}$U_{i\bar\alpha}$  
& $-1$ & 0 & 0  \\ 
& 0 & $-\sqrt{7}/4$ & $3/4$ \\
& 0 & $3/4$ & $\sqrt{7}/4$ \\
\end{tabular}
\end{ruledtabular}
\end{table*}

\begin{table*}[ht]
\caption{\label{tab:171MQDT_G92} Three-channel MQDT model parameters for the $\ket{\nu,\ell=4,F=9/2}$ series of $^{171}$Yb, obtained by fitting to spectroscopic data presented in Tab.~\ref{tab:171Spec_G92}.}
\begin{ruledtabular}
\begin{tabular}{cccc}
$i$,$\bar\alpha$,$\alpha$ & 1 $(F_c = 1)$ & 2 $(F_c = 1)$ & 3 $(F_c = 0)$\\
\colrule
\rule{0pt}{3.0ex}$\ket{i}$ & $(6s_{1/2})(ng_{9/2})$& $(6s_{1/2})(ng_{7/2})$ & $(6s_{1/2})(ng_{9/2})$ \\ 
$I_i$ ($\mathrm{cm}^{-1}$) & $50\,443.217463$ & $50\,443.217463$ & $50\,442.795744$ \\
$\ket{\bar\alpha}$ & $6sng\,^+G_4$ & $6sng\,^-G_4$ & $6sng\, ^3G_5$\\
$\mu_\alpha^{(0)}$ & 0.02627822 & 0.02547095 & 0.02536987\\
$\mu_\alpha^{(2)}$ & $-0.13288481$ & $-0.11668088$ & $-0.2007948$\\
\colrule
\rule{0pt}{3.0ex}$V_{\bar\alpha\alpha}$  & $\theta_{12} = -0.08222676$ & & \\  
\colrule
\rule{0pt}{3.0ex}$U_{i\bar\alpha}$  
& $\sqrt{11/20}$ & 0 & $-\sqrt{9/20}$  \\ 
& 0 & 1 & 0 \\
& $\sqrt{9/20}$ & 0 & $\sqrt{11/20}$ \\
\end{tabular}
\end{ruledtabular}
\end{table*}

\clearpage
\onecolumngrid
\section{Spectroscopy Data}
\label{sec:specdata}

\subsection{$^{174}$Yb $^{1,3}F_3$}

\begin{longtable*}{@{\extracolsep{\fill}}ccS[table-format=6.4]S[table-format=5.6]S[table-format=5.6]S[table-format=1.2]c} 
\caption{Spectroscopic data of $6snf\,^{1,3}F_3\leftrightarrow6sn'd\,^{1,3}D_2$ Rydberg-Rydberg transitions obtained by microwave spectroscopy in an atomic beam setup described in Sec.~\ref{sec:expsetup} and from Ref.~\cite{Lehec2017PhD}.} \label{tab:174Spec_Microwave_13F3} \\

\hline\hline\\\multirow{2}{*}{$f$} & \multirow{2}{*}{$i$}& {\multirow{2}{*}{$\nu_{f\leftarrow i}$ (MHz)}} & {\multirow{2}{*}{$\tilde\nu_\mathrm{exp.}$ ($\mathrm{cm}^{-1}$ )}}& {\multirow{2}{*}{$\tilde\nu_\mathrm{th.}$ ($\mathrm{cm}^{-1}$ )}} &  {$E_\mathrm{exp.}-E_\mathrm{th.}$ }  &\multirow{2}{*}{Ref.}\\ 
    & & & & & {($h\cdot\mathrm{MHz}$)} & \\\hline \\ 
 \endfirsthead

 {{\tablename\ \thetable{} (\textit{continued})}} \\
\hline\hline\\\multirow{2}{*}{$f$} & \multirow{2}{*}{$i$}& {\multirow{2}{*}{$\nu_{f\leftarrow i}$ (MHz)}} & {\multirow{2}{*}{$\tilde\nu_\mathrm{exp.}$ ($\mathrm{cm}^{-1}$ )}}& {\multirow{2}{*}{$\tilde\nu_\mathrm{th.}$ ($\mathrm{cm}^{-1}$ )}} &  {$E_\mathrm{exp.}-E_\mathrm{th.}$ }  &\multirow{2}{*}{Ref.}\\ 
    & & & & & {($h\cdot\mathrm{MHz}$)} & \\\hline \\ 
 \endhead

\hline
 \endfoot

 \hline
 \endlastfoot

$6s29f\,^1F_3$& $6s31d\,^1D_2$& -164206.72(10)& 50300.470140& 50300.470142&-0.05&This work\\
$6s29f\,^3F_3$& $6s31d\,^1D_2$& -99421.74(10)& 50302.631134& 50302.631134&0.00&This work\\
$6s31f\,^3F_3$& $6s33d\,^3D_2$& -73096.89(10)& 50320.720998& 50320.720996&0.06&This work\\
$6s31f\,^1F_3$& $6s33d\,^1D_2$& -134171.16(10)& 50318.978603& 50318.978607&-0.12&This work\\
$6s31f\,^3F_3$& $6s33d\,^1D_2$& -81935.64(10)& 50320.720992& 50320.720996&-0.12&This work\\
$6s32f\,^1F_3$& $6s34d\,^1D_2$& -121856.46(10)& 50326.911906& 50326.911903&0.07&This work\\
$6s32f\,^3F_3$& $6s34d\,^1D_2$& -74583.66(10)& 50328.488756& 50328.488757&-0.03&This work\\
$6s33f\,^1F_3$& $6s35d\,^1D_2$& -110995.57(10)& 50334.108980& 50334.108975&0.16&This work\\
\hline\hline
\end{longtable*}

\begin{longtable*}{@{\extracolsep{\fill}}ccS[table-format=6.4]S[table-format=5.6]S[table-format=5.6]S[table-format=1.2]c} 
\caption{Spectroscopic data of $6snf\,^{3}F_3\leftrightarrow6sn'f\,^{3}F_3$ Rydberg-Rydberg transitions obtained by two-photon microwave spectroscopy in an atomic beam setup described in Sec.~\ref{sec:expsetup}.} \label{tab:174Spec_Microwave_13F3_twophoton} \\

\hline\hline\\\multirow{2}{*}{$f$} & \multirow{2}{*}{$i$}& {\multirow{2}{*}{$\nu_{f\leftarrow i}$ (MHz)}} & {\multirow{2}{*}{$\tilde\nu_\mathrm{exp.}$ ($\mathrm{cm}^{-1}$ )}}& {\multirow{2}{*}{$\tilde\nu_\mathrm{th.}$ ($\mathrm{cm}^{-1}$ )}} &  {$E_\mathrm{exp.}-E_\mathrm{th.}$ }  &\multirow{2}{*}{Ref.}\\ 
    & & & & & {($h\cdot\mathrm{MHz}$)} & \\\hline \\ 
 \endfirsthead

 {{\tablename\ \thetable{} (\textit{continued})}} \\
\hline\hline\\\multirow{2}{*}{$f$} & \multirow{2}{*}{$i$}& {\multirow{2}{*}{$\nu_{f\leftarrow i}$ (MHz)}} & {\multirow{2}{*}{$\tilde\nu_\mathrm{exp.}$ ($\mathrm{cm}^{-1}$ )}}& {\multirow{2}{*}{$\tilde\nu_\mathrm{th.}$ ($\mathrm{cm}^{-1}$ )}} &  {$E_\mathrm{exp.}-E_\mathrm{th.}$ }  &\multirow{2}{*}{Ref.}\\ 
    & & & & & {($h\cdot\mathrm{MHz}$)} & \\\hline \\ 
 \endhead

\hline
 \endfoot

 \hline
 \endlastfoot

$6s35f\,^3F_3$& $6s35d\,^3F_3$& -160978.75(10)& 50347.829382& 50347.829377&0.14&This work\\
$6s37f\,^3F_3$& $6s37d\,^3F_3$& -135934.43(10)& 50358.127560& 50358.127566&-0.17&This work\\

\hline\hline
\end{longtable*}

\begin{longtable*}{@{\extracolsep{\fill}}c  S[table-format=5.6]S[table-format=5.6]S[table-format=2.2]c}
\caption{\label{tab:174YbSpec_13F3} Laser spectroscopic data of the $6snp\,^{1,3}F_3$ Rydberg series included in the determination of the MQDT model. The state energies and respective uncertainties are taken from Refs.~\cite{Aymar1984three,Ali1999Two}.}\\

\hline\hline\\
 [-6pt]{\multirow{2}{*}{State label}} & {\multirow{2}{*}{$\tilde\nu_\mathrm{exp.}$ ($\mathrm{cm}^{-1}$)}} & {\multirow{2}{*}{$\tilde\nu_\mathrm{th.}$ ($\mathrm{cm}^{-1}$ )}} & {$E_\mathrm{exp.}-E_\mathrm{th.}$}&{\multirow{2}{*}{Ref.}}\\
 & & & {($h\cdot\mathrm{MHz}$)}  &\\[3pt]
\hline\\ [-5pt] 
 \endfirsthead

 {{\tablename\ \thetable{} (\textit{continued})}} \\
\hline\hline\\
 [-6pt]{\multirow{2}{*}{State label}} & {\multirow{2}{*}{$\tilde\nu_\mathrm{exp.}$ ($\mathrm{cm}^{-1}$)}} & {\multirow{2}{*}{$\tilde\nu_\mathrm{th.}$ ($\mathrm{cm}^{-1}$ )}} & {$E_\mathrm{exp.}-E_\mathrm{th.}$}&{\multirow{2}{*}{Ref.}}\\
 & & & {($h\cdot\mathrm{MHz}$)}  &\\[3pt]
\hline\\ [-5pt] 
 \endhead

\hline
 \endfoot

 \hline
 \endlastfoot

$4f^{13}5d^26s\, ^3D_3$& 48584.1(5)& 48584.114897&-447&\cite{Ali1999Two}\\
$6s9f\,^1F_3$& 48669.2(5)& 48685.533000&-489651&\cite{Ali1999Two}\\
$6s9f\,^3F_3$& 48702.0(5)& 48706.677201&-140219&\cite{Ali1999Two}\\
$6s10f\,^1F_3$& 49032.2(5)& 49033.291629&-32726&\cite{Ali1999Two}\\
$6s10f\,^3F_3$& 49063.7(5)& 49066.998105&-98875&\cite{Ali1999Two}\\
$4f^{13}5d^26s\, ^1F_3$& 49122.4(5)& 49121.830010&17088&\cite{Ali1999Two}\\
$6s11f\,^1F_3$& 49308.2(5)& 49314.602047&-191929&\cite{Ali1999Two}\\
$6s11f\,^3F_3$& 49339.1(5)& 49331.711343&221506&\cite{Ali1999Two}\\
\hline\hline
\end{longtable*}

\clearpage

\subsection{$^{174}$Yb $^3F_2$}

\begin{longtable*}{@{\extracolsep{\fill}}ccS[table-format=6.4]S[table-format=5.6]S[table-format=5.6]S[table-format=1.2]} 
\caption{Spectroscopic data of $6snf\,^3F_2 \leftrightarrow 6sn'd\, ^{1, 3}D_2$ Rydberg-Rydberg transitions obtained by microwave spectroscopy in an atomic beam setup described in Sec.~\ref{sec:expsetup}.} \label{tab:174Spec_Microwave_3F2} \\

\hline\hline\\\multirow{2}{*}{$f$} & \multirow{2}{*}{$i$}& {\multirow{2}{*}{$\nu_{f\leftarrow i}$ (MHz)}} & {\multirow{2}{*}{$\tilde\nu_\mathrm{exp.}$ ($\mathrm{cm}^{-1}$ )}}& {\multirow{2}{*}{$\tilde\nu_\mathrm{th.}$ ($\mathrm{cm}^{-1}$ )}} &  {$E_\mathrm{exp.}-E_\mathrm{th.}$ } \\ 
    & & & & & {($h\cdot\mathrm{MHz}$)} \\\hline \\ 
 \endfirsthead

 {{\tablename\ \thetable{} (\textit{continued})}} \\
\hline\hline\\\multirow{2}{*}{$f$} & \multirow{2}{*}{$i$}& {\multirow{2}{*}{$\nu_{f\leftarrow i}$ (MHz)}} & {\multirow{2}{*}{$\tilde\nu_\mathrm{exp.}$ ($\mathrm{cm}^{-1}$ )}}& {\multirow{2}{*}{$\tilde\nu_\mathrm{th.}$ ($\mathrm{cm}^{-1}$ )}} &  {$E_\mathrm{exp.}-E_\mathrm{th.}$ }\\ 
    & & & & & {($h\cdot\mathrm{MHz}$)} \\\hline \\ 
 \endhead

\hline
 \endfoot

 \hline
 \endlastfoot

$6s29f\,^3F_2$& $6s31d\,^1D_2$& -106543.58(10)& 50302.393575& 50302.393583&-0.22\\
$6s31f\,^3F_2$& $6s33d\,^3D_2$& -77781.96(10)& 50320.564722& 50320.564713&0.24\\
$6s31f\,^3F_2$& $6s33d\,^1D_2$& -86620.74(10)& 50320.564714& 50320.564713&0.02\\
$6s32f\,^3F_2$& $6s34d\,^1D_2$& -78514.61(10)& 50328.357634& 50328.357632&0.06\\
$6s33f\,^3F_2$& $6s35d\,^1D_2$& -71392.41(10)& 50335.429999& 50335.429997&0.06\\
$6s34f\,^3F_2$& $6s36d\,^1D_2$& -65108.14(10)& 50341.867973& 50341.867970&0.09\\
$6s36f\,^3F_2$& $6s38d\,^1D_2$& -54592.62(10)& 50353.124987& 50353.124984&0.11\\
$6s38f\,^3F_2$& $6s41d\,^1D_2$& -168201.53(10)& 50362.603235& 50362.603234&0.04\\
\hline\hline
\end{longtable*}

\begin{longtable*}{@{\extracolsep{\fill}}ccS[table-format=6.4]S[table-format=5.6]S[table-format=5.6]S[table-format=1.2]} 
\caption{Spectroscopic data of $6snf\,^3F_2 \leftrightarrow 6sn'f\, ^3F_2$ Rydberg-Rydberg transitions obtained by two-photon microwave spectroscopy in an atomic beam setup desribed in Sec.~\ref{sec:expsetup}.} \label{tab:174Spec_Microwave_3F2_twophoton} \\

\hline\hline\\\multirow{2}{*}{$f$} & \multirow{2}{*}{$i$}& {\multirow{2}{*}{$\nu_{f\leftarrow i}$ (MHz)}} & {\multirow{2}{*}{$\tilde\nu_\mathrm{exp.}$ ($\mathrm{cm}^{-1}$ )}}& {\multirow{2}{*}{$\tilde\nu_\mathrm{th.}$ ($\mathrm{cm}^{-1}$ )}} &  {$E_\mathrm{exp.}-E_\mathrm{th.}$ } \\ 
    & & & & & {($h\cdot\mathrm{MHz}$)} \\\hline \\ 
 \endfirsthead

 {{\tablename\ \thetable{} (\textit{continued})}} \\
\hline\hline\\\multirow{2}{*}{$f$} & \multirow{2}{*}{$i$}& {\multirow{2}{*}{$\nu_{f\leftarrow i}$ (MHz)}} & {\multirow{2}{*}{$\tilde\nu_\mathrm{exp.}$ ($\mathrm{cm}^{-1}$ )}}& {\multirow{2}{*}{$\tilde\nu_\mathrm{th.}$ ($\mathrm{cm}^{-1}$ )}} &  {$E_\mathrm{exp.}-E_\mathrm{th.}$ }\\ 
    & & & & & {($h\cdot\mathrm{MHz}$)} \\\hline \\ 
 \endhead

\hline
 \endfoot

 \hline
 \endlastfoot

$6s35f\,^3F_2$& $6s35f\,^3F_2$& -161281.47(10)& 50347.745213& 50347.745208&0.15\\
$6s37f\,^3F_2$& $6s37f\,^3F_3$& -137903.16(10)& 50358.061891& 50358.061891&0.00\\
$6s38f\,^3F_2$& $6s38f\,^3F_2$& -125521.25(10)& 50362.603230& 50362.603234&-0.12\\

\hline\hline
\end{longtable*}

\clearpage

\subsection{$^{174}$Yb $^3F_4$}

\begin{longtable*}{@{\extracolsep{\fill}}ccS[table-format=6.4]S[table-format=5.6]S[table-format=5.6]S[table-format=1.2]} 
\caption{Spectroscopic data of $6snf\,^3F_4 \leftrightarrow 6sn'd\, ^{3}D_3$ Rydberg-Rydberg transitions obtained by microwave spectroscopy in an atomic beam setup described in Sec.~\ref{sec:expsetup}.} \label{tab:174Spec_Microwave_3F4} \\

\hline\hline\\\multirow{2}{*}{$f$} & \multirow{2}{*}{$i$}& {\multirow{2}{*}{$\nu_{f\leftarrow i}$ (MHz)}} & {\multirow{2}{*}{$\tilde\nu_\mathrm{exp.}$ ($\mathrm{cm}^{-1}$ )}}& {\multirow{2}{*}{$\tilde\nu_\mathrm{th.}$ ($\mathrm{cm}^{-1}$ )}} &  {$E_\mathrm{exp.}-E_\mathrm{th.}$ } \\ 
    & & & & & {($h\cdot\mathrm{MHz}$)} \\\hline \\ 
 \endfirsthead

 {{\tablename\ \thetable{} (\textit{continued})}} \\
\hline\hline\\\multirow{2}{*}{$f$} & \multirow{2}{*}{$i$}& {\multirow{2}{*}{$\nu_{f\leftarrow i}$ (MHz)}} & {\multirow{2}{*}{$\tilde\nu_\mathrm{exp.}$ ($\mathrm{cm}^{-1}$ )}}& {\multirow{2}{*}{$\tilde\nu_\mathrm{th.}$ ($\mathrm{cm}^{-1}$ )}} &  {$E_\mathrm{exp.}-E_\mathrm{th.}$ }\\ 
    & & & & & {($h\cdot\mathrm{MHz}$)} \\\hline \\ 
 \endhead

\hline
 \endfoot

 \hline
 \endlastfoot

$6s46f\,^3F_4$& $6s46d\,^3D_3$& 126444.06(10)& 50388.680141& 50388.680145&-0.09\\
$6s47f\,^3F_4$& $6s47d\,^3D_3$& 118209.09(10)& 50391.023243& 50391.023247&-0.13\\
$6s48f\,^3F_4$& $6s48d\,^3D_3$& 110674.37(10)& 50393.218148& 50393.218147&0.05\\
$6s49f\,^3F_4$& $6s49d\,^3D_3$& 103766.85(10)& 50395.277090& 50395.277083&0.23\\
$6s50f\,^3F_4$& $6s50d\,^3D_3$& 97422.14(10)& 50397.211055& 50397.211056&-0.03\\
$6s51f\,^3F_4$& $6s51d\,^3D_3$& 91584.75(10)& 50399.029981& 50399.029979&0.07\\
$6s52f\,^3F_4$& $6s52d\,^3D_3$& 86204.42(10)& 50400.742794& 50400.742797&-0.11\\
$6s53f\,^3F_4$& $6s52d\,^3D_3$& 134615.34(10)& 50402.357608& 50402.357607&0.05\\
$6s54f\,^3F_4$& $6s53d\,^3D_3$& 126930.25(10)& 50403.881744& 50403.881744&0.02\\

\hline\hline
\end{longtable*}

\clearpage

\subsection{$^{174}$Yb $^3G_3$}

\begin{longtable*}{@{\extracolsep{\fill}}ccS[table-format=6.4]S[table-format=5.6]S[table-format=5.6]S[table-format=1.2]} 
\caption{Spectroscopic data of $6sng\,^3G_3 \leftrightarrow 6sn'd\, ^{1, 3}D_2$ Rydberg-Rydberg transitions. The transitions are obtained by two-step microwave spectroscopy via an intermediate $6sn''f\, ^3F_3$ state in an atomic beam setup described in Sec.~\ref{sec:expsetup}.} \label{tab:174Spec_Microwave_3G3} \\

\hline\hline\\\multirow{2}{*}{$f$} & \multirow{2}{*}{$i$}& {\multirow{2}{*}{$\nu_{f\leftarrow i}$ (MHz)}} & {\multirow{2}{*}{$\tilde\nu_\mathrm{exp.}$ ($\mathrm{cm}^{-1}$ )}}& {\multirow{2}{*}{$\tilde\nu_\mathrm{th.}$ ($\mathrm{cm}^{-1}$ )}} &  {$E_\mathrm{exp.}-E_\mathrm{th.}$ } \\ 
    & & & & & {($h\cdot\mathrm{MHz}$)} \\\hline \\ 
 \endfirsthead

 {{\tablename\ \thetable{} (\textit{continued})}} \\
\hline\hline\\\multirow{2}{*}{$f$} & \multirow{2}{*}{$i$}& {\multirow{2}{*}{$\nu_{f\leftarrow i}$ (MHz)}} & {\multirow{2}{*}{$\tilde\nu_\mathrm{exp.}$ ($\mathrm{cm}^{-1}$ )}}& {\multirow{2}{*}{$\tilde\nu_\mathrm{th.}$ ($\mathrm{cm}^{-1}$ )}} &  {$E_\mathrm{exp.}-E_\mathrm{th.}$ }\\ 
    & & & & & {($h\cdot\mathrm{MHz}$)} \\\hline \\ 
 \endhead

\hline
 \endfoot

 \hline
 \endlastfoot

$6s34g\, ^3G_3$& $6s38d\,^1D_2$& -208325.48(10)& 50347.997011& 50347.997013&-0.07\\
$6s37g\, ^3G_3$& $6s41d\,^1D_2$& -162327.74(10)& 50362.799164& 50362.799162&0.06\\
$6s41g\, ^3G_3$& $6s45d\,^1D_2$& -119860.50(10)& 50377.706716& 50377.706713&0.10\\
$6s46g\, ^3G_3$& $6s51d\,^1D_2$& -145576.92(10)& 50391.151078& 50391.151080&-0.06\\

\hline\hline
\end{longtable*}

\begin{longtable*}{@{\extracolsep{\fill}}c S[table-format=1.3] S[table-format=1.3]}
	\caption{\label{tab:171Spec_3G3_gfactors} Predicted $g_\mathrm{th}$ and experimental $g_\mathrm{exp}$ factors of the $6sng\, ^3G_3$ Rydberg states obtained from microwave spectroscopy in an atomic beam setup described in Sec.~\ref{sec:expsetup}.}\\
	
	\hline\hline\\
	[-6pt]{State Label} & {$g_\mathrm{exp}$}   & {$g_\mathrm{th}$}    \\[3pt]
	\hline\\ [-5pt] 
	\endfirsthead
	
	{{\tablename\ \thetable{} (\textit{continued})}} \\
	\hline\hline \\
	[-6pt]{State Label} & {$g_\mathrm{exp}$}  & {$g_\mathrm{th}$}   \\[3pt]
	\hline\\ [-5pt] 
	\endhead
	
	\hline
	\endfoot
	
	\hline
	\endlastfoot

$6s37g\, ^3G_3$& 0.746(4)&0.75\\
$6s41g\, ^3G_3$& 0.746(3)&0.75\\

	\hline\hline
\end{longtable*}

\clearpage

\subsection{$^{174}$Yb $^\pm G_4$}

\begin{longtable*}{@{\extracolsep{\fill}}ccS[table-format=6.4]S[table-format=5.6]S[table-format=5.6]S[table-format=1.2]c} 
\caption{Spectroscopic data of $6sng\,^\pm G_4 \leftrightarrow 6sn'd\, ^{1, 3}D_2$ Rydberg-Rydberg transitions. These transitions are obtained by two-step microwave spectroscopy via an intermediate $6sn''f\, ^3F_3$ state in an atomic beam setup described in Sec.~\ref{sec:expsetup}.} \label{tab:174Spec_Microwave_pmG4} \\

\hline\hline\\\multirow{2}{*}{$f$} & \multirow{2}{*}{$i$}& {\multirow{2}{*}{$\nu_{f\leftarrow i}$ (MHz)}} & {\multirow{2}{*}{$\tilde\nu_\mathrm{exp.}$ ($\mathrm{cm}^{-1}$ )}}& {\multirow{2}{*}{$\tilde\nu_\mathrm{th.}$ ($\mathrm{cm}^{-1}$ )}} &  {$E_\mathrm{exp.}-E_\mathrm{th.}$ } & \multirow{2}{*}{Ref.} \\ 
    & & & & & {($h\cdot\mathrm{MHz}$)} & \\\hline \\ 
 \endfirsthead

 {{\tablename\ \thetable{} (\textit{continued})}} \\
\hline\hline\\\multirow{2}{*}{$f$} & \multirow{2}{*}{$i$}& {\multirow{2}{*}{$\nu_{f\leftarrow i}$ (MHz)}} & {\multirow{2}{*}{$\tilde\nu_\mathrm{exp.}$ ($\mathrm{cm}^{-1}$ )}}& {\multirow{2}{*}{$\tilde\nu_\mathrm{th.}$ ($\mathrm{cm}^{-1}$ )}} &  {$E_\mathrm{exp.}-E_\mathrm{th.}$ } & \multirow{2}{*}{Ref.}\\ 
    & & & & & {($h\cdot\mathrm{MHz}$)} &  \\\hline \\ 
 \endhead

\hline
 \endfoot

 \hline
 \endlastfoot

$6s28g\, ^+G_4$& $6s31d\,^1D_2$& -93198.90(10)& 50302.838706& 50302.838700&0.17&\cite{Niyaz2019Microwave}\\
$6s29g\, ^+G_4$& $6s32d\,^1D_2$& -83854.12(10)& 50312.351404& 50312.351395&0.27&\cite{Niyaz2019Microwave}\\
$6s30g\, ^+G_4$& $6s33d\,^1D_2$& -75730.23(10)& 50320.927982& 50320.927988&-0.18&\cite{Niyaz2019Microwave}\\
$6s31g\, ^+G_4$& $6s34d\,^1D_2$& -68628.92(10)& 50328.687385& 50328.687393&-0.22&\cite{Niyaz2019Microwave}\\
$6s32g\, ^+G_4$& $6s35d\,^1D_2$& -62392.36(10)& 50335.730209& 50335.730219&-0.32&\cite{Niyaz2019Microwave}\\
$6s33g\, ^+G_4$& $6s36d\,^1D_2$& -56891.84(10)& 50342.142039& 50342.142061&-0.65&\cite{Niyaz2019Microwave}\\
$6s34g\, ^+G_4$& $6s38d\,^1D_2$& -208352.33(10)& 50347.996116& 50347.996100&0.45&This work\\
$6s34g\, ^-G_4$& $6s38d\,^1D_2$& -208219.13(10)& 50348.000559& 50348.000558&0.01&This work\\
\hline\hline
\end{longtable*}

\begin{longtable*}{@{\extracolsep{\fill}}c S[table-format=1.3] S[table-format=1.5]}
	\caption{\label{tab:171Spec_pmG4_gfactors} Predicted $g_\mathrm{th}$ and experimental $g_\mathrm{exp}$ factors of the $6sng\, ^\pm G_4$ Rydberg states obtained from microwave spectroscopy in an atomic beam setup described in Sec.~\ref{sec:expsetup}.}\\
	
	\hline\hline\\
	[-6pt]{State Label} & {$g_\mathrm{exp}$}   & {$g_\mathrm{th}$}    \\[3pt]
	\hline\\ [-5pt] 
	\endfirsthead
	
	{{\tablename\ \thetable{} (\textit{continued})}} \\
	\hline\hline \\
	[-6pt]{State Label} & {$g_\mathrm{exp}$}  & {$g_\mathrm{th}$}   \\[3pt]
	\hline\\ [-5pt] 
	\endhead
	
	\hline
	\endfoot
	
	\hline
	\endlastfoot

$6s34g\, ^-G_4$& 1.021(4)&1.0327\\
$6s37g\, ^-G_4$& 1.031(4)&1.0327\\
$6s41g\, ^-G_4$& 1.027(5)&1.0327\\
$6s34g\, ^+G_4$& 1.012(9)&1.0174\\
$6s37g\, ^+G_4$& 1.018(4)&1.0174\\
$6s41g\, ^+G_4$& 1.013(4)&1.0174\\

	\hline\hline
\end{longtable*}

\clearpage

\subsection{$^{174}$Yb $^3D_3$}

\begin{longtable*}{@{\extracolsep{\fill}}ccS[table-format=6.4]S[table-format=5.6]S[table-format=5.6]S[table-format=1.2]} 
\caption{Spectroscopic data of $6snd\,^3D_3 \leftrightarrow 6sn'd\, ^{3}D_2$ Rydberg-Rydberg transitions obtained by two-photon microwave spectroscopy in an atomic beam setup described in Sec.~\ref{sec:expsetup}.} \label{tab:174Spec_Microwave_3D3} \\

\hline\hline\\\multirow{2}{*}{$f$} & \multirow{2}{*}{$i$}& {\multirow{2}{*}{$\nu_{f\leftarrow i}$ (MHz)}} & {\multirow{2}{*}{$\tilde\nu_\mathrm{exp.}$ ($\mathrm{cm}^{-1}$ )}}& {\multirow{2}{*}{$\tilde\nu_\mathrm{th.}$ ($\mathrm{cm}^{-1}$ )}} &  {$E_\mathrm{exp.}-E_\mathrm{th.}$ } \\ 
    & & & & & {($h\cdot\mathrm{MHz}$)} \\\hline \\ 
 \endfirsthead

 {{\tablename\ \thetable{} (\textit{continued})}} \\
\hline\hline\\\multirow{2}{*}{$f$} & \multirow{2}{*}{$i$}& {\multirow{2}{*}{$\nu_{f\leftarrow i}$ (MHz)}} & {\multirow{2}{*}{$\tilde\nu_\mathrm{exp.}$ ($\mathrm{cm}^{-1}$ )}}& {\multirow{2}{*}{$\tilde\nu_\mathrm{th.}$ ($\mathrm{cm}^{-1}$ )}} &  {$E_\mathrm{exp.}-E_\mathrm{th.}$ }\\ 
    & & & & & {($h\cdot\mathrm{MHz}$)} \\\hline \\ 
 \endhead

\hline
 \endfoot

 \hline
 \endlastfoot

$6s46d\,^3\!D_3$& $6s45d\,^3\!D_2$& 85831.61(10)& 50384.462457& 50384.462422&1.05\\
$6s47d\,^3\!D_3$& $6s46d\,^3\!D_2$& 80078.09(10)& 50387.080168& 50387.080212&-1.32\\
$6s48d\,^3\!D_3$& $6s47d\,^3\!D_2$& 74831.41(10)& 50389.526471& 50389.526449&0.66\\
$6s49d\,^3\!D_3$& $6s48d\,^3\!D_2$& 70028.11(10)& 50391.815708& 50391.815801&-2.79\\
$6s50d\,^3\!D_3$& $6s49d\,^3\!D_2$& 65632.47(10)& 50393.961388& 50393.961403&-0.44\\
$6s52d\,^3\!D_3$& $6s51d\,^3\!D_2$& 57884.10(10)& 50397.867372& 50397.867324&1.44\\

\hline\hline
\end{longtable*}

\clearpage
\subsection{$^{174}$Yb $^{1,3}D_2$}

\begin{longtable*}{@{\extracolsep{\fill}}c S[table-format=2.5]S[table-format=2.3]S[table-format=2.5]S[table-format=2.5]}
\caption{Experimental ($\alpha_{0,\mathrm{exp.}}$) and predicted ($\alpha_{0,\mathrm{theo.}}$) static dipole polarizabilities of $6snd\,^{1}D_2$ Rydberg states of $^{174}$Yb for $m_J = 0, 1$, scaled by $\nu_{6s}^7$. The experimental values for $\alpha_{0,\mathrm{exp.}}$ are obtained by measurement of quadratic Stark shifts. Prediction for the static dipole polarizability $\alpha_{0,\mathrm{theo.}}$ are obtained from a non-perturbative diagonalization of the atomic Hamiltonian including electric fields \cite{zimmerman1979stark}.}\label{tab:174Spec_Pol_1D2} \\

\hline\hline\\
 [-6pt]{\multirow{3}{*}{State label}} &  \multicolumn{2}{c}{$m_J=0$} & \multicolumn{2}{c}{$m_J=1$} \\ 
 \cline{2-3} \cline{4-5}
 \rule{0pt}{3.0ex}& \multicolumn{1}{c}{$\alpha_{0,\mathrm{exp.}}/\nu_{6s}^7$} & \multicolumn{1}{c}{$\alpha_{0,\mathrm{theo.}}/\nu_{6s}^7$} & \multicolumn{1}{c}{$\alpha_{0,\mathrm{exp.}}/\nu_{6s}^7$} & \multicolumn{1}{c}{$\alpha_{0,\mathrm{theo.}}/\nu_{6s}^7$} \\
  & {$\left(h\cdot\mathrm{mHz}/\left(\mathrm{V}/\mathrm{cm}\right)^2\right)$} &{$\left(h\cdot\mathrm{mHz}/\left(\mathrm{V}/\mathrm{cm}\right)^2\right)$} & {$\left(h\cdot\mathrm{mHz}/\left(\mathrm{V}/\mathrm{cm}\right)^2\right)$} & {$\left(h\cdot\mathrm{mHz}/\left(\mathrm{V}/\mathrm{cm}\right)^2\right)$}\\[3pt]
\hline\\ [-5pt] 
 \endfirsthead

 {{\tablename\ \thetable{} (\textit{continued})}} \\
\hline\hline\\
 [-6pt]{\multirow{3}{*}{State label}} &  \multicolumn{2}{c}{$m_J=0$} & \multicolumn{2}{c}{$m_J=1$} \\ 
 \cline{2-5}
 & \multicolumn{1}{c}{$\alpha_{0,\mathrm{exp.}}$} & \multicolumn{1}{c}{$\alpha_{0,\mathrm{theo.}}$} & \multicolumn{1}{c}{$\alpha_{0,\mathrm{exp.}}$} & \multicolumn{1}{c}{$\alpha_{0,\mathrm{theo.}}$} \\
  & {$\left(h\cdot\mathrm{mHz}/\left(\mathrm{V}/\mathrm{cm}\right)^2\right)$} &{$\left(h\cdot\mathrm{mHz}/\left(\mathrm{V}/\mathrm{cm}\right)^2\right)$} & {$\left(h\cdot\mathrm{mHz}/\left(\mathrm{V}/\mathrm{cm}\right)^2\right)$} & {$\left(h\cdot\mathrm{mHz}/\left(\mathrm{V}/\mathrm{cm}\right)^2\right)$}\\[3pt]
\hline\\ [-5pt] 
 \endhead

\hline
 \endfoot

 \hline
 \endlastfoot

$6s40d\,^1D_2$& -0.82(6)& -0.826& -0.58(5)& -0.612\\
$6s41d\,^1D_2$& -0.81(10)& -0.823& -0.61(7)& -0.610\\
$6s42d\,^1D_2$& -0.81(2)& -0.820& -0.59(2)& -0.608\\
$6s43d\,^1D_2$& -0.79(4)& -0.816& -0.59(5)& -0.605\\
$6s44d\,^1D_2$& {\textemdash} & {\textemdash} & -0.61(3)& -0.603\\
$6s45d\,^1D_2$& -0.77(3)& -0.809& -0.60(2)& -0.600\\
$6s47d\,^1D_2$& -0.80(1)& -0.799& -0.60(3)& -0.593\\
$6s49d\,^1D_2$& -0.76(7)& -0.789& -0.60(8)& -0.585\\
\hline\hline
\end{longtable*}

\begin{longtable*}{@{\extracolsep{\fill}}c S[table-format=2.5]S[table-format=2.5]S[table-format=2.5]S[table-format=2.5]}
\caption{Experimental ($\alpha_{0,\mathrm{exp.}}$) and predicted ($\alpha_{0,\mathrm{theo.}}$) static dipole polarizabilities of $6snd\,^{3}D_2$ Rydberg states of $^{174}$Yb for $m_J = 0, 2$, scaled by $\nu_{6s}^7$. The experimental values for $\alpha_{0,\mathrm{exp.}}$ are obtained by measurement of quadratic Stark shifts. Prediction for the static dipole polarizability $\alpha_{0,\mathrm{theo.}}$ are obtained from a non-perturbative diagonalization of the atomic Hamiltonian including electric fields \cite{zimmerman1979stark}.}\label{tab:174Spec_Pol_3D2} \\

\hline\hline\\
 [-6pt]{\multirow{3}{*}{State label}} &  \multicolumn{2}{c}{$m_J=0$} & \multicolumn{2}{c}{$m_J=2$} \\ 
 \cline{2-3}\cline{4-5}
 \rule{0pt}{3.0ex}& \multicolumn{1}{c}{$\alpha_{0,\mathrm{exp.}}/\nu_{6s}^7$} & \multicolumn{1}{c}{$\alpha_{0,\mathrm{theo.}}/\nu_{6s}^7$} & \multicolumn{1}{c}{$\alpha_{0,\mathrm{exp.}}/\nu_{6s}^7$} & \multicolumn{1}{c}{$\alpha_{0,\mathrm{theo.}}/\nu_{6s}^7$} \\
  & {$\left(h\cdot\mathrm{mHz}/\left(\mathrm{V}/\mathrm{cm}\right)^2\right)$} &{$\left(h\cdot\mathrm{mHz}/\left(\mathrm{V}/\mathrm{cm}\right)^2\right)$} & {$\left(h\cdot\mathrm{mHz}/\left(\mathrm{V}/\mathrm{cm}\right)^2\right)$} & {$\left(h\cdot\mathrm{mHz}/\left(\mathrm{V}/\mathrm{cm}\right)^2\right)$}\\[3pt]
\hline\\ [-5pt] 
 \endfirsthead

 {{\tablename\ \thetable{} (\textit{continued})}} \\
\hline\hline\\
 [-6pt]{\multirow{3}{*}{State label}} &  \multicolumn{2}{c}{$m_J=0$} & \multicolumn{2}{c}{$m_J=2$} \\ 
 \cline{2-3}\cline{4-5}
 & \multicolumn{1}{c}{$\alpha_{0,\mathrm{exp.}}$} & \multicolumn{1}{c}{$\alpha_{0,\mathrm{theo.}}$} & \multicolumn{1}{c}{$\alpha_{0,\mathrm{exp.}}$} & \multicolumn{1}{c}{$\alpha_{0,\mathrm{theo.}}$} \\
  & {$\left(h\cdot\mathrm{mHz}/\left(\mathrm{V}/\mathrm{cm}\right)^2\right)$} &{$\left(h\cdot\mathrm{mHz}/\left(\mathrm{V}/\mathrm{cm}\right)^2\right)$} & {$\left(h\cdot\mathrm{mHz}/\left(\mathrm{V}/\mathrm{cm}\right)^2\right)$} & {$\left(h\cdot\mathrm{mHz}/\left(\mathrm{V}/\mathrm{cm}\right)^2\right)$}\\[3pt]
\hline\\ [-5pt] 
 \endhead

\hline
 \endfoot

 \hline
 \endlastfoot

$6s50d\,^3D_2$& 0.50(2)& 0.505& -0.60(2)& -0.617\\
$6s53d\,^3D_2$& 0.82(1)& 0.879 & -0.64(9)& -0.618\\
$6s56d\,^3D_2$& 1.20(12)& 1.356& -0.63(6)& -0.618\\
$6s59d\,^3D_2$& 1.87(10)& 1.981& -0.55(3)& -0.618\\
$6s62d\,^3D_2$& 2.83(21)& 2.836& -0.54(3)& -0.618\\
\hline\hline
\end{longtable*}

\clearpage

\subsection{$^{171}$Yb $P$ and $F$, $F=3/2$}

\begin{longtable*}{@{\extracolsep{\fill}}S[table-format=2.7] S[table-format=2.7] S[table-format=7.4] S[table-format=5.6]S[table-format=5.6]S[table-format=2.2] c}
\caption{\label{tab:171Spec_PF32} Spectroscopic data of the odd-parity $F=3/2$ Rydberg series obtained from microwave spectroscopy in an atomic beam setup described in Sec.~\ref{sec:expsetup} and from Ref.~\cite{peper2025spectroscopy}. The transitions are from $i = \ket{\nu, S, F=1/2}$ ($^a$) and $i = \ket{\nu, D, F=5/2}$ ($^b$).}\\

\hline\hline\\
[-6pt]{\multirow{2}{*}{$\nu_{F_c=1}^f$} }& {\multirow{2}{*}{$\nu_{F_c=1}^i$}}& {\multirow{2}{*}{$\nu_{f\leftarrow i}$ (MHz)}} & {\multirow{2}{*}{$\tilde\nu_\mathrm{exp.}$ ($\mathrm{cm}^{-1}$ )}}& {\multirow{2}{*}{$\tilde\nu_\mathrm{th.}$ ($\mathrm{cm}^{-1}$ )}} & {$E_\mathrm{exp.}-E_\mathrm{th.}$} & {\multirow{2}{*}{Ref.} } \\ 
&&&&& {($h\cdot\mathrm{MHz}$)}&\\[3pt]
\hline\\ [-5pt]
 \endfirsthead

 {{\tablename\ \thetable{} (\textit{continued})}} \\
 \hline\hline\\
[-6pt]{\multirow{2}{*}{$\nu_{F_c=1}^f$} }& {\multirow{2}{*}{$\nu_{F_c=1}^i$}}& {\multirow{2}{*}{$\nu_{f\leftarrow i}$ (MHz)}} & {\multirow{2}{*}{$\tilde\nu_\mathrm{exp.}$ ($\mathrm{cm}^{-1}$ )}}& {\multirow{2}{*}{$\tilde\nu_\mathrm{th.}$ ($\mathrm{cm}^{-1}$ )}} & {$E_\mathrm{exp.}-E_\mathrm{th.}$} & {\multirow{2}{*}{Ref.} } \\ 
&&&&& {($h\cdot\mathrm{MHz}$)}&\\[3pt]
\hline\\ [-5pt]
 \endhead

\hline
 \endfoot

\hline\\
 \endlastfoot

28.0308208& 27.7206600 $^a$& 94221.66(10)& 50303.554600& 50303.554509&2.73&\cite{peper2025spectroscopy}\\
28.0712201& 27.7206600 $^a$& 106260.55(10)& 50303.956173& 50303.956217&-1.31&\cite{peper2025spectroscopy}\\
29.8585903& 29.7179208 $^a$& 35019.09(10)& 50320.129965& 50320.129880&2.55&\cite{peper2025spectroscopy}\\
30.8406323& 30.7166005 $^a$& 27991.62(10)& 50327.843979& 50327.843904&2.26&\cite{peper2025spectroscopy}\\
31.0158421& 30.7166005 $^a$& 66958.10(10)& 50329.143761& 50329.143722&1.16&\cite{peper2025spectroscopy}\\
31.0684240& 30.7166005 $^a$& 78522.04(10)& 50329.529492& 50329.529525&-0.98&\cite{peper2025spectroscopy}\\
32.8005057& 33.7127811 $^a$& -163250.19(10)& 50341.219493& 50341.219442&1.54&\cite{peper2025spectroscopy}\\
33.0043773& 32.7140335 $^a$& 53847.84(10)& 50342.475682& 50342.475656&0.77&\cite{peper2025spectroscopy}\\
\hline\hline
\end{longtable*}

\clearpage
\begin{longtable*}{@{\extracolsep{\fill}}S[table-format=3.7]  S[table-format=1.3] S[table-format=1.3]c}
	\caption{\label{tab:171Spec_PF32_gfactors} Predicted $g_\mathrm{th}$ and experimental $g_\mathrm{exp}$ factors of odd-parity $F = 3/2$ Rydberg states obtained from microwave spectroscopy in an atomic beam setup described in Sec.~\ref{sec:expsetup} and from Ref.~\cite{peper2025spectroscopy}.}\\
	
	\hline\hline\\
	[-6pt]{$\nu_{F_c=1}$} & {$g_\mathrm{exp}$}   & {$g_\mathrm{th}$} & {Ref.}   \\[3pt]
	\hline\\ [-5pt] 
	\endfirsthead
	
	{{\tablename\ \thetable{} (\textit{continued})}} \\
	\hline\hline \\
	[-6pt]{$\nu_{F_c=1}$} & {$g_\mathrm{exp}$}  & {$g_\mathrm{th}$} & {Ref.}  \\[3pt]
	\hline\\ [-5pt] 
	\endhead
	
	\hline
	\endfoot
	
	\hline
	\endlastfoot

28.0308208& 1.05(2)&1.037& \cite{peper2025spectroscopy}\\
28.0712201& 1.48(2)&1.461& \cite{peper2025spectroscopy}\\
29.8585903& 0.96(2)&0.954& \cite{peper2025spectroscopy}\\
30.8406323& 0.97(2)&0.953& \cite{peper2025spectroscopy}\\
31.0158421& 1.12(2)&1.115& \cite{peper2025spectroscopy}\\
31.0684240& 1.39(2)&1.380& \cite{peper2025spectroscopy}\\
33.0043773& 1.13(10)&1.159& This work\\
33.7782931& 0.97(2)&0.951& \cite{peper2025spectroscopy}\\
\hline\hline
\end{longtable*}

\clearpage

\subsection{$^{171}$Yb $P$ and $F$, $F=5/2$}

\begin{longtable*}{@{\extracolsep{\fill}}S[table-format=2.7] S[table-format=2.7] S[table-format=7.4] S[table-format=5.6]S[table-format=5.6]S[table-format=2.2]}
\caption{\label{tab:171Spec_PF52} Spectroscopic data of the odd-parity $F=5/2$ Rydberg series obtained from microwave spectroscopy in an atomic beam setup described in Sec.~\ref{sec:expsetup}. The transitions are from $i = \ket{\nu, D, F=3/2}$ ($^a$) and $i = \ket{\nu, D, F=5/2}$ ($^b$).}\\

\hline\hline\\
[-6pt]{\multirow{2}{*}{$\nu_{F_c=1}^f$} }& {\multirow{2}{*}{$\nu_{F_c=1}^i$}}& {\multirow{2}{*}{$\nu_{f\leftarrow i}$ (MHz)}} & {\multirow{2}{*}{$\tilde\nu_\mathrm{exp.}$ ($\mathrm{cm}^{-1}$ )}}& {\multirow{2}{*}{$\tilde\nu_\mathrm{th.}$ ($\mathrm{cm}^{-1}$ )}} & {$E_\mathrm{exp.}-E_\mathrm{th.}$} \\ 
&&&&& {($h\cdot\mathrm{MHz}$)}\\[3pt]
\hline\\ [-5pt]
 \endfirsthead

 {{\tablename\ \thetable{} (\textit{continued})}} \\
 \hline\hline\\
[-6pt]{\multirow{2}{*}{$\nu_{F_c=1}^f$} }& {\multirow{2}{*}{$\nu_{F_c=1}^i$}}& {\multirow{2}{*}{$\nu_{f\leftarrow i}$ (MHz)}} & {\multirow{2}{*}{$\tilde\nu_\mathrm{exp.}$ ($\mathrm{cm}^{-1}$ )}}& {\multirow{2}{*}{$\tilde\nu_\mathrm{th.}$ ($\mathrm{cm}^{-1}$ )}} & {$E_\mathrm{exp.}-E_\mathrm{th.}$}  \\ 
&&&&& {($h\cdot\mathrm{MHz}$)}\\[3pt]
\hline\\ [-5pt]
 \endhead

\hline
 \endfoot

\hline\\
 \endlastfoot

29.7236865& 30.2830951 $^b$& -136300.49(10)& 50319.010053& 50319.010054&-0.04\\
30.8948730& 31.2828519 $^b$& -84962.57(10)& 50328.248683& 50328.248661&0.65\\
30.8948730& 31.2873408 $^a$& -85927.51(10)& 50328.248671& 50328.248661&0.30\\
30.9406098& 31.2828519 $^b$& -74783.86(10)& 50328.588208& 50328.588306&-2.95\\
31.8909357& 32.2826465 $^b$& -78022.77(10)& 50335.318251& 50335.318243&0.24\\
31.9396221& 32.2826465 $^b$& -68170.47(10)& 50335.646888& 50335.646939&-1.53\\
32.7142362& 33.2824703 $^b$& -104067.63(10)& 50340.680805& 50340.680782&0.68\\
32.8868658& 33.2824703 $^b$& -71881.43(10)& 50341.754421& 50341.754427&-0.18\\
\hline\hline
\end{longtable*}

\begin{longtable*}{@{\extracolsep{\fill}}S[table-format=3.7]  S[table-format=1.3] S[table-format=1.3]}
	\caption{\label{tab:171Spec_PF52_gfactors} Predicted $g_\mathrm{th}$ and experimental $g_\mathrm{exp}$ factors of odd-parity $F = 5/2$ Rydberg states obtained from microwave spectroscopy in an atomic beam setup described in Sec.~\ref{sec:expsetup}.}\\
	
	\hline\hline\\
	[-6pt]{$\nu_{F_c=1}$} & {$g_\mathrm{exp}$}   & {$g_\mathrm{th}$}    \\[3pt]
	\hline\\ [-5pt] 
	\endfirsthead
	
	{{\tablename\ \thetable{} (\textit{continued})}} \\
	\hline\hline \\
	[-6pt]{$\nu_{F_c=1}$} & {$g_\mathrm{exp}$}  & {$g_\mathrm{th}$}   \\[3pt]
	\hline\\ [-5pt] 
	\endhead
	
	\hline
	\endfoot
	
	\hline
	\endlastfoot

37.9365633& 1.052(6)&1.059\\
40.0792597& 1.194(3)&1.200\\
40.8508844& 0.761(2)&0.764\\
41.6750499& 1.089(2)&1.096\\
41.9356366& 1.044(2)&1.043\\
43.6621221& 1.078(10)&1.081\\
43.8370847& 0.793(10)&0.794\\
43.9353117& 1.033(3)&1.038\\
46.6384759& 1.048(9)&1.055\\
47.8200539& 0.833(10)&0.838\\
49.9346386& 1.043(10)&1.027\\
51.5857859& 1.007(3)&1.008\\
51.8054911& 0.876(2)&0.882\\
55.5305637& 0.978(2)&0.974\\
55.7935270& 0.917(10)&0.920\\
62.4052777& 0.932(7)&0.930\\
62.7770959& 0.977(10)&0.968\\
62.9338935& 1.027(6)&1.015\\
73.0758403& 1.186(13)&1.200\\
73.1365777& 0.885(6)&0.897\\
74.0757528& 1.173(4)&1.200\\
74.1082585& 0.890(10)&0.897\\
75.0750680& 1.125(11)&1.155\\
75.0800529& 0.937(3)&0.941\\
76.0758916& 1.203(4)&1.199\\
77.0757889& 1.200(10)&1.200\\

	\hline\hline
\end{longtable*}

\clearpage

\subsection{$^{171}$Yb $F$, $F=7/2$}

\begin{longtable*}{@{\extracolsep{\fill}}S[table-format=2.7] S[table-format=2.7] S[table-format=7.4] S[table-format=5.6]S[table-format=5.6]S[table-format=2.2]}
\caption{\label{tab:171Spec_F72} Spectroscopic data of the $\ket{\nu, \ell = 3, F=7/2}$ Rydberg series obtained from microwave spectroscopy in an atomic beam setup described in Sec.~\ref{sec:expsetup}. The transitions are from $i = \ket{\nu, D, F=5/2}$ states.}\\

\hline\hline\\
[-6pt]{\multirow{2}{*}{$\nu_{F_c=1}^f$} }& {\multirow{2}{*}{$\nu_{F_c=1}^i$}}& {\multirow{2}{*}{$\nu_{f\leftarrow i}$ (MHz)}} & {\multirow{2}{*}{$\tilde\nu_\mathrm{exp.}$ ($\mathrm{cm}^{-1}$ )}}& {\multirow{2}{*}{$\tilde\nu_\mathrm{th.}$ ($\mathrm{cm}^{-1}$ )}} & {$E_\mathrm{exp.}-E_\mathrm{th.}$}\\ 
&&&&& {($h\cdot\mathrm{MHz}$)}\\[3pt]
\hline\\ [-5pt]
 \endfirsthead

 {{\tablename\ \thetable{} (\textit{continued})}} \\
 \hline\hline\\
[-6pt]{\multirow{2}{*}{$\nu_{F_c=1}^f$} }& {\multirow{2}{*}{$\nu_{F_c=1}^i$}}& {\multirow{2}{*}{$\nu_{f\leftarrow i}$ (MHz)}} & {\multirow{2}{*}{$\tilde\nu_\mathrm{exp.}$ ($\mathrm{cm}^{-1}$ )}}& {\multirow{2}{*}{$\tilde\nu_\mathrm{th.}$ ($\mathrm{cm}^{-1}$ )}} & {$E_\mathrm{exp.}-E_\mathrm{th.}$}  \\ 
&&&&& {($h\cdot\mathrm{MHz}$)}\\[3pt]
\hline\\ [-5pt]
 \endhead

\hline
 \endfoot

\hline\\
 \endlastfoot

30.8850837& 31.2828519& -87145.34(10)& 50328.175873& 50328.175768&3.16\\
30.9429059& 31.2828519& -74269.75(10)& 50328.605357& 50328.605317&1.18\\
31.8813744& 32.2826465& -79965.79(10)& 50335.253439& 50335.253514&-2.27\\
31.9416448& 32.2826465& -67759.09(10)& 50335.660610& 50335.660563&1.42\\
32.7115967& 33.2824703& -104564.56(10)& 50340.664229& 50340.664234&-0.15\\
32.8775585& 33.2824703& -73606.58(10)& 50341.696876& 50341.696973&-2.91\\
32.9406224& 32.2826465& 124849.84(10)& 50342.085352& 50342.085317&1.06\\
33.7080475& 34.2823173& -96190.18(10)& 50346.637805& 50346.637814&-0.28\\
\hline\hline
\end{longtable*}

\begin{longtable*}{@{\extracolsep{\fill}}S[table-format=3.7]  S[table-format=1.3] S[table-format=1.3]}
	\caption{\label{tab:171Spec_F72_gfactors} Predicted $g_\mathrm{th}$ and experimental $g_\mathrm{exp}$ factors of $\ket{\nu, \ell = 3, F = 7/2}$ Rydberg states obtained from microwave spectroscopy in an atomic beam setup described in Sec.~\ref{sec:expsetup}.}\\
	
	\hline\hline\\
	[-6pt]{$\nu_{F_c=1}$} & {$g_\mathrm{exp}$}   & {$g_\mathrm{th}$}    \\[3pt]
	\hline\\ [-5pt] 
	\endfirsthead
	
	{{\tablename\ \thetable{} (\textit{continued})}} \\
	\hline\hline \\
	[-6pt]{$\nu_{F_c=1}$} & {$g_\mathrm{exp}$}  & {$g_\mathrm{th}$}   \\[3pt]
	\hline\\ [-5pt] 
	\endhead
	
	\hline
	\endfoot
	
	\hline
	\endlastfoot

32.7115967& 0.844(6)&0.868\\
33.7080475& 0.864(9)&0.871\\
37.8571681& 1.295(10)&1.265\\
37.9374692& 1.013(4)&1.024\\
41.6696716& 0.886(7)&0.909\\
41.9361033& 1.032(4)&1.035\\
45.6399140& 0.940(6)&0.942\\
45.8241623& 1.172(2)&1.189\\
45.9352055& 1.028(3)&1.043\\
49.8100502& 1.136(3)&1.147\\
49.9345732& 1.037(3)&1.049\\
54.5352667& 1.014(4)&1.022\\
54.7956343& 1.098(4)&1.098\\
54.9340036& 1.046(5)&1.055\\

	\hline\hline
\end{longtable*}

\clearpage

\subsection{$^{171}$Yb $G$, $F=5/2$}

\begin{longtable*}{@{\extracolsep{\fill}}S[table-format=2.7] S[table-format=2.7] S[table-format=7.4] S[table-format=5.6]S[table-format=5.6]S[table-format=2.2]}
\caption{\label{tab:171Spec_G52} Spectroscopic data of the $\ket{\nu, \ell = 4, F=5/2}$ Rydberg series obtained from microwave spectroscopy in an atomic beam setup described in Sec.~\ref{sec:expsetup}. The transitions are two-step: from a $i = \ket{\nu, D, F=5/2}$ state and via an intermediate $\ket{\nu, F, F=7/2}$ state.}\\

\hline\hline\\
[-6pt]{\multirow{2}{*}{$\nu_{F_c=1}^f$} }& {\multirow{2}{*}{$\nu_{F_c=1}^i$}}& {\multirow{2}{*}{$\nu_{f\leftarrow i}$ (MHz)}} & {\multirow{2}{*}{$\tilde\nu_\mathrm{exp.}$ ($\mathrm{cm}^{-1}$ )}}& {\multirow{2}{*}{$\tilde\nu_\mathrm{th.}$ ($\mathrm{cm}^{-1}$ )}} & {$E_\mathrm{exp.}-E_\mathrm{th.}$}\\ 
&&&&& {($h\cdot\mathrm{MHz}$)}\\[3pt]
\hline\\ [-5pt]
 \endfirsthead

 {{\tablename\ \thetable{} (\textit{continued})}} \\
 \hline\hline\\
[-6pt]{\multirow{2}{*}{$\nu_{F_c=1}^f$} }& {\multirow{2}{*}{$\nu_{F_c=1}^i$}}& {\multirow{2}{*}{$\nu_{f\leftarrow i}$ (MHz)}} & {\multirow{2}{*}{$\tilde\nu_\mathrm{exp.}$ ($\mathrm{cm}^{-1}$ )}}& {\multirow{2}{*}{$\tilde\nu_\mathrm{th.}$ ($\mathrm{cm}^{-1}$ )}} & {$E_\mathrm{exp.}-E_\mathrm{th.}$}  \\ 
&&&&& {($h\cdot\mathrm{MHz}$)}\\[3pt]
\hline\\ [-5pt]
 \endhead

\hline
 \endfoot

\hline\\
 \endlastfoot

34.9739907& 36.2820652& -190437.84(10)& 50353.503011& 50353.503007&0.13\\
37.9739744& 39.2817828& -149380.70(10)& 50367.118156& 50367.118160&-0.13\\
40.9739616& 42.2815767& -119329.68(10)& 50377.853730& 50377.853730&-0.01\\

 \hline\hline
\end{longtable*}

\clearpage

\subsection{$^{171}$Yb $G$, $F=7/2$}

\begin{longtable*}{@{\extracolsep{\fill}}S[table-format=2.7] S[table-format=2.7] S[table-format=7.4] S[table-format=5.6]S[table-format=5.6]S[table-format=2.2]}
\caption{\label{tab:171Spec_G72} Spectroscopic data of the $\ket{\nu, \ell = 4, F=7/2}$ Rydberg series obtained from microwave spectroscopy in an atomic beam setup described in Sec.~\ref{sec:expsetup}. The transitions are two-step: from a $i = \ket{\nu, D, F=5/2}$ state and via an intermediate $\ket{\nu, F, F=7/2}$ state.}\\

\hline\hline\\
[-6pt]{\multirow{2}{*}{$\nu_{F_c=1}^f$} }& {\multirow{2}{*}{$\nu_{F_c=1}^i$}}& {\multirow{2}{*}{$\nu_{f\leftarrow i}$ (MHz)}} & {\multirow{2}{*}{$\tilde\nu_\mathrm{exp.}$ ($\mathrm{cm}^{-1}$ )}}& {\multirow{2}{*}{$\tilde\nu_\mathrm{th.}$ ($\mathrm{cm}^{-1}$ )}} & {$E_\mathrm{exp.}-E_\mathrm{th.}$}\\ 
&&&&& {($h\cdot\mathrm{MHz}$)}\\[3pt]
\hline\\ [-5pt]
 \endfirsthead

 {{\tablename\ \thetable{} (\textit{continued})}} \\
 \hline\hline\\
[-6pt]{\multirow{2}{*}{$\nu_{F_c=1}^f$} }& {\multirow{2}{*}{$\nu_{F_c=1}^i$}}& {\multirow{2}{*}{$\nu_{f\leftarrow i}$ (MHz)}} & {\multirow{2}{*}{$\tilde\nu_\mathrm{exp.}$ ($\mathrm{cm}^{-1}$ )}}& {\multirow{2}{*}{$\tilde\nu_\mathrm{th.}$ ($\mathrm{cm}^{-1}$ )}} & {$E_\mathrm{exp.}-E_\mathrm{th.}$}  \\ 
&&&&& {($h\cdot\mathrm{MHz}$)}\\[3pt]
\hline\\ [-5pt]
 \endhead

\hline
 \endfoot

\hline\\
 \endlastfoot

34.8924287& 36.2820652& -203026.45(10)& 50353.083100& 50353.083097&0.10\\
34.9738315& 36.2820652& -190462.24(10)& 50353.502197& 50353.502190&0.21\\
34.9742708& 36.2820652& -190394.75(10)& 50353.504449& 50353.504444&0.14\\
37.8695406& 39.2817828& -161981.08(10)& 50366.697853& 50366.697859&-0.21\\
37.9738150& 39.2817828& -149399.95(10)& 50367.117514& 50367.117521&-0.23\\
37.9742550& 39.2817828& -149346.93(10)& 50367.119282& 50367.119285&-0.08\\
40.9738020& 42.2815767& -119345.14(10)& 50377.853214& 50377.853221&-0.20\\
40.9742426& 42.2815767& -119302.90(10)& 50377.854623& 50377.854627&-0.11\\
\hline\hline
\end{longtable*}

\clearpage

\subsection{$^{171}$Yb $G$, $F=9/2$}

\begin{longtable*}{@{\extracolsep{\fill}}S[table-format=2.7] S[table-format=2.7] S[table-format=7.4] S[table-format=5.6]S[table-format=5.6]S[table-format=2.2]}
\caption{\label{tab:171Spec_G92} Spectroscopic data of the $\ket{\nu, \ell = 4, F=9/2}$ Rydberg series obtained from microwave spectroscopy in an atomic beam setup described in Sec.~\ref{sec:expsetup}. The transitions are two-step: from a $i = \ket{\nu, D, F=5/2}$ state and via an intermediate $\ket{\nu, F, F=7/2}$ state.}\\

\hline\hline\\
[-6pt]{\multirow{2}{*}{$\nu_{F_c=1}^f$} }& {\multirow{2}{*}{$\nu_{F_c=1}^i$}}& {\multirow{2}{*}{$\nu_{f\leftarrow i}$ (MHz)}} & {\multirow{2}{*}{$\tilde\nu_\mathrm{exp.}$ ($\mathrm{cm}^{-1}$ )}}& {\multirow{2}{*}{$\tilde\nu_\mathrm{th.}$ ($\mathrm{cm}^{-1}$ )}} & {$E_\mathrm{exp.}-E_\mathrm{th.}$}\\ 
&&&&& {($h\cdot\mathrm{MHz}$)}\\[3pt]
\hline\\ [-5pt]
 \endfirsthead

 {{\tablename\ \thetable{} (\textit{continued})}} \\
 \hline\hline\\
[-6pt]{\multirow{2}{*}{$\nu_{F_c=1}^f$} }& {\multirow{2}{*}{$\nu_{F_c=1}^i$}}& {\multirow{2}{*}{$\nu_{f\leftarrow i}$ (MHz)}} & {\multirow{2}{*}{$\tilde\nu_\mathrm{exp.}$ ($\mathrm{cm}^{-1}$ )}}& {\multirow{2}{*}{$\tilde\nu_\mathrm{th.}$ ($\mathrm{cm}^{-1}$ )}} & {$E_\mathrm{exp.}-E_\mathrm{th.}$}  \\ 
&&&&& {($h\cdot\mathrm{MHz}$)}\\[3pt]
\hline\\ [-5pt]
 \endhead

\hline
 \endfoot

\hline\\
 \endlastfoot

34.8924458& 36.2820652& -203023.95(10)& 50353.083184& 50353.083185&-0.05\\
34.9742630& 36.2820652& -190396.26(10)& 50353.504398& 50353.504404&-0.17\\
34.9746256& 36.2820652& -190340.14(10)& 50353.506270& 50353.506264&0.19\\
37.8695523& 39.2817828& -161979.41(10)& 50366.697908& 50366.697907&0.05\\
37.9742421& 39.2817828& -149348.46(10)& 50367.119231& 50367.119233&-0.06\\
37.9746110& 39.2817828& -149304.28(10)& 50367.120705& 50367.120712&-0.20\\
40.9742257& 42.2815767& -119304.20(10)& 50377.854580& 50377.854573&0.21\\
40.9745995& 42.2815767& -119268.84(10)& 50377.855759& 50377.855765&-0.18\\
\hline\hline
\end{longtable*}

\clearpage

\subsection{$^{171}$Yb $D$, $F=3/2$}

\begin{longtable*}{@{\extracolsep{\fill}}c S[table-format=2.5]S[table-format=2.5]S[table-format=2.5]S[table-format=2.5]}
\caption{Experimental ($\alpha_{0,\mathrm{exp.}}$) and predicted ($\alpha_{0,\mathrm{theo.}}$) static dipole polarizabilities of $\ket{\nu, \ell = 2, F=3/2}$ Rydberg states of $^{171}$Yb, scaled by $\nu_{F_c=1}^7$. The experimental values for $\alpha_{0,\mathrm{exp.}}$ are obtained by measurement of quadratic Stark shifts. Prediction for the static dipole polarizability $\alpha_{0,\mathrm{theo.}}$ are obtained from a non-perturbative diagonalization of the atomic Hamiltonian including electric fields. }\label{tab:171Spec_Pol_D32} \\

\hline\hline\\
 [-6pt]{\multirow{3}{*}{$\nu_{F_c=1}$}} &  \multicolumn{2}{c}{$m_F=1/2$} & \multicolumn{2}{c}{$m_F=3/2$} \\ 
 \cline{2-3}\cline{4-5}
 \rule{0pt}{3.0ex}& \multicolumn{1}{c}{$\alpha_{0,\mathrm{exp.}}/\nu_{F_c=1}^7$} & \multicolumn{1}{c}{$\alpha_{0,\mathrm{theo.}}/\nu_{F_c=1}^7$} & \multicolumn{1}{c}{$\alpha_{0,\mathrm{exp.}}/\nu_{F_c=1}^7$} & \multicolumn{1}{c}{$\alpha_{0,\mathrm{theo.}}/\nu_{F_c=1}^7$} \\
  & {$\left(h\cdot\mathrm{mHz}/\left(\mathrm{V}/\mathrm{cm}\right)^2\right)$} &{$\left(h\cdot\mathrm{mHz}/\left(\mathrm{V}/\mathrm{cm}\right)^2\right)$} & {$\left(h\cdot\mathrm{mHz}/\left(\mathrm{V}/\mathrm{cm}\right)^2\right)$} & {$\left(h\cdot\mathrm{mHz}/\left(\mathrm{V}/\mathrm{cm}\right)^2\right)$}\\[3pt]
\hline\\ [-5pt] 
 \endfirsthead

 {{\tablename\ \thetable{} (\textit{continued})}} \\
\hline\hline\\
 [-6pt]{\multirow{3}{*}{$\nu_{F_c=1}$}} &  \multicolumn{2}{c}{$m_F=1/2$} & \multicolumn{2}{c}{$m_F=3/2$} \\ 
 \cline{2-3}\cline{4-5}
 & \multicolumn{1}{c}{$\alpha_{0,\mathrm{exp.}}$} & \multicolumn{1}{c}{$\alpha_{0,\mathrm{theo.}}$} & \multicolumn{1}{c}{$\alpha_{0,\mathrm{exp.}}$} & \multicolumn{1}{c}{$\alpha_{0,\mathrm{theo.}}$} \\
  & {$\left(h\cdot\mathrm{mHz}/\left(\mathrm{V}/\mathrm{cm}\right)^2\right)$} &{$\left(h\cdot\mathrm{mHz}/\left(\mathrm{V}/\mathrm{cm}\right)^2\right)$} & {$\left(h\cdot\mathrm{mHz}/\left(\mathrm{V}/\mathrm{cm}\right)^2\right)$} & {$\left(h\cdot\mathrm{mHz}/\left(\mathrm{V}/\mathrm{cm}\right)^2\right)$}\\[3pt]
\hline\\ [-5pt] 
 \endhead

\hline
 \endfoot

 \hline
 \endlastfoot

29.2877938& -0.70(8) & -0.722 & {\textemdash} & {\textemdash} \\ 
31.2873408& -0.70(10)& -0.709& -0.16(5)& -0.133\\
33.2870187& -0.67(10)& -0.689& -0.14(4)& -0.135\\
35.2867706& -0.67(10)& -0.661& -0.16(10)& -0.137\\
39.2864065& -0.55(2)& -0.555& -0.19(10)& -0.138\\
40.2863344& -0.50(10)& -0.505& -0.17(10)& -0.138\\
41.2862680& -0.45(3)& -0.435& -0.19(10)& -0.138\\
42.2862067& -0.33(10)& -0.328& -0.17(3)& -0.137\\
\hline\hline
\end{longtable*}

\clearpage

\subsection{$^{171}$Yb $D$, $F=5/2$}

\begin{longtable*}{@{\extracolsep{\fill}}c S[table-format=2.5]S[table-format=2.5]S[table-format=2.5]S[table-format=2.5]S[table-format=2.5]S[table-format=2.5]}
\caption{Experimental ($\alpha_{0,\mathrm{exp.}}$) and predicted ($\alpha_{0,\mathrm{theo.}}$) static dipole polarizabilities of $\ket{\nu, \ell = 2, F=5/2}$ Rydberg states of $^{171}$Yb, scaled by $\nu_{F_c=1}^7$. The experimental values for $\alpha_{0,\mathrm{exp.}}$ are obtained by measurement of quadratic Stark shifts. Prediction for the static dipole polarizability $\alpha_{0,\mathrm{theo.}}$ are obtained from a non-perturbative diagonalization of the atomic Hamiltonian including electric fields. }\label{tab:171Spec_Pol_D52} \\

\hline\hline\\
 [-6pt]{\multirow{3}{*}{$\nu_{F_c=1}$}} &  \multicolumn{2}{c}{$m_F=1/2$} & \multicolumn{2}{c}{$m_F=3/2$} & \multicolumn{2}{c}{$m_F=5/2$}\\ 
 \cline{2-3}\cline{4-5}\cline{6-7}
 \rule{0pt}{3.0ex}& \multicolumn{1}{c}{$\alpha_{0,\mathrm{exp.}}/\nu_{F_c=1}^7$} & \multicolumn{1}{c}{$\alpha_{0,\mathrm{theo.}}/\nu_{F_c=1}^7$} & \multicolumn{1}{c}{$\alpha_{0,\mathrm{exp.}}/\nu_{F_c=1}^7$} & \multicolumn{1}{c}{$\alpha_{0,\mathrm{theo.}}/\nu_{F_c=1}^7$} & \multicolumn{1}{c}{$\alpha_{0,\mathrm{exp.}}/\nu_{F_c=1}^7$} & \multicolumn{1}{c}{$\alpha_{0,\mathrm{theo.}}/\nu_{F_c=1}^7$}\\
  & \multicolumn{2}{c}{$\left(h\cdot\mathrm{mHz}/\left(\mathrm{V}/\mathrm{cm}\right)^2\right)$} &\multicolumn{2}{c}{$\left(h\cdot\mathrm{mHz}/\left(\mathrm{V}/\mathrm{cm}\right)^2\right)$} & \multicolumn{2}{c}{$\left(h\cdot\mathrm{mHz}/\left(\mathrm{V}/\mathrm{cm}\right)^2\right)$}\\[3pt]
\hline\\ [-5pt] 
 \endfirsthead

 {{\tablename\ \thetable{} (\textit{continued})}} \\
\hline\hline\\
 [-6pt]{\multirow{3}{*}{$\nu_{F_c=1}$}} &  \multicolumn{2}{c}{$m_F=1/2$} & \multicolumn{2}{c}{$m_F=3/2$} & \multicolumn{2}{c}{$m_F=5/2$}\\ 
 \cline{2-3}\cline{4-5}
 & \multicolumn{1}{c}{$\alpha_{0,\mathrm{exp.}}$} & \multicolumn{1}{c}{$\alpha_{0,\mathrm{theo.}}$} & \multicolumn{1}{c}{$\alpha_{0,\mathrm{exp.}}$} & \multicolumn{1}{c}{$\alpha_{0,\mathrm{theo.}}$} & \multicolumn{1}{c}{$\alpha_{0,\mathrm{exp.}}$} & \multicolumn{1}{c}{$\alpha_{0,\mathrm{theo.}}$}\\
  & \multicolumn{2}{c}{$\left(h\cdot\mathrm{mHz}/\left(\mathrm{V}/\mathrm{cm}\right)^2\right)$} &\multicolumn{2}{c}{$\left(h\cdot\mathrm{mHz}/\left(\mathrm{V}/\mathrm{cm}\right)^2\right)$} & \multicolumn{2}{c}{$\left(h\cdot\mathrm{mHz}/\left(\mathrm{V}/\mathrm{cm}\right)^2\right)$}\\[3pt]
\hline\\ [-5pt] 
 \endhead

\hline
 \endfoot

 \hline
 \endlastfoot

27.2840767& -0.78(9)& -0.858 & {\textemdash} & {\textemdash} & {\textemdash} & {\textemdash}\\
30.2830568& -0.89(10)& -0.880 & -0.56(3)& -0.616 & {\textemdash} & {\textemdash}\\
33.2824685& -0.87(10)& -0.894& -0.63(3)& -0.630& -0.104(4)& -0.100\\
36.2820774& -0.94(3)& -0.903& -0.62(3)& -0.638& -0.11(10)& -0.110\\
39.2818018& -0.90(2)& -0.906& -0.64(3)& -0.643& -0.13(2)& -0.116\\
42.2815999& -0.90(2)& -0.906& -0.59(5)& -0.644& -0.13(4)& -0.121\\
45.2814471&  -0.91(2)& -0.901& -0.66(5)& -0.642& -0.14(6)& -0.123\\
48.2813279& -0.88(10)& -0.891& -0.62(5)& -0.635 & {\textemdash} & {\textemdash}\\
\hline\hline
\end{longtable*}

\clearpage

\end{document}